\documentclass[onecolumn]{aa}
\usepackage[varg]{txfonts}
\newif\ifgrayscale
\grayscaletrue 

\newif\ifincludeplots
\includeplotstrue 

\usepackage{graphicx,natbib,url,twoopt}

\usepackage{breqn}
\usepackage{morefloats}
\newcommand{\eg}{{\it e.g.}}

\newcommand{\ie}{{\it i.e.}}
\newcommand{\au}{\,\mathrm{au}}
\newcommand{\myr}{\,\mathrm{Myr}}
\newcommand{\gpcmc}{\,\mathrm{g \; cm^{-3}}}
\newcommand{\tdeg}{^{\circ}}
\newcommand{\pv}{p_V}
\newcommand{\x}{ \; \mathrm{x} \;}
\newcommand{\mps}{\,\mathrm{m\;s^{-1}}}
\newcommand{\FYnospace}{{1993~FY$_{12}$}}

\begin{document}

\title{Size-dependent modification of asteroid family Yarkovsky V-shapes}

\author{
B. T. Bolin\inst{1,2,3}
\and A. Morbidelli\inst{1}
\and K. J. Walsh\inst{4}
}

\institute{Laboratoire Lagrange, Universit\'e C\^ote d'Azur, Observatoire de la C\^ote
d'Azur, CNRS, Blvd. de l'Observatoire, CS 34229, 06304 Nice cedex 4,
France \email{[bryce.bolin;morby;marco.delbo]@oca.eu}
\and Department of Astronomy, University of Washington, 3910 15th Ave NE, Seattle, WA 98195 
\and B612 Asteroid Institute,  20 Sunnyside Ave,  Suite 427, Mill
Valley, CA, 94941, United States 
\and Southwest Research Institute, 1050 Walnut St. Suite 300, Boulder, CO 80302, United States \email{kwalsh@boulder.swri.edu}}


\begin{abstract}
{The thermal properties of the surfaces of asteroids determine the magnitude of the drift rate cause by the Yarkovsky force. In the general case of Main Belt asteroids, the Yarkovsky force is indirectly proportional to the thermal inertia, $\Gamma$. } {Following the proposed relationship between $\Gamma$ and asteroid diameter $D$, we find that asteroids' Yarkovsky drift rates might have a more complex size dependence than previous thought, leading to a curved family V-shape boundary in semi-major axis, a, vs. 1/$D$ space. This implies that asteroids are drifting faster at larger sizes than previously considered decreasing on average the known ages of asteroid families.} {The V-Shape curvature is determined for $>$25 families located throughout the Main Belt to quantify the Yarkovsky size-dependent drift rate.} {We find that there is no correlation between family age and V-shape curvature. In addition, the V-shape curvature decreases for asteroid families with larger heliocentric distances suggesting that the relationship between $\Gamma$ and $D$ is weaker in the outer MB possibly due to homogenous surface roughness among family members.} {}
\end{abstract}


\maketitle


\section{Introduction}
\label{s.Introduction}
Imaging of small scale surface features on the 17 km asteroid Eros during the NEAR-Shoemaker mission revealed that it was covered mostly with mm-sized grains \citep[][]{Veverka2001} whereas cm-sized grains where observed on the surface of the 350 m asteroid Itokawa during the Hayabusa mission \citep[][]{Yano2006}. Thermal inertia values determined from thermal modeling of mid-infrared observations of asteroids \citep[][]{Delbo2015}, combined with the regolith model of \citet[][]{Gundlach2013} confirmed the sizes of the surface regolith of Eros and Itokawa observed by spacecraft missions \citep[][]{Mueller2012a, Muller2014a}. For particles larger than several hundred $\mu$m, $\Gamma$ increases with surface particle size because heat transfer within a grain is much more efficient than the heat transfer by radiation and gas diffusion among grains \citep[][]{Gundlach2012,Delbo2015}. As a result, coarse surface regolith found on small asteroids is a better conductor of heat compared to the finer surface regolith found on larger asteroids.

One explanation for the difference between the corse regolith of small asteroids and the fine regolith of large asteroids is that larger asteroids have more gravity and are able to retain more fine dust during disruption events compared to smaller asteroids \citep[][]{Michel2015}. Additionally, larger asteroids have a longer collisional lifetime compared to smaller asteroids \citep[][]{Farinella1998, Bottke2005b} and as a result have longer surface ages allowing more time for fine regolith production resulting from comminution or thermal cracking of coarse regolith into more fine regolith \citep[][]{Horz1997,Delbo2014}. As a result of the retention of fine regolith and the relatively poor thermal conductivity of fine regolith compared to coarse regolith, large asteroids have lower $\Gamma$ values compared to smaller asteroids. In turn, $\Gamma$ is a determining factor in an asteroid's Yarkovsky drift rate \citep[][]{Delbo2007, Vokrouhlicky2015}.

\subsection{Yarkovsky drift of family fragments}
\label{s.YarkoRateandVShape}
\subsubsection{Yarkovsky drift rate of a single asteroid}
The Yarkovsky force causes the modification of an asteroid's semi-major axis, $a$, eccentricity $e$ and inclination $i$ \citep[][]{Vokrouhlicky2015}. The effect of the Yarkovsky force on an asteroid's $e$ and $i$ are indistinguishable compared to perturbative effects on $e$ and $i$ whereas the effect of the Yarkovsky force on an asteroid's $a$ is distinct on secular timescales \citep[][]{Bottke2000,Spitale2002}. The Yarkovsky force has a secular effect in evolving $e$ in cases where asteroids are in a mean motion resonances (MMR) with Jupiter such as the population of Hilda asteroids that are in a 3:2 MMR with Jupiter as discussed in8 \citet[][]{Bottke2002b} and \citet[][]{Milani2016}, but we focus on the general case where an asteroid is not in a MMR and the Yarkovsky force causes secular evolution only in $a$.

The Yarkovsky force has diurnal and seasonal components, but the seasonal component has a much smaller effect on $a$ than the diurnal component, so we assume the following form for the orbit-averaged $\frac{da}{dt}$ from \citep[][]{Rubincam1995, Farinella1998, Vokrouhlicky1999a}. 
\begin{equation}
\label{dadt}
\left ( \frac{da}{dt} \right ) \; = \; - \frac{8}{9} \; \frac{1-A \;  \Phi}{n} W \left ( R_f, \Theta_f \right ) \mathrm{cos} \; \gamma
\end{equation}
where $A$ is the bond albedo defined by \citep[][]{Bowell1988} and $\Phi \; = \; \pi R^2 F(a) / (mc)$, where $R$ is the radius of the asteroid, $F(a)$ is the solar flux at semi-major axis $a$, equal to $\frac{F_{\mathrm{1au}}}{r^2}$ ($F_{\mathrm{1au}} \; = \; 1360$ $\frac{\mathrm{W}}{\mathrm{m}^2}$, $r$ is the heliocentric distance of the object), $m$ is the mass of the asteroid, $c$ is the speed of light, $\gamma$ is the obliquity of the asteroid and $f$ is the rotation frequency of the asteroid.

The thermal parameter $W$ from  is defined as
\begin{equation}
\label{W}
W \left ( R_f, \Theta_f \right ) \; = \;  - \frac{k_1(R_f)\Theta_f}{1 + 2k_2(R_f)\Theta_f + k_3(R_f)\Theta_f^2}
\end{equation}
where $R_f$ is equal to $R/l_f$, $l_f \; = \; \sqrt{K / \rho C f}$ and $K$ is the surface conductivity of the asteroid, $C$ is the surface heat capacity of the asteroid, $\rho$ is the surface density. Moreover, $\Theta \; = \; \Gamma \sqrt{f} / (\epsilon \sigma T^3_*)$, $\epsilon$ is the surface thermal emissivity, $\sigma$ is the Stefan-Botlzmann constant, $T_*$, the instantaneous sub-solar temperature in thermal equilibrium, which is equal to $\sqrt[4]{(1 - A) F(a)/\epsilon \sigma} $

\subsubsection{Yarkovsky drift modification caused by $D$ dependence of $\Gamma$}
\label{s.YarkodriftMod}
\begin{figure}
\centering
\hspace*{-1.0cm}
\ifincludeplots
\includegraphics[scale=0.55]{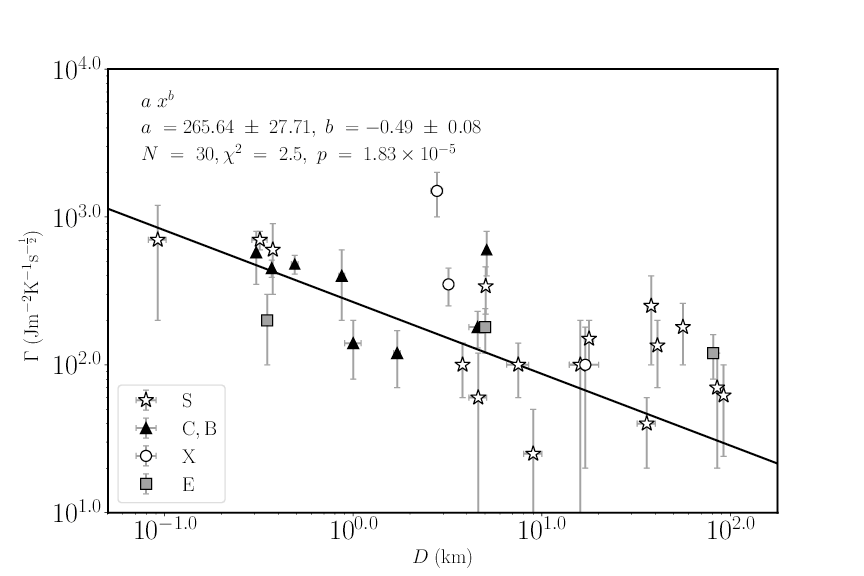}
\else
I am not enabling plots.
\fi
\caption{$D$ vs.$\Gamma$ for Near Earth and Main Belt Asteroids of S, C, B, X and E types and dynamical classes with $D <$ 100 km. The data are fit to the function $y \; = \; a \; x^b$ shown as the dark line using orthogonal distance regression \citep[][]{Boggs1990}. Measurements of $D$ and $\Gamma$ are taken from \citet[][]{Delbo2003, Lamy2008, Delbo2009, Masiero2011, Muller2011, Wolters2011, Marchis2012, Muller2012, Muller2013, Emery2014, AliLagoa2014, Muller2014, Rozitis2014a,Hanus2015, Naidu2015, Hanus2016a}.}
\label{fig.DvsGammaDlt100AllTax}
\end{figure}

\begin{figure}
\centering
\hspace*{-1.0cm}
\ifincludeplots
\includegraphics[scale=0.55]{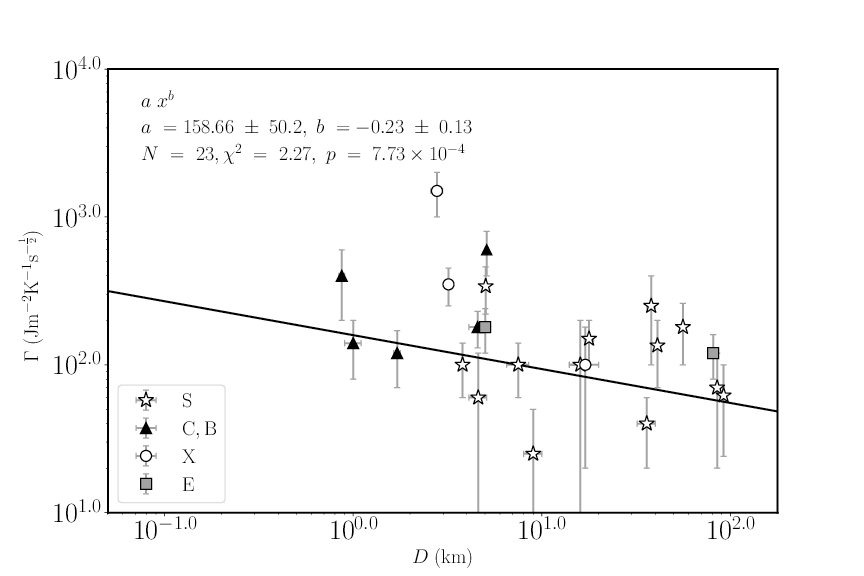}
\else
I am not enabling plots.
\fi
\caption{The same as Fig.~\ref{fig.DvsGammaDlt100AllTax} for Near Earth and Main Belt Asteroids of S, C, B, X and E types and dynamical classes with 0.5 km $< \; D < $ 100 km. Measurements of $D$ and $\Gamma$ are taken from \citet[][]{Delbo2003, Lamy2008, Delbo2009, Masiero2011, Muller2011, Wolters2011, Marchis2012, Rozitis2014a,Hanus2015, Naidu2015, Hanus2016a}.}
\label{fig.DvsGammaD0.5lt100AllTax}
\end{figure}

Eq.~\ref{W} can be re-written assuming $k_1, \; k_2, \; k_3 \; = \; 0.5$ for asteroids with $D$ on the km scale and larger as seen in Fig.~2 of \citet[][]{Peterson1976,Vokrouhlicky1998c,Vokrouhlicky1999a}
\begin{equation}
\label{WLarge}
W \simeq W \left (\Theta_f \right ) \; = \;  - \frac{0.5\Theta_f}{1 + \Theta_f + 0.5\Theta_f^2}
\end{equation}
Approximating Eqs.~\ref{dadt} and \ref{WLarge} for asteroids with identical $A$, $F(a)$, $n$, $\gamma$, $a$, $r$, $f$, $\epsilon$ and $\Theta_f \gg 1$
\begin{equation}
\label{dadtproptoTheta}
\left ( \frac{da}{dt} \right ) \; \propto \; \frac{1}{D\Theta_f}
\end{equation}

where $\Theta_f$$\propto \Gamma$. While 1 $\lesssim$ $\Theta_f$ $\lesssim$ 2, for Near Earth Asteroids \citep[][]{Greenberg2017a}, $\Theta_f \gg 1$ holds true in general for Main Belt asteroids with $D$ $<$ 40 km that have an average thermal inertia $>$ 100 J m$^{-2}$ s$^{-0.5}$ K$^{-1}$ \citep[][]{Delbo2015}, and where rotation frequencies of km-scale asteroids are typically  $f \; \simeq \; 1\times 10^{-4}$ \citep{Pravec2002}. Combining $\Theta_f$$\propto \Gamma$ with Eq.~\ref{dadtproptoTheta}
\begin{equation}
\label{dadtproptoGamma}
\left ( \frac{da}{dt} \right ) \; \propto \; \frac{1}{D\Gamma}
\end{equation}
Recent $\Gamma$ measurements for MBAs and NEOs with $D \; <$ 100 km suggests that $D$ is related to $\Gamma$ by the relationship $\Gamma \; = \; a \; D^{b}$ where $a \simeq 265$ $\mathrm{\frac{J}{m^2 s^{0.5} K}}$ and $b \simeq -0.50\pm0.08$ \citep[][]{Delbo2015} as seen in Fig.~\ref{fig.DvsGammaDlt100AllTax}. The slope of the correlation $b$ increases to $\sim-0.2\pm0.13$ for MBAs and NEOs in the range 0.5 km $<$ $D$ $<$ 100 km, which is the $D$ range of currently observable asteroids in the Main Belt \citep[][]{Jedicke2015} as seen in Fig.~\ref{fig.DvsGammaD0.5lt100AllTax}.

Approximating Eq.~\ref{dadtproptoGamma} with $\Gamma \; = \; D^{b}$ and $\alpha \; = \; b + 1$ we obtain
\begin{equation}
\label{dadtproptoxi}
\left ( \frac{da}{dt} \right ) \; \propto \; D^{-\alpha}
\end{equation}
Notice this gives a curvature to the V-shape of the family in $a$ vs. $D_r$ space if $\alpha \; \ne$ 1.0. Therefore, following the formulation of the Yarkovsky drift rate from \citet[][]{Spoto2015} conbined with Eq.~\ref{dadtproptoxi}, we compute $\frac{da}{dt}$ with the following formula
\begin{equation}
\label{eqn.yarkorate_final}
\frac{da}{dt} (D,\alpha,a,e,N,A) \; =\;  \left ( \frac{da}{dt} \right )_{0} \; \frac{\sqrt{a}_{0}(1-e^2_{0})}{\sqrt{a}(1-e^2)} \; \left ( \frac{D_0}{D} \right )^{\alpha} \; \left ( \frac{\rho_0}{\rho} \right ) \; \left ( \frac{1 - A}{1 - A_{0}} \right )\; \left ( \frac{\au}{\myr} \right ) \; \frac{cos(\theta)}{cos(\theta_0)}
\end{equation} 
where $\left ( \frac{da}{dt} \right )_{0} $ $\sim$ 4.7 $\times \; 10^{-5} \; \frac{\au}{\myr}$ for an asteroid in the inner Main Belt with $a_0 \; = \; $ 2.37 au, $e_0 \; = \; $ 0.2, $D_0$ = 5 km, $\rho_0$  = 2.5 $\gpcmc$, $A_0 \; =$ 0.1, surface conductivity equal to 0.01 $\mathrm{\frac{W}{m}}$ and obliquity, $\theta_0 \; = \; $ 0$\tdeg$ \citep[][]{Bottke2006,Vokrouhlicky2015}. Using the values for $\left ( \frac{da}{dt} \right )_{0} $, $a_0$, $e_0$, $D_0$, $\rho_0$, $A_0$, surface conductivity and $\theta_0$ from \citet[][]{Bottke2006,Vokrouhlicky2015} is appropriate because the value of $\alpha$ is not affected by the values of these variables. For asteroids with 0.5 km $<$ $D$ $<$ 100 km, the value of $\alpha$ that characterizes the curvature of the V-shape of the family should be $\sim$0.8 because $b \sim -0.2$ as indicated by $D$ vs. $\Gamma$ data plotted in Fig.~\ref{fig.DvsGammaD0.5lt100AllTax}. Eq.~\ref{eqn.yarkorate_final} is more appropriate for asteroids at smaller sizes for smaller asteroid families because asteroids at larger sizes that drifted at maximum speed over the entire family age are probably rare and difficult to identify relative to the background. The information in the family V-shape in smaller families is mostly contributed to by smaller asteroids because asteroids at larger sizes that drifted at a maximum drift rate described by Eq.~\ref{eqn.yarkorate_final} over the entire family age are probably rare and difficult to identify relative to the background. Whereas for larger families, asteroids up to a larger size contain most of the information in the V-shape because the increase in the number of larger asteroids results in the leading edge of the family V-shape becoming adequately populated by asteroids that traveled at the maximum drift rate than when compared to V-shapes of smaller asteroid families.

\subsubsection{Yarkovsky V-shapes}
\label{s.YarkoRateandVShape}
As described in \citet[][]{Bolin2017}, asteroid families, whose members' proper elements $e$ and $i$ have become too dispersed due to chaotic diffusion can be identified by searching for correlations in $a$ vs. $\frac{1}{D}$, $H$ space. The size-dependent Yarkovsky force gives a family the V-shape in $a$ vs. $\frac{1}{D}$,$H$ distribution on Myr time-scales. In practice, it is possible for a family to obtain a V-shape on shorter timescales due to the contribution of the initial velocity field \citep[][]{Bolin2017a}.

In the standard case of \citet[][]{Nesvorny2003} and \citet[][]{Vokrouhlicky2006b} the sides of the V-shape in $a$ vs. $\frac{1}{D}$ space is
\begin{equation}
\label{eqn.Davsdadt}
\Delta a \; = \; \frac{da}{dt}(D) \; \Delta t
\end{equation}
$\Delta a$ is defined as $a \; - \; a_c$ where $a_c$ is the family center,  $\frac{da}{dt}(D)$ is the size-dependent maximal Yarkovsky semi-major axis drift rate and $\Delta t$ is the age of the family. The drift rate can be recalculated for different bulk and surface densities, orbit, rotation period, obliquity and thermal properties \citep[][]{Bottke2006, Chesley2014, Spoto2015}. We define the drift rate $\frac{da}{dt}(D)$ as formulated in Eq.~\ref{eqn.yarkorate_final}.

The width of the V-shape in $a$ vs. $1/D$ space can be defined by the constant $C$ assuming the case where $\alpha \; = \; 1.0$ is
\begin{equation}
\label{eqn.Cpvdadtvsdeltat}
C \; = \; \Delta t \left ( \sqrt \pv \; \left ( \frac{da}{dt} \right )_0 \right )
\end{equation}
from \citet[][]{Vokrouhlicky2006b} where $\pv$ is the geometric albedo, which is assumed to be the average albedo for family members (an assumption well supported by observations; \citep[][]{Masiero2013}. $\left ( \frac{da}{dt} \right )_0$ is the same as in Eq.~\ref{eqn.yarkorate_final}. Typical $\pv$ values used are 0.05 and 0.15 are used for C- and S-type asteroids, respectively \citep[][]{Masiero2011, Masiero2015a}.

Combining Eqs.~\ref{eqn.Davsdadt}, \ref{eqn.yarkorate_final} and \ref{eqn.Cpvdadtvsdeltat}, we define the border of the V-shape in reciprocal diameter, $\frac{1}{D}$ or $D_r$, space as
\begin{equation}
\label{eqn.apvDvsCmod}
D_r(a,a_c,C_{\alpha},\pv,\alpha) \; = \frac{\sqrt{\pv}}{1329}\; \left(\frac{\left | \Delta a \right |}{ C_{\alpha}}\right)^{\frac{1}{\alpha}}
\end{equation}
where $C_\alpha$ in Eq.~\ref{eqn.apvDvsCmod} is normalized to a value of $C$ with the following factor
\begin{equation}
\label{Cnorm}
C = C_\alpha\left( \frac{\sqrt{pv}}{1329} \right )^{1-\alpha}
\end{equation}
Eq.~\ref{eqn.apvDvsCmod} is re-written in terms of $(a,a_c,C,\pv,\alpha)$ by using Eq.~\ref{Cnorm}
\begin{equation}
\label{eqn.apvDvsCfinal}
D_r(a,a_c,C,\pv,\alpha) \; = \; \left(\frac{\left | \Delta a \right | \sqrt{\pv}}{ 1329 \; C}\right)^{\frac{1}{\alpha}}
\end{equation}

\begin{figure}
\centering
\hspace*{-1.1cm}
\ifincludeplots
\includegraphics[scale=0.55]{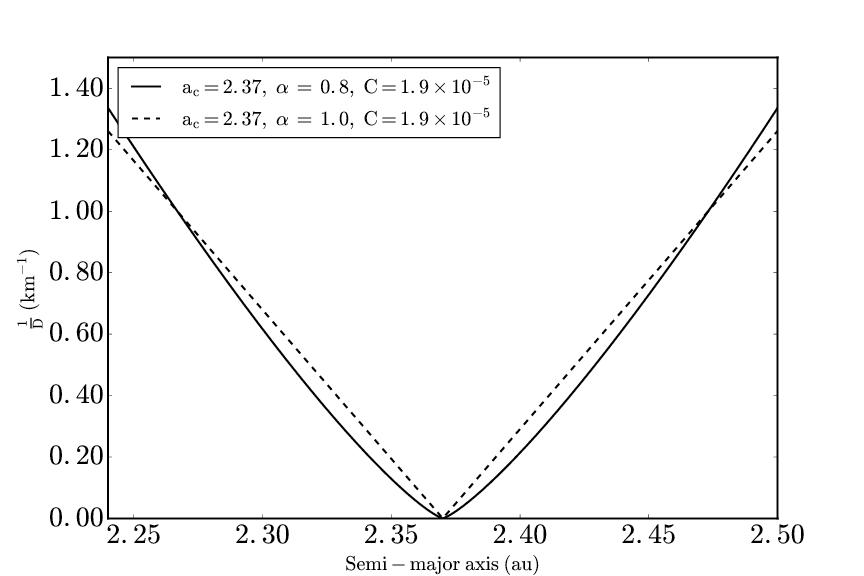}
\else
I am not enabling plots.
\fi
\caption{Two generic family V-shapes with $a_c \; = \; 2.37$ au and $C \; = \; 1.9 \times 10^{-5}$ au and $\alpha \; = \; 0.8$, 1.0.}
\label{fig.GenericVShapes}
\end{figure}

Two generic V-shapes are plotted in Fig.~\ref{fig.GenericVShapes} using Eq.~\ref{eqn.apvDvsCfinal} and 1.0 and $\sim$ 0.8 for the value of $\alpha$ and shows how the curved V-shape line crosses the straight V-shape border at $D_r \; = \; 1.0$.

\subsection{V-shape identification technique and measurement of $\alpha$} 
\label{s.v-shapeidentificationandalpha}
Family V-shapes' $a_c$, $C$ and $\alpha$ are measured according to Eq.~\ref{eqn.apvDvsCfinal} in $a$ vs. $D_r$ space with the V-shape border method described in \citet[][]{Bolin2017,Bolin2017a}. That is, the location of the border of the V-shape described by $a_c$, $C$ and $\alpha$ andÊEq.~\ref{eqn.apvDvsCfinal} are determined by maximizing the ratio of the number of objects inside of a V-shape border, $N_{in}$ to the number of objects located outside the border $N_{out}$, where $N_{in}$ and $N_{out}$ are described with the following equations
\begin{equation} 
\label{eq.border_method_N_inner}
  N_{in}(a_c,C,dC,\pv,\alpha)
    = \; \frac{\Sigma_j 
       \; w(D_j) \;
       \int\limits_{a_1}^{a_2} da
    	\int\limits_{D_r(a, a_c,C, \pv,\alpha)}^{D_r(a, a_c,C_-, \pv,\alpha)} dD_r
       \; \delta(a_{j}-a)
       \; \delta( D_{r,j}-D_r )}{{\int\limits_{a_1}^{a_2}  da
    	\int\limits_{D_r(a, a_c,C, \pv,\alpha)}^{D_r(a, a_c,C, \pv,\alpha)} \; dD_r}}
\end{equation}
\begin{equation} 
\label{eq.border_method_N_outer}
  N_{out}(a_c,C,dC,\pv,\alpha)
    = \; \frac{\Sigma_j 
       \; w(D_j) \;
       \int\limits_{a_1}^{a_2} da
    	\int\limits_{D_r(a, a_c,C_+, \pv,\alpha)}^{D_r(a, a_c,C, \pv,\alpha)} dD_r
       \; \delta(a_{j}-a)
       \; \delta( D_{r,j}-D_r )}{{\int\limits_{a_1}^{a_2}  da
    	\int\limits_{D_r(a, a_c,C, \pv,\alpha)}^{D_r(a, a_c,C, \pv,\alpha)} \; dD_r}}
\end{equation}

Eqs.~\ref{eq.border_method_N_outer} and \ref{eq.border_method_N_inner} are normalized the area in $a$ vs. $D_r$ between the nominal and outer V-shapes defined by $D_r(a, a_c,C_+, \pv,\alpha)$ and $D_r(a, a_c,C, \pv,\alpha)$ in the denominator for Eq.~\ref{eq.border_method_N_outer} and between the nominal and inner V-shapes defined by defined by $D_r(a, a_c,C, \pv,\alpha)$ and $D_r(a, a_c,C_-, \pv,\alpha)$ in the denominator for Eq.~\ref{eq.border_method_N_inner}.

The symbol $\Sigma_j$ in Eqs.\ref{eq.border_method_N_outer} and \ref{eq.border_method_N_inner} indicates summation on the asteroids of the catalog, with semi-major axis $a_j$ and reciprocal diameter $D_{r,j}$. The symbol $\delta$ indicates Dirac's function, and $a_1$ and $a_2$ are the low and high semi-major axis range in which the asteroid catalog is considered.  The function $w(D)$ weighs the right-side portions of Eqs.~\ref{eq.border_method_N_outer} and \ref{eq.border_method_N_inner} by their size so that the location of the V-shape in $a$ vs. $D_r$ space will be weighted towards its larger members. The exponent 2.5 is used for $w(D) = D^{2.5}$, in agreement with the cumulative size distribution of collisionally relaxed populations and with the observed distribution for MBAs in the $H$ range $12\; < \;H \;< \;16$ \citep[][]{Jedicke2002}.

\citet{Walsh2013} found that the borders of the V-shapes of the Eulalia and new Polana  family could be identified by the peak in the ratio $\frac{N_{in}}{N_{out}}$ where $N_{in}$ and  ${N_{out}}$ are the number of asteroids falling between the curves defined by Eq.~\ref{eqn.apvDvsCfinal} for values $C$ and $C_-$ and  $C$ and $C_+$, respectively, with $C_-=C-dC$ and $C_+=C+dC$. We extend our technique to search for a peak in the ratio $\frac{N_{in}^2}{N_{out}}$, which corresponds to weighting the ratio of $\frac{N_{in}}{N_{out}}$ by the value of $N_{in}$. This approach has been shown to provide sharper results \citep[][]{Delbo2017}. We consider only asteroids in the border of the V-shape because the functional form of the V-shape may become distorted interior to the V-shape border due to varying obliquity reorientation rates with asteroid size \citep[][]{Paolicchi2016}.

\begin{figure}
\centering
\hspace*{-0.7cm}
\ifincludeplots
\includegraphics[scale=0.455]{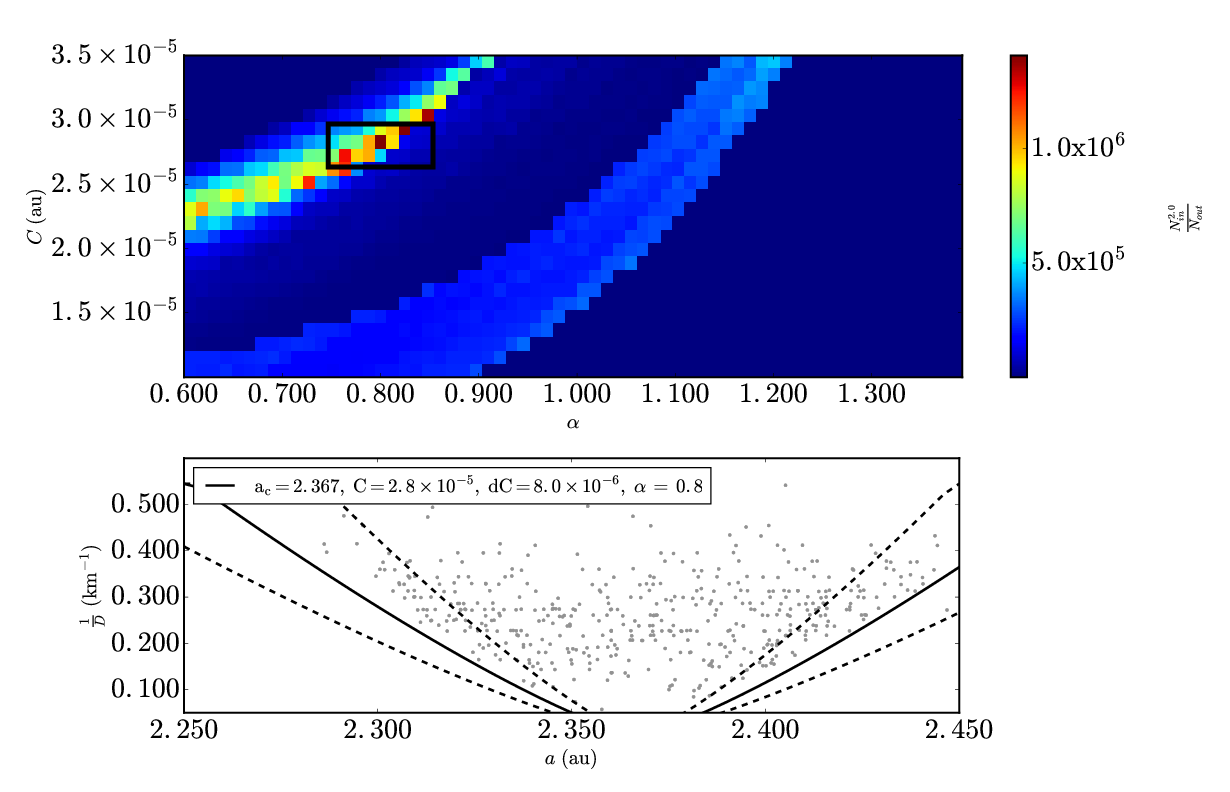}
\else
I am not enabling plots.
\fi
\caption{Application of the V-shape identification to synthetic asteroid family data at Time = 200 Myrs. (Top panel) The ratio of $N_{out}(a_c,C,dC,\pv,\alpha)^2$ to $N_{in}(a_c,C,dC,\pv,\alpha)$ ratio in the $\alpha$-$C$ range, ($a_c\pm \frac{\Delta \alpha}{2}$,$C\pm \frac{\Delta C}{2}$) where $\Delta \alpha$ is equal to $1.2 \times 10^{-2}$ au and $\Delta C$, not to be confused with $dC$, is equal to $1.0 \times 10^{-6}$ au, for the single synthetic family. The box marks the peak value in $\frac{N_{in}(a_c,C,dC,\pv,\alpha)^2}{N_{out}(a_c,C,dC,\pv,\alpha)}$ for the synthetic family V-shape. (Bottom Panel) $D_r(a,a_c,C,\pv, \alpha)$ is plotted for the peak values with the primary V-shape as a solid line where $\pv = 0.05$. The dashed lines mark the boundaries for the area in $a$ vs. $D_r$ space for $N_{in}$ and $N_{out}$ using Eq.~\ref{eqn.apvDvsCfinal},  $D_r(a,a_c,C\pm dC,\pv,\alpha)$ where $a_c$ = 2.367 au and $dC \; = \; 8.0 \x 10^{-6}$ au.}
\label{fig.synErig200Myrs}
\end{figure}

 Whereas \citet[][]{Walsh2013} and \citet[][]{Delbo2017} assumed $\alpha \; = \; 1.0$ and searched for a maximum of $\frac{N_{in}^2}{N_{out}}$ in the space $a_c$, $C$, here we perform the maximum search in three dimensions, in the space $a_c$, $C$ and $\alpha$. The three dimensional search is tested on a synthetic family generated as described in Section~\ref{s.synfamily} producing a peak value in $\frac{N_{in}(a_c,C,dC,\pv,\alpha)^2}{N_{out}(a_c,C,dC,\pv,\alpha)}$ indicated by the black rectangle in the top panel of Fig.~\ref{fig.synErig200Myrs}. For simplicity, in the top panel of the figure we plot the value of the ratio on the $\alpha$, $C$ plane for the value of $a_c$ that maximizes the ratio in each cell. The V-shape that maximizes the ratio is plotted in the bottom panel, together with the asteroids of the family using Eq.~\ref{eqn.apvDvsCfinal} as seen in the bottom panel of Fig.~\ref{fig.synErig200Myrs}.

The value of $dC$ is used similarly as in \citet[][]{Bolin2017} and \citet[][]{Bolin2017a}. The value of dC used depends on the density of asteroids on the family V-shape edge. The value of dC can be a few 10$\%$ of the V-shape's $C$ value if the density of asteroids on the V-shape edge is high such as the case of the Massalia family as discussed in Section~\ref{s.Massalia} and more, up to 40$\sim$50$\%$ if the V-shape edge is more diffuse such as in the case of the Dora(2) sub-family as discussed in Section~\ref{s.Dora2} \citep[][]{Nesvorny2015a}. The inner and outer V-shapes must be wide enough to include enough asteroids in the inner V-shape and measure a $N_{in}^2$ to $N_{out}$ ratio high enough to identify the family V-shape. Only asteroids that belong to the nominal HCM classification of the family are used instead of the full catalogue of asteroids. The nominal family classification can include interlopers that the V-shape technique can include if the value of $dC$ that is used is too large \citep[][]{Nesvorny2015a, Radovic2017a}. Asteroids that fall out of the best fit V-shape with a reasonable value of $dC$ may be true interlopers even if nominal members of the HCM-defined family.

\begin{figure}
\centering
\hspace*{-0.7cm}
\ifincludeplots
\includegraphics[scale=0.455]{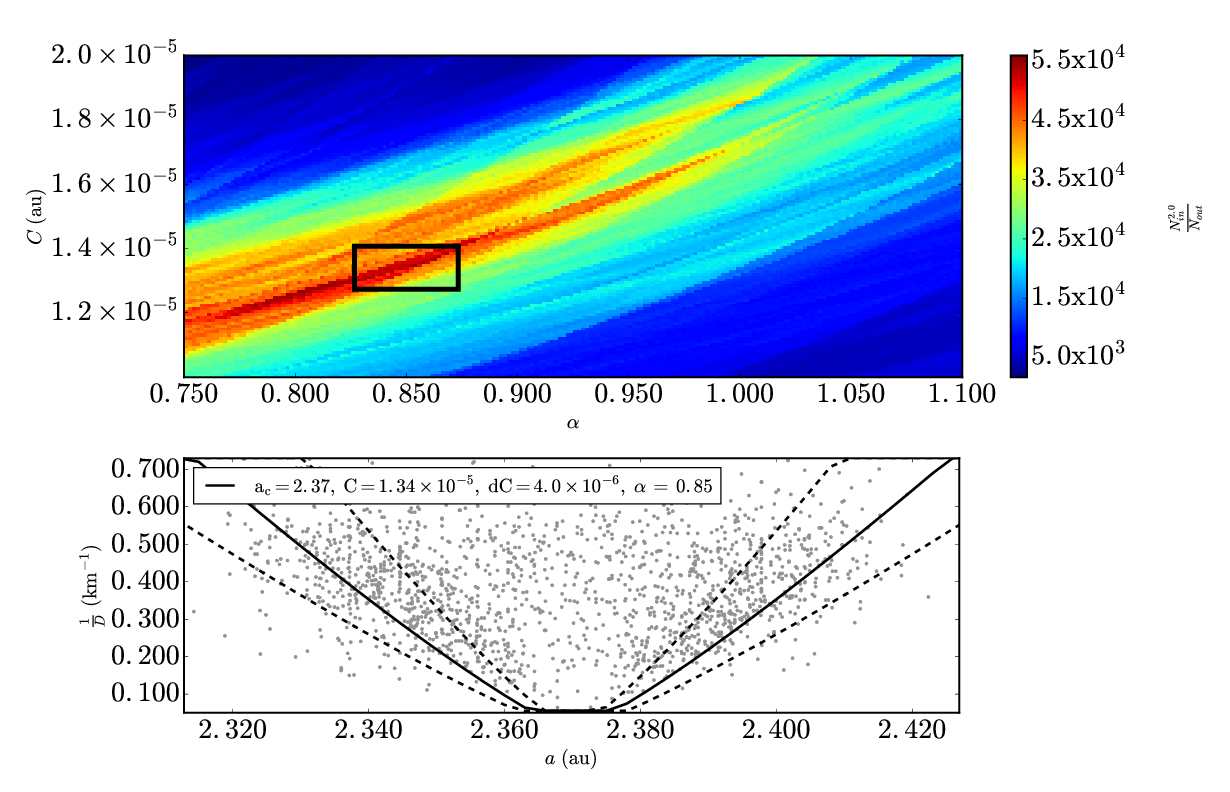}
\else
I am not enabling plots.
\fi
\caption{The same as Fig.~\ref{fig.synErig200Myrs} for the Erigone asteroid family data from \citet[][]{Nesvorny2015a}. (Top panel) $\Delta \alpha$ is equal to $1.8 \times 10^{-3}$ au and $\Delta C$, is equal to $9.0 \times 10^{-8}$ au. (Bottom Panel) $D_r(a,a_c,C\pm dC,\pv,\alpha)$ is plotted with $\pv = 0.05$, $a_c$ = 2.796 au and $dC \; = \; 7.5 \x 10^{-6}$ au.}
\label{fig.ErigoneBorderDensity1}
\end{figure}

\begin{figure}
\centering
\hspace*{-0.7cm}
\ifincludeplots
\includegraphics[scale=0.455]{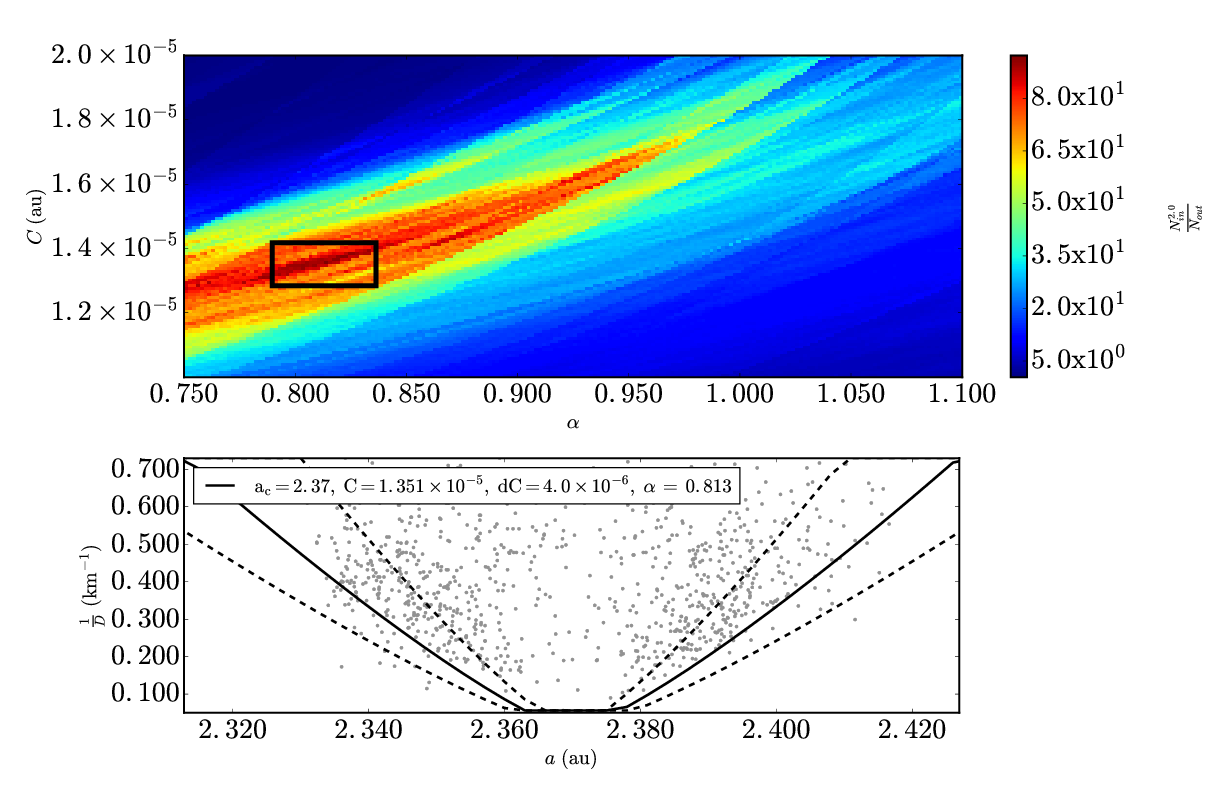}
\else
I am not enabling plots.
\fi
\caption{Same as Fig.~\ref{fig.ErigoneBorderDensity1}, but repeated for the Erigone family defined by \citet[][]{Milani2014}.}
\label{fig.ErigoneBorderDensity1Milani}
\end{figure}

 The V-shape identification technique was tested on families identified by both \citet[][]{Nesvorny2015a} and \citet[][]{Milani2014} such as the Erigone family to verify that the V-shape $a_c$, $C$ and $\alpha$ determination on family membership definitions from either database produces the same results as seen in Figs.~\ref{fig.ErigoneBorderDensity1} and \ref{fig.ErigoneBorderDensity1Milani} and discussed for the Erigone, Massalia, Agnia, and Maria in Sections~\ref{s.erigone}, ~\ref{s.Massalia}, \ref{s.Agnia} and \ref{s.Maria}.

\subsection{Data set and uncertainties of $\alpha$  measurements}
\subsubsection{Data set}
The data used to measure the V-shapes of asteroid families were taken from the Asteroid Dynamic Site\footnote{\tt{http://hamilton.dm.unipi.it/astdys/}} (AstDys) for the $H$ magnitudes. The offset between $H$ magnitudes from the MPC and individually calibrated magnitudes from \citet[][]{Pravec2012} and \citet[][]{Veres2015} is assumed to be constant for objects in the range 12$< \; H \; <$ 18 \citep[][]{Veres2015}. Family definitions were taken from \citet[][]{Nesvorny2015a}. Asteroid family data for the Erigone, Massalia, Agnia, Eunomia, Hoffmeister, Maria and Ursula families were used from both \citet[][]{Milani2014} and \citet[][]{Nesvorny2015a}. Family visual albedo, $\pv$, data from \citet[][]{Masiero2013} and \citet[][]{Spoto2015} were used to calibrate the conversion from $H$ magnitudes to asteroid $D$ using the relation $D = 2.99 \x 10^8 \; \frac{10^{0.2 \; (m_\odot  \; - \; H)}}{\sqrt{\pv}}$ \citep[][]{Bowell1988} where $m_\odot \; = \; -26.76$ \citep[][]{Pravec2007}. Numerically and analytically calculated MBA proper elements were taken from AstDys \citep[][]{Knezevic2003}. Numerically calculated proper elements were used preferentially and analytical proper elements were used for asteroids, that had numerically calculated elements as of September 2017.

\subsubsection{Uncertainty of $\alpha$}
\label{s.montecarlo}
The value of $\alpha$ located where $\frac{N_{in}(a_c,C,dC,\pv,\alpha)^2}{N_{out}(a_c,C,dC,\pv,\alpha)}$ peaks in $\alpha$ vs. $C$ space represents the best estimate of the $\alpha$ of a asteroid family's V-shape using the nominal $a$ and $D_r$ asteroid values. Differences between family members in their physical properties cause a spread in possible $\alpha$ values that measured together increase the uncertainty in the measured value of $\alpha$. Variation between family members' $D$ is directly caused by uncertainy in their $H$ magnitude and $\pv$ measurements. In addition to uncertainty in asteroids' $D$,  lack of complete information about the true population of asteroids within a family and outliers contribution to a family's $a$ vs. $D_r$ distribution can increase the range of uncertainty in $\alpha$ values compatible with the family V-shape. We devise the following Monte Carlo procedure to quantify the variation in $\alpha$ taking into account these affects.

At least 1,200 Monte Carlo trials are completed per family. Some families have significantly more than 1,200 Monte Carlo trials as described in the Appendix if additional CPU time was available. In each trial, the location of the peak $\frac{N_{in}(a_c,C,dC,\pv,\alpha)^2}{N_{out}(a_c,C,dC,\pv,\alpha)}$ value in $\alpha$ vs. $C$ is recorded. Three steps are completed to randomize the asteroid family data from the original $a$ vs. $D_r$ distribution per trial. The first step is to create a resampled data set of family fragments by removing $\sqrt{N}$ objects randomly where $N$ is the number of objects in $a$ vs. $D_r$ space in order to include variations caused by incomplete knowledge of the asteroid family's population. Incompleteness of asteroid family fragments increases for smaller fragments and is more pronounced in the middle and outer portions of the main belt \citep[][]{Jedicke1998,Jedicke2002}, and the increased incompleteness and greater number of smaller Main Belt asteroids in the asteroid family catalogues causes the variation in $\alpha$ to be weighted towards smaller fragments than larger fragments.

A second step is taken to determine the variation caused by incomplete information in the family fragment population by resampling the fragments' $a$  by their own $a$ distribution. In this step, a continuous distribution interpolating the $a$ values of family fragments per $D_r$ bin is generated and used to reassign the fragments new $a$ values. The bin size of $D_r$ used is 0.001 km$^{-1}$ for all asteroid families.

The third step is to randomize the measurements of $H$ and $\pv$ of the asteroids by their known uncertainties. Asteroid $H$ values were varied per Monte Carlo run by adding a random value between -0.25 and 0.25 magnitude equal to the known uncertainties for $H$ values from the MPC catalogue \citep[][]{Oszkiewicz2011, Pravec2012}. Asteroid fragments' $H$ were converted to $D$ after their $H$ values are randomized using the following equation
\begin{equation}
\label{eq.HtoD}
D = 2.99 \x 10^8 \; \frac{10^{0.2 \; (m_\odot  \; - \; H)}}{\sqrt{\pv}}
 \end{equation}
 from \citet[][]{Harris2002}, and a value of $\pv$ chosen at random for each asteroid using central values and uncertainties per asteroid family from \citet[][]{Masiero2013} and \citet[][]{Spoto2015}.

\begin{figure}
\centering
\hspace*{-0.9cm}
\ifincludeplots
\includegraphics[scale=0.3225]{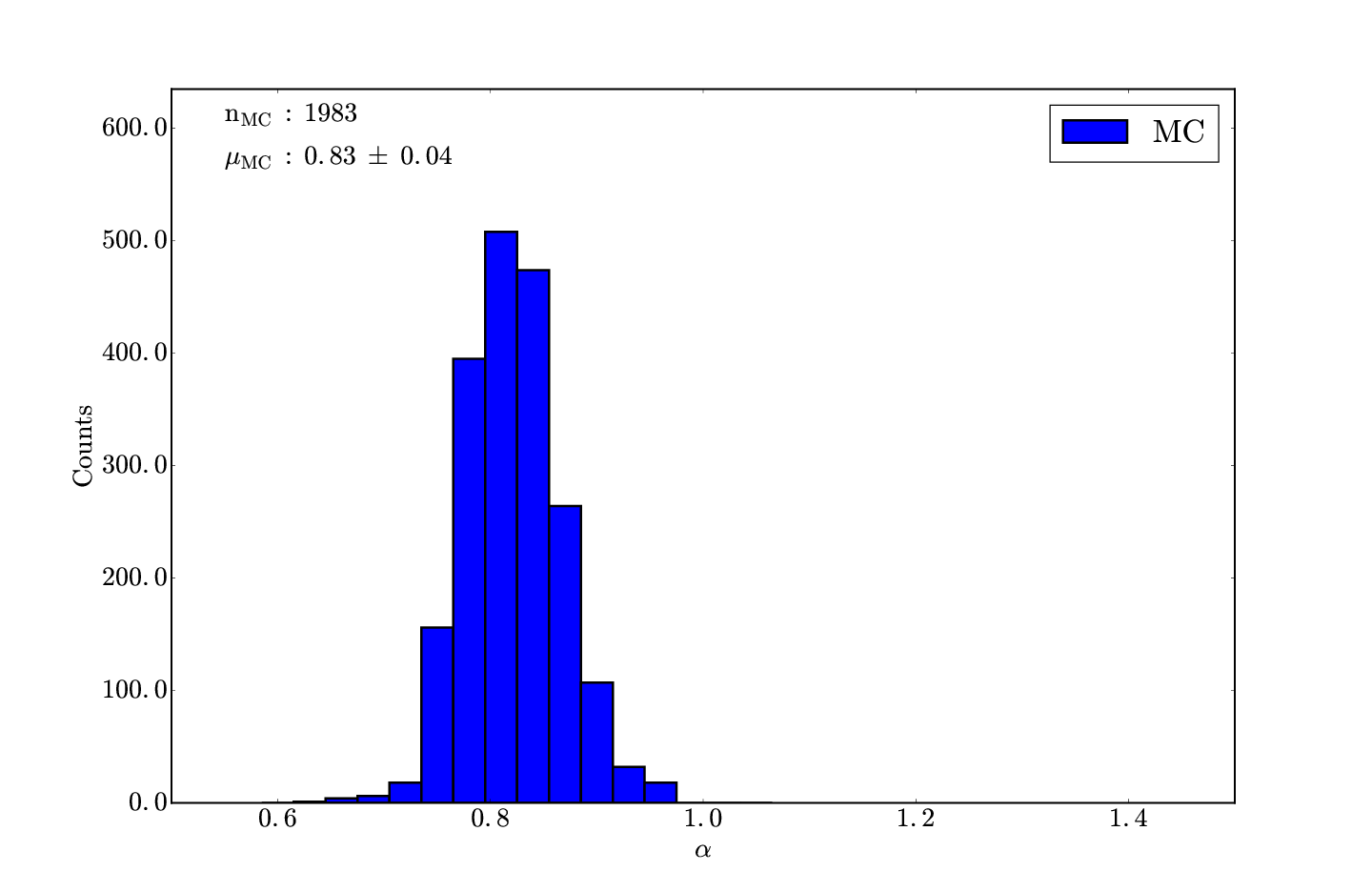}
\else
I am not enabling plots.
\fi
\caption{Histogram of $\alpha$ located at the peak value of $N_{out}(a_c,C,dC,\pv,\alpha)^2$ to $N_{in}(a_c,C,dC,\pv,\alpha)$ in each of the $\sim$2,000 trials repeating the V-shape technique for the Erigone family. The mean of the distribution is centered at $\alpha$ = 0.83 $\pm$ 0.04 and the bin size in the histogram is 0.03.}
\label{fig.ErigoneMC}
\end{figure}

\begin{figure}
\centering
\hspace*{-0.9cm}
\ifincludeplots
\includegraphics[scale=0.3225]{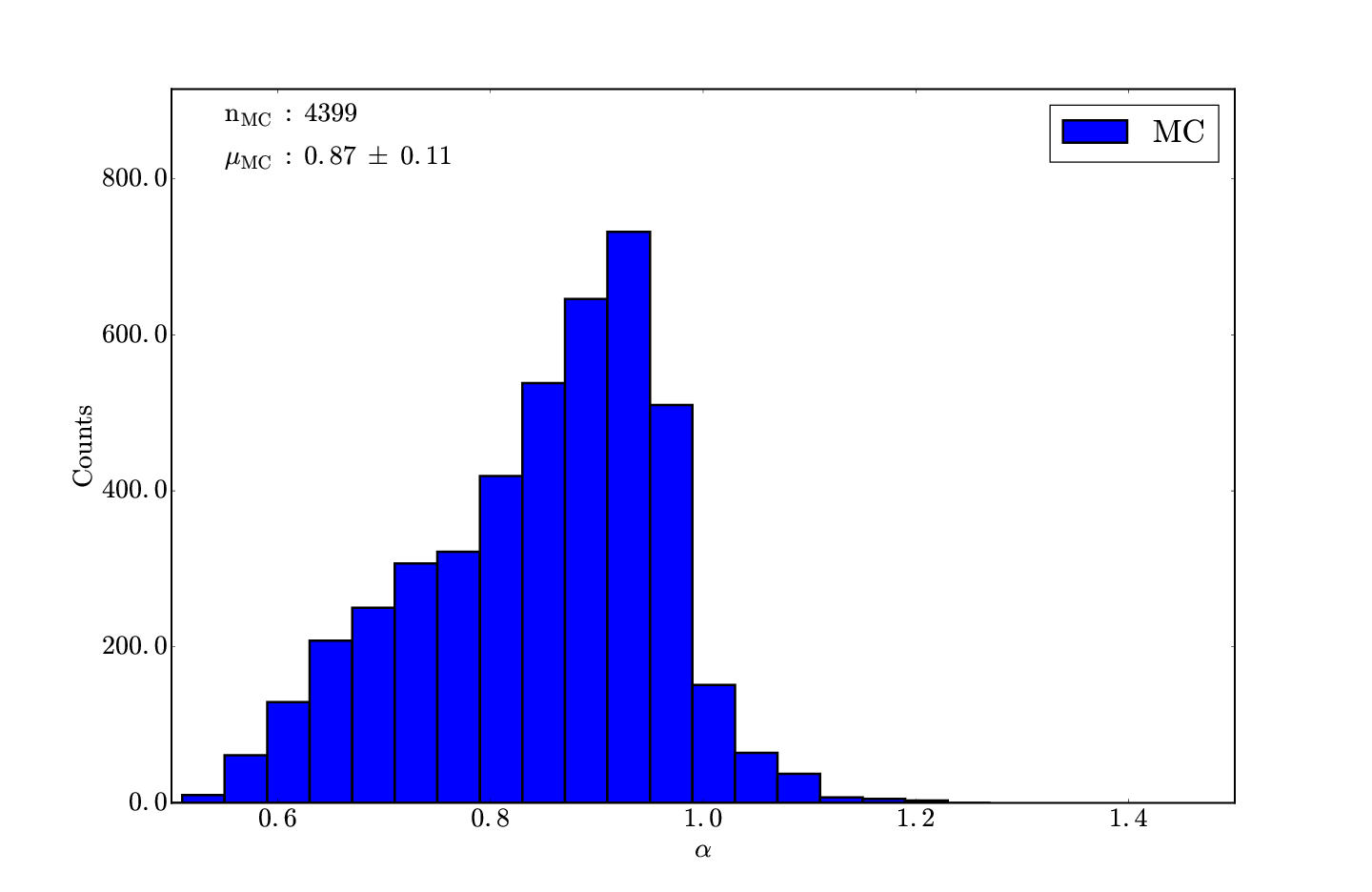}
\else
I am not enabling plots.
\fi
\caption{The same as Fig.~\ref{fig.ErigoneMC} with $\sim$4,400 trials repeating the V-shape technique for the Misa family. The mean of the distribution is centered at $\alpha$ = 0.87 $\pm$ 0.11 and the bin size in the histogram is 0.04.}
\label{fig.MisaSubMC}
\end{figure}

The mean and root mean square (RMS) uncertainty of $\alpha$  was determined from the distribution of the values of $\alpha$ determined in each of the Monte Carlo trials. Having more fragments and a well defined-V-shape causes the Monte Carlo technique to produce a narrower distribution in $\alpha$ (E.g.,~ for the Erigone family, $\alpha \; = \; 0.83 \pm 0.04$, Fig.~\ref{fig.ErigoneMC}), while having fewer fragments and a more diffuse V-shape  results in a broader $\alpha$ distribution (\eg,~for the Misa sub \ family, $\alpha \; = \; 0.87 \pm 0.11$, Fig.~\ref{fig.MisaSubMC}).

\subsection{Family ages}
\label{s.ages}
The time of travel for an individual asteroid with diameter $D$ is
\begin{equation}
\label{eqn.traveltime}
t_{travel} (D) \; = \; \frac{\Delta a (D)}{\frac{da}{dt}(D)}
\end{equation}
The asteroid travel time as a function of C, asteroid diameter, D, $\alpha$, the V-shape center $a_c$, average eccentricity of asteroid family members $e_{\mu}$, the average family member density $\rho_{\mu}$, the average family member visual albedo $p_{V,\mu}$ and the average family member bond albedo $A_{\mu}$ is found by combining Eqs.~\ref{eqn.apvDvsCfinal} and \ref{eqn.traveltime} and expanding the denominator of \ref{eqn.traveltime} to include $\frac{da}{dt} $ as a function of $D,\alpha,a_c,e_{\mu},\rho_{\mu}$ and $A_{\mu}$
\begin{equation}
\label{eqn.familyageCDrPv}
t_{travel}(C,D,\alpha,a_c,e_{\mu},\rho_{\mu},p_{V,\mu},A_{\mu}) \; = \; \frac{1329 \; C\; \left ( \frac{1}{D} \right )^{\alpha}}{\sqrt{p_{V,\mu}} \;  \frac{da}{dt} (D,\alpha,a_{c},e_{\mu},\rho_{\mu},A_{\mu}) }
\end{equation}

Eq.~\ref{eqn.familyageCDrPv} assumes no initial dispersion in the asteroid family due to the initial ejection velocity field. In reality such initial dispersion exists. The value of $C$ measured from the distribution of asteroid family members in reality includes the contribution of the initial ejection velocity field from Eq.~\ref{eqn.VEVvsCalphaFinal} and the contribution to $C$ from the Yarkovsky effect \citep[][]{Vokrouhlicky2006b,Nesvorny2015a}
\begin{equation}
\label{eqn.Ccombo}
C \; = \; C_{YE}  \; + \; C_{EV}
\end{equation}
where $C_{YE}$ is the width of the V-shape due to the Yarkovsky effect described by Eq.~\ref{eqn.Cpvdadtvsdeltat} and $C_{EV}$ is the width of the V-shape due to the initial ejection velocity of fragments described by \citep[][]{Vokrouhlicky2006b,Bolin2017a}
\begin{equation}
\label{eqn.VEVvsCalphaFinal}
C_{EV}(n, V_{EV}, \pv) \; = \; \frac{2}{n} \; V_{EV} \; \sqrt{\pv}
\end{equation}
where $n$ is the mean motion of the parent body, $V_{EV}$ is a parameter describing the width of the fragment velocity distribution \citep[][]{Michel2004, Nesvorny2006a, Durda2007}. The nominal value of $C_{EV}$ can exceed more than 50$\%$ of $C$ for asteroid families younger than 100 Myrs \citep[][]{Nesvorny2015a,Carruba2016d}. The error on the estimate of the initial ejection velocity field can be large enough so that there is a possibility that $C \; - C_{EV} \; \lesssim \; 0$ for young asteroid families, but we assume asteroid families considered in this study are all at least $\gtrsim$20 Myrs old and will have $C \; - C_{EV} \; > \; 0$. 

The contribution of the initial ejection of the fragments must be subtracted from the calculation of $t_{travel}$ \citep[][]{Nesvorny2002b, Carruba2016d, Carruba2016e}. The value of $C$ includes contribution from the spread in $a$ of the fragments caused by their initial ejection velocities in addition to their spread caused by the Yarkovsky effect
\begin{equation}
\label{eqn.Ccombo}
C_{YE}  \; = \;C \; - \; C_{EV}
\end{equation}

The age of the family is found by using $C_{YE}$ and expanding $\frac{da}{dt} (D,\alpha,a,e,\rho,A) $ from Eq.~\ref{eqn.yarkorate_final} in the denominator of Eq.~\ref{eqn.familyageCDrPv} and then simplifying
\begin{equation}
\label{eqn.familyagenoejection}
t_{age}(C_{YE} ,\alpha,a_c,e_{\mu},\rho_{\mu},p_{V,\mu},A_{\mu}) \; = \; \frac{1329 \; C_{YE}  \; \sqrt{a_{c}} \; (1 - e_{\mu}^2) \;  \rho_{\mu}  \; (1 - A_0)}{\sqrt{p_{V,\mu} \; a_{0}} \;  (1 - e^2) D_0^{\alpha} \rho_{0} (1 - A_{\mu}) \left ( \frac{da}{dt} \right )_0}
\end{equation}
We simplify $A$ in Eq.~\ref{eqn.familyagenoejection} according to $A \; = \;  \pv \; (0.290 + 0.684 \; G)$ from \citet[][]{Harris2002} to only include $\pv$ and $G$ in the formulation
\begin{equation}
\label{eqn.familyagenoejectionpvg}
t_{age}(C_{YE},\alpha,a_c,e_{\mu},\rho_{\mu},p_{V,\mu},p_{V,\mu},G_{\mu}) \; = \; \frac{1329 \; C_{YE} \; \sqrt{a_{c}} \; (1 - e_{\mu}^2) \;  \rho_{\mu}  \; (1 - ( p_{V,0} \; (0.290 + 0.684 \; G_{0}))}{\sqrt{p_{V,\mu} \; a_{0}} \;  (1 - e^2) \; D_0^{\alpha} \; \rho_{0} \; (1 - ( p_{V,\mu} \; (0.290 + 0.684 \; G_{\mu})))\;  \left ( \frac{da}{dt} \right )_0}
\end{equation}
where $p_{V,0} \; = \; 0.2$ and $G_0 \; = \; 0.24$.

The ages determined by Eq.~\ref{eqn.familyagenoejectionpvg} are valid for families $<2$ Gyrs old because the Sun's luminosity varied by $\lesssim 10\%$ over the last 2 Gyrs \citep[][]{Bertotti2003}. Eq.~\ref{eqn.familyagenoejectionpvg} is modified to include changing luminosity of the Sun for families older than 2 Gyrs \citep[][]{Vokrouhlicky2006b,Carruba2016c}
\begin{equation}
\label{eqn.familyagenoejectionpvgsolar}
t_{age>2Gyrs}(C_{YE},\alpha,a_c,e_{\mu},\rho_{\mu},p_{V,\mu},p_{V,\mu},G_{\mu}, t_1) \;  = \; \frac{t_{age}(C_{YE}, \alpha,a_c,e_{\mu},\rho_{\mu},p_{V,\mu},p_{V,\mu},G_{\mu}) }{ \left (\frac{1}{4.57 \;  \mathrm{Gyrs} \; - \; t_1}\int\limits_{t_1}^{4.57 \; \mathrm{Gyrs}}\left[1.3 + \frac{0.3 t}{4.57 \; \mathrm{Gyrs}}  \right ]^{-1} dt \right )}
\end{equation}
where $t_1$ is the epoch of the family's formation in Gyrs measured from the beginning of the solar system. Eq.~\ref{eqn.familyagenoejectionpvgsolar} does not include the evolution of fragments' $\pv$ caused by space weathering on secular timescales \citep[\eg][]{Jedicke2004,Vernazza2009}.

Family ages calculated with Eqs.~\ref{eqn.familyagenoejectionpvg} and\ref{eqn.familyagenoejectionpvgsolar} using the values of $a_c$, $C$ and $\alpha$ determined by the V-shape determination technique described in Section~\ref{s.v-shapeidentificationandalpha} can differ from the previous family ages obtained assuming $\alpha \; = \; 1.0$ \citep[][]{Vokrouhlicky2006b,Broz2013b, Spoto2015}. Unlike in the case where family ages that are recomputed with the stochastic YORP model are always younger compared to their previously determined ages \citep[][]{Bottke2015a,Carruba2016c}, family ages calculated with variable $\alpha$, $t(C_{YE,\alpha},\alpha)$ or $t_{age>2Gyrs}(C_{YE,\alpha},\alpha)$, can be smaller, the same or larger compared to $t(C_{y, \alpha = 1})$ or $t_{age>2Gyrs}(C_{y, \alpha = 1})$ when $\frac{D^{\alpha -1} \; C_{YE}}{C_{YE,\alpha}} \; > \; 1$, $\frac{D^{\alpha -1} \; C_{YE}}{C_{YE,\alpha}} \; = \; 1$ or $\frac{D^{\alpha -1} \; C_{YE}}{C_{YE,\alpha}} \; < \;1$ respectively. 

\section{Results}
\label{s.results}
\subsection{Synthetic family}
\label{s.synfamily}
The V-shape $a_c$, $C$ and $\alpha$ determination technique is tested on a synthetic asteroid family where fragments are initially dispersed simulating the disruption of a parent body and then are allowed to evolve for several 100 Myrs under planetary perturbations and the Yarkovsky effect. The V-shape determination technique is applied to the Synthetic asteroid family's V-shape to measure its $a_c$, $C$ and $\alpha$ and ensure they match the values assumed for the generation of the synthetic family in the simulation.

The break up of a synthetic asteroid family and its fragments' subsequent evolution due to the Yarkovsky effect is simulated by using 650 particles at $\left (a ,e , \sin i \right ) \; = \; \left (2.37, 0.21, 0.08 \right )$ and distributed in $a$ vs. $D_r$ space according to
\begin{equation}
\label{eqn.deltaaejection_a_vs_D_r}
D_r(a,a_c,n,V_{EV},\alpha_{EV}) \; = \; \frac{1}{D_0} \; \left ( \frac{\left | a - a_c \right | \; n}{2 \; V_{EV}} \right )^{\frac{1}{\alpha_{EV}} }
\end{equation}
from \citet[][]{Vokrouhlicky2006b,Bolin2017a} where $\alpha_{EV}$ is the exponent scaling $V_{EV}$ with $D$ \citep[][]{Cellino1999}. A value of $\alpha_{EV} \; = \; 1.0$ was used based on recent work on ejection velocity V-shapes of young asteroid families indicating that $\alpha_{EV} \; \simeq$ 1 \citep[][]{Bolin2017a}. $D_0$ = 5 km and $V_{EV}$ = 30 $\mps$ using fragments with $2 \mathrm{km} \; < \; D \; < \; 75 \mathrm{km}$ distributed according to the known members of the Erigone family defined by \citet{Nesvorny2015a}.  The eccentricity and inclination distributions of the ejected fragments were determined by using Gaussian scaling described in \citet[][]{Zappala2002}. $V_{EV}$ = 30 $\mps$ corresponds to a typical initial displacement of $\sim 7.0 \times 10^{-3}$ where given $V_{EV}$, the displacement in $a$ is size-independent. 

The Yarkovsky drift rates were defined with Eq.~\ref{eqn.yarkorate_final} with $\left ( \frac{da}{dt} \right )_{0} $ $\sim$ 4.7 $\times \; 10^{-5} \; \au \; \myr^{-1}$, $a_0 \; = \; $ 2.37 au, $e_0 \; = \; $ 0.2, $D_0$ = 5 km, $\rho_0$  = 2.5 $\gpcmc$ and Bond albedo, $A_0$, is equal to 0.1, surface conductivity equal to 0.01 $\mathrm{W \; m^{-1} \; K^{-1}}$ and $\theta_0 \; =$ 0$\tdeg$ \citep[][]{Bottke2006,Vokrouhlicky2015}. For the synthetic family, $\rho$ = 2.3 $\gpcmc$, $A \; = \; 0.02$ and $cos (\theta)$ is uniformly distributed between -1 and 1. The Yarkovsky drift rate was scaled with $D^{-\alpha \; \simeq \; -0.8}$ (now defined as $\alpha_{YE}$ for the Yarkovsky effect) as suggested by the relationship between $D$ and $\Gamma$ data for current asteroid data with 0.5 km $<$ $D$ $<$ 100 km discussed in Section \ref{s.YarkodriftMod}. The particles were evolved with the Yarkovsky effect and gravitational perturbations from Mercury, Venus, Earth, Mars, Juputer and Saturn using the $SWIFT$\_$RMVS$ code \citep[][]{Levison1994}. Particles are removed from the simulation if they collide with one of the planets or evolve on to orbits that have a perihelion of 0.1 au. YORP rotational and spin-axis variation are not included in the simulation. 

The V-shape identification technique was applied on the synthetic family data at Time = 200 Myrs using the techniques in Section~\ref{s.v-shapeidentificationandalpha}. As discussed in \citet[][]{Bolin2017a}, the time it takes for the V-shape of the synthetic Erigone family to transition from having its $ \alpha \; = \; \alpha_{EV}$ equal to 1.0 to $\alpha$ equal to $\alpha_{YE} \; \simeq \; 0.8$ is $\sim$20 Myrs. Measuring a V-shape's $\alpha$ using synthetic family data from time steps after 20 Myrs will be measuring the V-shape's $\alpha_{YE}$. Eqs.~\ref{eq.border_method_N_outer} and \ref{eq.border_method_N_inner} are integrated using the interval ($-\infty$,$\infty$) for the Dirac delta function $\delta(a_{j}-a)$ and the interval [$0.04,0.60$] for the Dirac delta function $\delta(D_{r,j}-D_r )$. Eq.~\ref{eqn.apvDvsCfinal} is truncated to 0.04 km$^{-1}$ for $D_r$ $ <$ $0.04$ km$^{-1}$ and to 0.60 km$^{-1}$ for $D_r$ $>$ 0.60 km$^{-1}$. Asteroids with 0.04 $< \; D_r \; <$ 0.60 were chosen because the number of asteroids in this $D_r$ is large enough so that the leading edge of the V-shape is defined by asteroids with $\mathrm{cos}(\theta) \; = \; $1.0 or -1.0 according to Eq.~\ref{eqn.apvDvsCfinal}. 
 
 The V-shape identification technique located a peak at $(a_c, \; C, \; \alpha) \; = \; (2.367 \; \mathrm{au}, \; 2.8 \times 10^{-5} \; \mathrm{au}, \;  0.8)$ as seen in the top panel of Fig.~\ref{fig.synErig200Myrs}. The peak value of $\frac{N_{in}^2}{N_{out}}$ is  $\sim$8 standard deviations above the mean value of $\frac{N_{in}^2}{N_{out}}$ in the range 2.25 au $< \; a \;<$ 2.45 au, 1.0 $\times \; 10^{-5}$ au $< \; C \;<$ 3.5 $\times \; 10^{-5}$ au and 0.6 $< \; \alpha \;<$ 1.4.  A $dC \; = \; 8.0 \; \times 10^{-6}$ au was used. The concentration of the peak to one localized area in $\alpha$ vs. $C$ space is due to the sharpness of the synthetic family's V-shape border. The procedure was repeated again using only larger asteroids in the synthetic family with 5 km $<$ $D$ $<$ 10 km to determine if measuring $\alpha$ in V-shapes consisting of only larger asteroids resulted in a different values of $\alpha$ than in V-shapes consisting of a full range of smaller asteroids. We did not measure any significant difference between the values of $\alpha$ measured in the two cases.
 
\subsection{Main Belt asteroid families}
The V-shape $a_c$, $C$ and $\alpha$ determination technique was applied to 26 asteroid families located through the inner, central and outer Main Belt. Proximity to mean motion and secular resonances can remove asteroids from an asteroid family resulting in an incomplete V-shape. Asteroid families were divided into three categories: complete V-shapes, clipped V-shapes where one or both sides of a family V-shape do not form a full V and half V-shapes where only one side of the V-shape is complete. Completeness of the V-shape does not change the functional form of the V-shape technique described in Section~\ref{s.v-shapeidentificationandalpha}, but affects the range of asteroids used in the technique as will be described in the following sections. All families are assumed to be old enough so that $C \; - C_{EV} \; > \; 0$ and therefore their measured value of $\alpha \; = \; \alpha_{YE}$ as discussed in Section~\ref{s.ages}.

\subsubsection{Complete V-shape families}
\label{s.completeV-shape}
 An asteroid family with a complete V-shape is defined as having a complete V-shape in extent along the $a$ axis relative to the center of the V-shape in $a$ vs. $D_r$ space such as the Erigone family seen in the bottom panel of Fig.~\ref{fig.ErigoneBorderDensity1}. Eqs.~\ref{eq.border_method_N_outer} and \ref{eq.border_method_N_inner} are integrated using the interval ($-\infty$,$\infty$) for the Dirac delta function $\delta(a_{j}-a)$ due to their symmetric shape. The V-shapes of some families such as the Erigone family have been noted to be non-symmetric in the value of $C$ between the inward and outward halves of their V-shapes \citep[][]{Spoto2015}. We do not find significant differences in $C$ between the inner and outer V-shape halves of asteroid families and therefore fit the families with a unique value of $C$. The interval for the Dirac delta function $\delta(D_{r,j}-D_r )$ is chosen with respect to the range in $D_r$ that contains the complete V-shape of the family. 
 
 The measured values of $\alpha$ and their uncertainties, family ages and the physical properties assumed for each family in the measurement for four complete V-shape families using the techniques described in Sections~\ref{s.v-shapeidentificationandalpha}, \ref{s.montecarlo} and \ref{s.ages} for each of the complete V-shape families are summarized in Table~\ref{t.completefams}. A description of how the V-shape determination technique is implemented for each complete V-shape family is described in Section~\ref{s.appendixcompletevshape}.

\begin{table}[] 
\centering
\hspace*{-1.8cm}
\label{t.completefams}
\begin{tabular}{|l|l|l|l|l|l|l|l|l|l|}
\hline
Designation & Tax. & D$_{pb}$& $t_{age,\alpha=1}$& $t_{age}$& $N$ & $a_{c}$ & $\alpha$ & $\pv$ & $D_{s}$ - $D_{l}$  \\ \hline
 &  & (km) & (Gyr) & (Gyr) &  & (au) &    &  &  (km) \\ \hline
Erigone     & C/X  & 96.5            & 0.13$\pm$0.07 & 0.09$\pm$0.04  & 1742                   & 2.370                & 0.83 $\pm$ 0.05         & 0.05 $\pm$ 0.01 & 1.5 - 23.7            \\ \hline
Massalia    & S    & 145.0           & 0.15$\pm$0.07&0.19$\pm$0.10 & 6414                   & 2.410                & 0.73 $\pm$ 0.06         & 0.24 $\pm$ 0.07 & 0.4 - 10.8            \\ \hline
Misa(2)  & C    & 27            &0.13$\pm$0.06   & 0.12$\pm$0.06&      427                  &         2.655            & 0.87 $\pm$ 0.11 &     0.1 $\pm$ 0.06          &       2.3 - 10.0              \\ \hline
Tamara  & C    &    106.0         &0.18$\pm$0.09   & 0.12$\pm$0.06&        111               &      2.310               & 0.7 $\pm$ 0.04 &     0.06 $\pm$ 0.02          &                      1.7 - 9.8\\ \hline
\end{tabular}
\caption{Complete V-shape families: Diameters for the parent body, $D_{pb}$, were taken from the means of asteroid family parent bodies in \citet[][]{Broz2013b} and \citet[][]{Durda2007} if $D_{pb}$ was available from both sources. The estimate of the Tamara parent body size is taken from \citet[][]{Novakovic2017}.}
\end{table}

\subsubsection{Clipped V-shape Families}
\label{s.clippedV-shape}

\begin{figure}
\centering
\hspace*{-0.7cm}
\ifincludeplots
\includegraphics[scale=0.455]{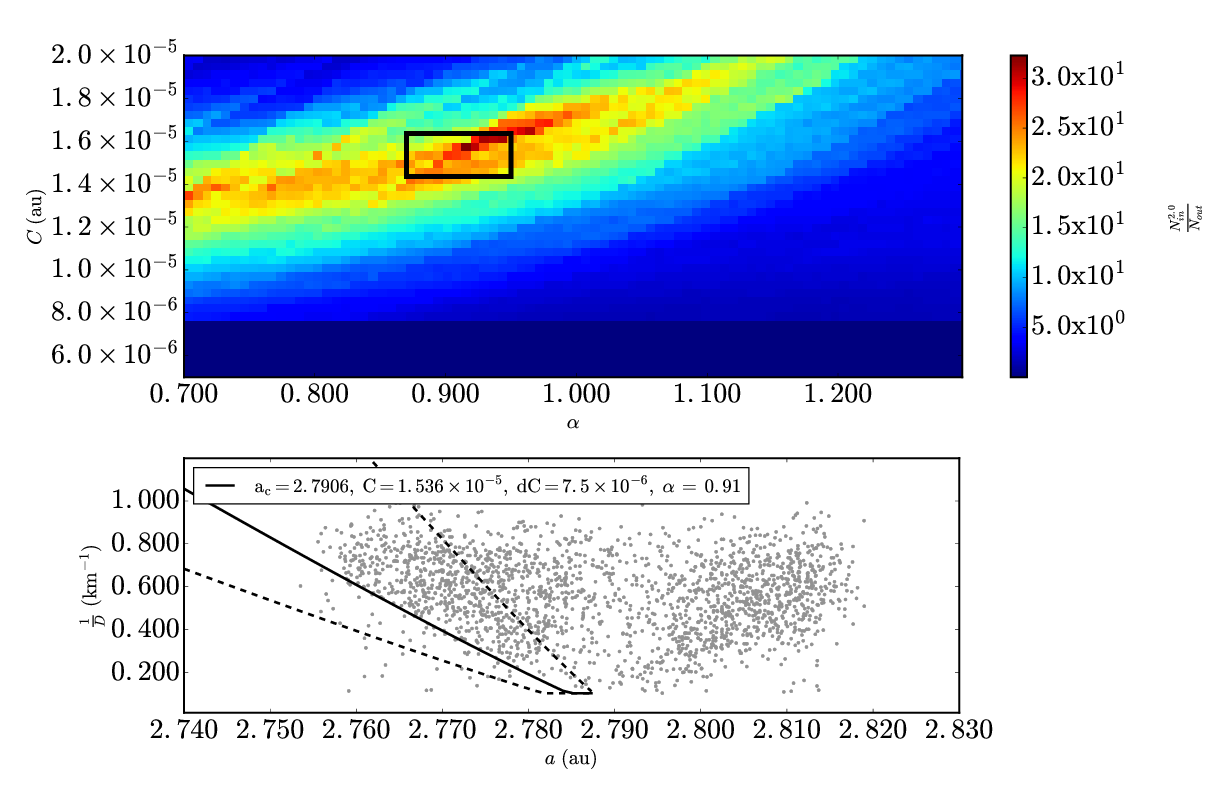}
\else
I am not enabling plots.
\fi
\caption{The same as Fig.~\ref{fig.synErig200Myrs} for Agnia asteroid family data from \citet[][]{Nesvorny2015a}. (Top panel) $\Delta \alpha$ is equal to $1.4 \times 10^{-2}$ au and $\Delta C$, is equal to $7.4 \times 10^{-7}$ au. (Bottom Panel) $D_r(a,a_c,C\pm dC,\pv,\alpha)$ is plotted with $\pv = 0.18$, $a_c$ = 2.791 au and $dC \; = \; 7.5 \x 10^{-6}$ au.}
\label{fig.AgniaAlpha}
\end{figure}

 An asteroid family with a clipped V-shape is defined as having at least one full V-shape half in addition to another partial V-shape such as the Agnia family seen in the bottom panel of Fig.~\ref{fig.AgniaAlpha} where the outer V-shape half is depleted of asteroids at about $D_r$ = 0.4 km$^{-1}$ because it is intersected by the 5:2 MMR with Jupiter at 2.82 au. Intervals used for integrating eqs.~\ref{eq.border_method_N_outer} and \ref{eq.border_method_N_inner} for the Dirac delta function $\delta(a_{j}-a)$ are ($-\infty$,$a_c$] when applying the V-shape technique to only the complete inner V-shape half, [$a_c$,$\infty$) when applying the technique to only the complete outer V-shape half, and ($-\infty$,$\infty$) when applying the technique to both the complete and incomplete halves. The interval used for $\delta(a_{j}-a)$ is determined by whether or not the number of asteroids in the complete V-shape borders is large enough to obtain a statistically robust determination of $\alpha$. 
 
As discussed in Section~\ref{s.completeV-shape}, we assume that the values of $C$ and $\alpha$ are the same on both V-shape halves. The measured values of $\alpha$ and their uncertainties, family ages and the physical properties assumed for each family are summarized in Table~\ref{t.clippedfams}. A description of how the V-shape determination technique is implemented for 12 clipped V-shape family is described in Section~\ref{s.appendixclippedvshape}.
 
\begin{table}[]
\centering
\hspace*{-1.8cm}
\label{t.clippedfams}
\begin{tabular}{|l|l|l|l|l|l|l|l|l|l|}
\hline
Designation & Tax. & D$_{pb}$& $t_{age,\alpha=1}$& $t_{age}$& $N$ & $a_{c}$ & $\alpha$ & $\pv$ & $D_{s}$ - $D_{l}$  \\ \hline
 &  & (km) & (Gyr) & (Gyr) &  & (au) &    &  &  (km) \\ \hline

Agnia       &  S    &   50.0            &  0.1$\pm$0.05        & 0.12$\pm$0.06 &              2123        &   2.791                  &   0.90 $\pm$ 0.03       &   0.18 $\pm$ 0.06  &       0.6 - 8.5              \\ \hline
Astrid      &   C   &      43.0         &       0.11$\pm$0.06        & 0.1$\pm$0.05 &                452      &  2.787                   &               0.81 $\pm$ 0.07        &      0.08  $\pm$    0.02   &           1.3 - 6.7          \\ \hline
Baptistina  &   X   &    35.0           &     0.18$\pm$0.09      & 0.2$\pm$0.1 &                    2450  &       2.263              &       0.83 $\pm$ 0.05    & 0.16 $\pm$ 0.03    & 0.5 - 19.9                    \\ \hline
Dora(2)        &   C   &    125.0           &     0.13$\pm$0.07      &   0.1$\pm$0.05  &               1223   &                2.796     &   0.86 $\pm$0.04       &  0.05 $\pm$ 0.01     & 1.2 - 22.7                    \\ \hline
Eos         &     K &       294.5        &      1.13$\pm$0.56    &  1.08$\pm$0.54 &                    6897 &     3.024                &       0.92 $\pm$ 0.02   &  0.13 $\pm$ 0.04     & 2.9 - 23.3                    \\ \hline
Eunomia     &   S   &      275.5         & 1.66$\pm$0.83   &  1.1$\pm$0.55 &                    1311 &       2.635              &     0.77 $\pm$ 0.03     &  0.19 $\pm$ 0.04    & 4.8 - 19.3                    \\ \hline
Hoffmeister &   CF   &        41.4$^{*}$       &    0.22$\pm$0.11       &  0.22$\pm$0.11 &               1773      &      2.785               &        0.84 $\pm$ 0.03  & 0.04 $\pm$ 0.01     & 1.9 - 27.1                    \\ \hline
Hungaria    &    E  &      25.0         &  0.3$\pm$0.15         & 0.27$\pm$0.13 &                  2337    &         1.943            &   0.79 $\pm$ 0.09       &   0.35 $\pm$ 0.1   & 0.8 - 3.3                    \\ \hline
Hygiea & CB  &      426.0       & 1.16$\pm$0.58    &  0.92$\pm$0.46    &                 553 &         3.157            &            0.92 $\pm$ 0.02           &  0.06 $\pm$ 0.02            &      8.4 - 31.9               \\ \hline
Koronis     &    S  &      148.5         &    1.57$\pm$0.79       &  1.94$\pm$0.97     &              516   &         2.883            &      0.93 $\pm$ 0.03    &   0.15 $\pm$ 0.04    & 6.8 - 27.3                    \\ \hline
Naema     &  C    &      79.0         &     0.07$\pm$0.03      & 0.04$\pm$0.02&                      281 &            2.939         &     0.81 $\pm$ 0.05     &  0.08 $\pm$ 0.02     & 1.8 - 19.9                    \\ \hline
Padua & C/X  & 83.5       & 0.33$\pm$0.17   &  0.31$\pm$0.16     &                558 &             2.744        &            0.89 $ \pm$ 0.11           &  0.1 $\pm$ 0.03             & 1.9 - 7.7                    \\ \hline

\end{tabular}
\caption{Clipped V-shape families: Diameters for the parent body, $D_{pb}$, were taken from the means of asteroid family parent bodies in \citet[][]{Broz2013b} and \citet[][]{Durda2007} if $D_{pb}$ was available from both sources. Ages of asteroid families were taken from \citet[][]{Broz2013b} and \citet[][]{Spoto2015}. The estimate of the Hoffmeister parent body size is determined using the method of \citet[][]{Tanga1999}. Ages for the Hoffmesiter and Hygiea family are taken from \citet[][]{Carruba2017a} and \citet[][]{Carruba2014}. Diameter of the parent body for the Dora(2) was determined using the method of \citet[][]{Tanga1999}. The $\alpha$ of the sub-family V-shape in the Dora family is measured.}
\end{table}

\subsubsection{Half V-shape families}
\label{s.halfV-shape}
\begin{figure}
\centering
\hspace*{-0.7cm}
\ifincludeplots
\includegraphics[scale=0.455]{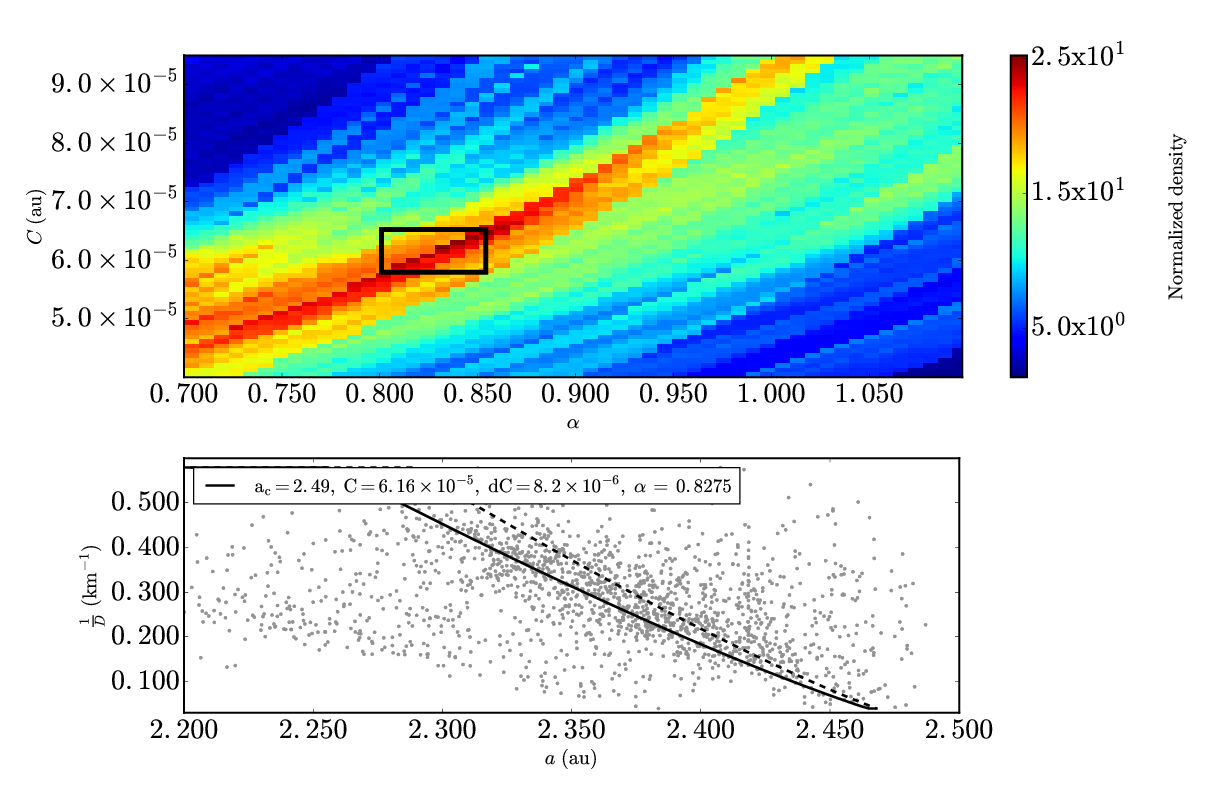}
\else
I am not enabling plots.
\fi
\caption{The same as Fig.~\ref{fig.synErig200Myrs} for the Eulalia asteroid family data from \citet[][]{Nesvorny2015a}. (Top panel) $\Delta \alpha$ is equal to $7.5 \times 10^{-3}$ au and $\Delta C$, is equal to $8.0 \times 10^{-7}$ au. (Bottom Panel) $D_r(a,a_c,C\pm dC,\pv,\alpha)$ is plotted with $\pv = 0.05$, $a_c$ = 2.49 au and $dC \; = \; 3.2 \x 10^{-6}$ au.}
\label{fig.EulaliaDensity1}
\end{figure}

An asteroid family with a half V-shape is comprised of being only one full V-shape in $a$ vs. $D_r$ space such as the Eulalia family seen in the bottom panel of Fig.~\ref{fig.EulaliaDensity1} where the family's V-shape center is located within the vicinity of the 3:1 MMR with Jupiter at 2.5 au. The Dirac delta function $\delta(a_{j}-a)$ in Eqs.~\ref{eq.border_method_N_outer} and \ref{eq.border_method_N_inner} is integrated using the interval ($-\infty$,$a_c$] or [$a_c$,$\infty$) for the half V-shape case.The interval for the Dirac delta function $\delta(D_{r,j}-D_r )$ is chosen to include the full $D_r$ range encompassing the half V-shape in $a$ vs. $D_r$ space. 

The measured values of $\alpha$ and their uncertainties, family ages and the physical properties assumed for each family are summarized in Table~\ref{t.halfvfams}. A description of how the V-shape determination technique is implemented for 10 clipped V-shape family is described in Section~\ref{s.appendixhalfvshape}.

\begin{table}[]
\centering
\hspace*{-1.8cm}
\label{t.halfvfams}
\begin{tabular}{|l|l|l|l|l|l|l|l|l|l|}
\hline
Designation & Tax. & D$_{pb}$& $t_{age,\alpha=1}$& $t_{age}$& $N$ & $a_{c}$ & $\alpha$ & $\pv$ & $D_{s}$ - $D_{l}$  \\ \hline
 &  & (km) & (Gyr) & (Gyr) &  & (au) &    &  &  (km) \\ \hline
Adeona      &   C   &    178.0           &1.4$\pm$0.7& 1.15$\pm$0.58&                 2152  &           2.705          &     0.83 $\pm$ 0.03     &   0.07 $\pm$ 0.02   &           1.6 - 14.4          \\ \hline
Eulalia     &   C   &         130.0$^{*}$      &0.84$\pm$0.42 &0.86$\pm$0.43&                  1818     &         2.490            &   0.78 $\pm$ 0.06       &   0.06 $\pm$ 0.02    & 1.9 - 22.2                    \\ \hline
Flora     &   S   &         155.0$^{*}$      &1.06$\pm$0.53 &1.16$\pm$0.58&                5362    &         2.200            &   0.83 $\pm$ 0.06      &  0.29 $\pm$ 0.09    &  1.4 - 7.4                    \\ \hline
Maria       &    S  &       116.0        &1.66$\pm$0.83 &1.16$\pm$0.58&             1144     &              2.584       &     0.87 $\pm$ 0.03     &       0.25 $\pm$ 0.06&           2.6 - 10.4   \\ \hline
Nemausa  &    C  &       70.0        &3.8$\pm$1.9 &4.3$\pm$2.1&             3949     &              2.37       &     0.92 $\pm$ 0.03     &       0.05 $\pm$ 0.02&           3.4 - 36.8   \\ \hline
Nemesis     & C    &        193.0      &0.15$\pm$0.08 &0.19$\pm$0.09&                1250    &                2.738     &         0.80 $\pm$ 0.03  &   0.05 $\pm$ 0.01  &         1.8 - 16.8            \\ \hline
New Polana      &  C    &      130.0$^{*}$         &2.56$\pm$1.28 &2.06$\pm$1.03&            1818      &         2.426            &     0.79 $\pm$ 0.06     &    0.06 $\pm$ 0.02   &      1.9 - 20.1               \\ \hline
Rafita      &    S  &        27.0       &0.38$\pm$0.19 &0.38$\pm$0.19&                  1251  &            2.549         &      0.79 $\pm$ 0.05    &   0.25 $\pm$ 0.06   &        0.6 - 16.4             \\ \hline
Sulamitis   &    C  &        65.0       &0.5$\pm$0.25 &0.47$\pm$0.23&              284  &                2.472     &      0.87 $\pm$ 0.02    &   0.04 $\pm$ 0.01    &  1.4 - 14.0                   \\ \hline
Ursula      &  CX    &        232.0       &2.67$\pm$1.33 &2.29$\pm$1.14&           1209  &            3.218         &     0.9 $\pm$ 0.02    & 0.06 $\pm$ 0.02      &       4.1 - 31.0              \\ \hline
\end{tabular}
\caption{Half V-shape families: Diameters for the parent body, $D_{pb}$, were taken from the means of asteroid family parent bodies in \citet[][]{Broz2013b} and \citet[][]{Durda2007} if $D_{pb}$ was available from both sources. The $D_{pb}$ for the Eulalia family was taken from \citet[][]{Walsh2013}. Ages of asteroid families were taken from \citet[][]{Broz2013b} and \citet[][]{Spoto2015}. The age of the Eulalia and New Polana families are taken from \citet[][]{Walsh2013}. The age of the Flora family was taken from \citet[][]{Vokrouhlicky2017b}.}
\end{table}

\section{Discussion and Conclusion}

The dependence of thermal inertial of asteroids on asteroids' physical sizes suggests that the size dependence of the the Yarkovsky semi-major axis drift should be proportional to $D^{\alpha}$ with $\alpha \; < \; 1$. We have analyzed the V-shape in the $a$ vs. $D_r$ distribution of 26 families and determined the value of $\alpha$ that best characterizes these shapes. We have analyzed the V-shapes of families located in the Inner, Middle and outer Main Belt and determined their ages. Although the 26 families used in this study represent only a quarter of the known $\sim$110 asteroid families defined by \citet[][]{Nesvorny2015a}, they constitute a representative sample of asteroid families in the Main Belt because the families in the sample are spread evenly through Inner, Central and outer Main Belt and cover the main taxonomic types of asteroids

The statistical uncertainty (standard deviation) on the determination of $\alpha$ was estimated using a Monte Carlo technique, as described in Section~\ref{s.montecarlo}. We found that the difference between the measured value of $\alpha$ and $\alpha \; =$1.0 is more than three times the standard deviations for the majority of family V-shapes. This suggests that curvature of family V-shapes in $a$ vs. $D_r$ space is real and widespread throughout families in the Main Belt. In a different study, \citet[][]{Bolin2017a}, focousing on young families dominated by the initial ejection velocity field, we determined that $\alpha \; = \; 1.0$. Thus, the values of $\alpha$ different from 1 that we obtain in this paper should be attributed to the non-trivial size dependence of the Yarkovsky drift speed. The average value of $\alpha$ of the 26 family Yarkovsky V-shapes in this study is 0.87 $\pm$ 0.01, or within in the Student's t-distribution 99.8$\%$ confidence interval  of 0.86 - 0.87. The average value of $\alpha$ when considering only the 8 S-type families and the 16 C-type families with $\alpha$ measurements separately are statistically indistinguishable.

\begin{figure}
\centering
\hspace*{-1.6cm}
\ifincludeplots
\includegraphics[scale=0.55]{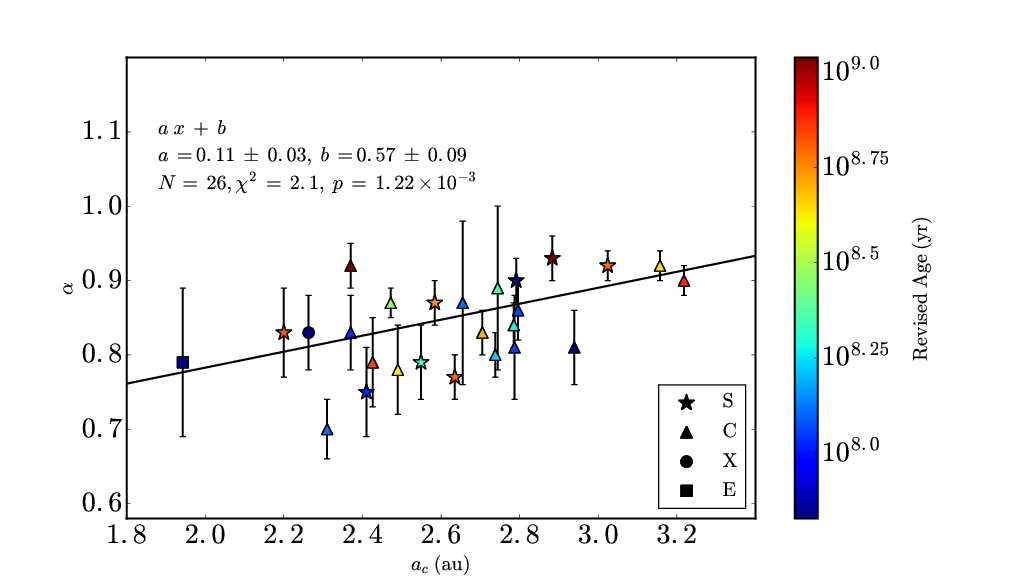}
\else
I am not enabling plots.
\fi
\caption{$a_c$ vs. $\alpha$ vs. Revised Age for asteroid families of all taxonomies. The Eos asteroid family is labeled as an S-type in this plot. The data are fit to the function $y \; = \; a \; x \; + \; b$ in $a_c$ vs. $\alpha$ space and is shown as the dark line using orthogonal distance regression \citep[][]{Boggs1990}.}
\label{fig.a_c_vs_alpha_all_tax}
\end{figure}

The measured values of $\alpha$ changes throughout the Main belt with the $\alpha$ of asteroid family V-shapes in the inner and Central Main Belt (defined as 1.8 au $<$ $a$ $<$ 2.5 au and 2.5 au $<$ $a$ $<$ 2.8 respectively) having a lower $\alpha$ value on average, $\alpha_{\mu} \; \simeq \; 0.84 \; \pm \; 0.01$, than that of families in the outer Belt (defined as 2.8 au $<$ $a$ $<$ 3.3 au, $\alpha_{\mu} \; \simeq \; 0.91 \; \pm \; 0.01$) as seen in Fig.~\ref{fig.a_c_vs_alpha_all_tax}. A linear fit to the results in $a_c$ vs. $\alpha$ space is significantly sloped with $\alpha \; = a \; x \; + \; b$ where $a$ = $ 0.1 \pm 0.03 \; \mathrm{au}^{-1}$,  $x$ =  $a_c$ and $b$ =  $0.57 \pm 0.09$ as seen in Fig.~\ref{fig.a_c_vs_alpha_all_tax}. There is some indication that this slope is somewhat larger if we restrict ourselves to the 8 S-type family v-shapes spread throughout the Inner, Central and outer Main Belt. In this case we find $\alpha \simeq0.2$ au$^{-1}$ $\pm$ 0.07 overlapping with the slope including families of all taxonomic types. Instead, there is no change in slope when considering only the 16 C-type families.

The inward curvature of V-shapes in $a$ vs. $D_r$ space with $\alpha$ $<$ 1.0 suggests that objects smaller than $\sim$1 km are drifting slower and larger objects are drifting faster compared to the case with $\alpha \; = \; 1.0$. A possible explanation for the inward curvature of family V-shapes and the slower drift rate of small asteroids is the dependence of thermal inertia on $D$, first described by \citet[][]{Delbo2007}. The average $\alpha \; = \; 0.87 \; \pm \; 0.01$ of family V-shapes overlaps with the value of $\alpha \; = \; $ 0.77 $\pm$ 0.13 expected from the relationship in $D$ vs. $\Gamma$ space for asteroids with 0.5 km $<$ $D$ $<$ 100 km as described in Section~\ref{s.YarkodriftMod}. Additionally, the planetary regolith model of \citet[][]{Gundlach2013}, which determines surface regolith size from an asteroid's $\Gamma$, surface temperature and taxonomic type predicts a slight increase in the linear slope of  $D$ vs. $\Gamma$ for the outer Main Belt families, which corresponds to a $\sim10\%$ higher $\alpha$ for asteroid families in the outer Main Belt compared to asteroid families in the inner Main Belt. This is in good agreement with the difference in the mean value of $\alpha$ between inner and outer Main Belt family V-shapes determined in this paper, although this difference is also comparable to the relative uncertainty of the model of \citet[][]{Gundlach2013}. 

Some caution must be used in comparing $\alpha$ measurements determined from asteroid family V-shapes with the $\alpha$ expected from the asteroid's  $D$ vs. $\Gamma$ relationship because the spin rate of an asteroid can affect its $\Gamma$. In fact, slower spinning asteroids, \ie those with rotation periods greater than 10 h, may have a higher $\Gamma$ than more quickly spinning asteroids possibly as a result of rapidly increasing material density and $\Gamma$ with surface depth \citep[][]{Harris2016a}. Additionally, thermal inertia is expected to have a heliocentric dependence as a result of its temperature dependence.

One explanation for the apparent increase of $\alpha$ towards 1.0 with heliocentric distance is that asteroid family members in the outer Belt have similar regolith properties. An increase in $\alpha$ to 1.0 for an asteroid family V-shape implies that there is no decrease in $\Gamma$ with increasing $D$ as observed in the general asteroid population if $\alpha$ is assumed to be an indication of the linear slope in $D$ vs. $\Gamma$ space for individual asteroid family members.  As discussed in Section~\ref{s.Introduction}, small and large asteroids have different surface regolith properties with small asteroids having coarser regolith resulting in larger values of $\Gamma$ compared to larger asteroids. A more even distribution in $\Gamma$ between larger and smaller asteroids would imply that small and large asteroids have either coarse or fine regolith. 

One possible source of surface regolith coarseness homogenization between small and large asteroids is that recent family creation are producing vast quantities of dust coating the surfaces of the members of other asteroid families in the vicinity. The outer Belt contains a higher proportion of young asteroid families that were created within the last 20 Myrs such as the \FYnospace, Brasilia, Iannini, Karin, K{\"o}nig, Koronis(2), Theobalda and Veritas asteroid families \citep[][]{Nesvorny2015a, Bolin2017a} some of which have been attributed as the source of the {\it IRAS} dust bands \citep[][]{Grogan2001,Nesvorny2003}. Enough dust would have to be produced and accreted onto a significant amount of asteroids within a family and homogenize the surface regolith properties between members to have a significant effect in changing the curvature of the family V-shape.

\begin{figure}
\centering
\hspace*{-1.0cm}
\ifincludeplots
\includegraphics[scale=0.55]{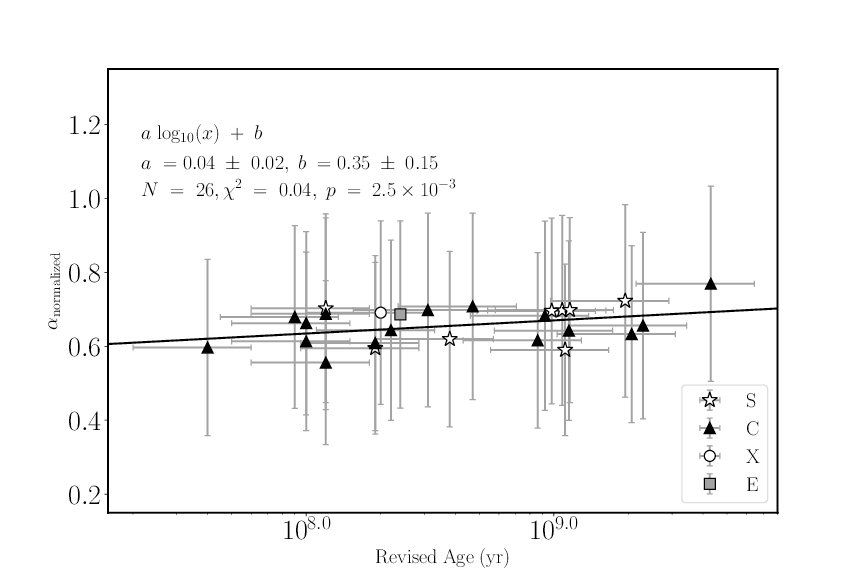}
\else
I am not enabling plots.
\fi
\caption{Age vs.$\alpha$ for asteroid families of all taxonomies. The data are fit to the function $y \; = \; a \; \mathrm{log_{10}}(x) \; + \; b$ shown as the dark line using orthogonal distance regression \citep[][]{Boggs1990}.}
\label{fig.age_vs_alpha_all_tax_age}
\end{figure}

The ages of asteroid families are calculated with Eqs.~\ref{eqn.familyagenoejectionpvg} and \ref{eqn.familyagenoejectionpvgsolar} with $\alpha$ determined by the V-shape technique and plotted in Fig.~\ref{fig.age_vs_alpha_all_tax_age}. The family V-shape $\alpha$ are normalized to 1 au with respect to $a_c$ according to $\alpha \; = \; 0.11\pm 0.03 \; a_c \; + \; 0.57 \pm 0.09$ determined from the linear fit in Fig.~\ref{fig.a_c_vs_alpha_all_tax}. The $\alpha_{normalized}$ of family V-shapes describes the relative amount of curvature of a family V-shape if all family V-shapes had the same $a_c$. Higher values of $\alpha_{normalized}$ corresponds to V-shapes with less curvature compared to lower values of $\alpha_{normalized}$. The resulting fit in Revised Age vs. $\alpha_{normalized}$ space is compatible with no trend in increasing or decreasing curvature with age. The slope is only 0.03 with a relatively large uncertainty of 0.01 due to large uncertainties in the linear slope of asteroid V-shape data in $a_c$ vs. $\alpha$ space and the large uncertainties on the age of asteroid families as discussed in Section~\ref{s.ages}. 

The curvature of the V-shape of asteroid families is similar to that produced by the so-called "stochastic YORP" effect \citep[][]{Bottke2015a}. The stochastic YORP model applied to asteroid family V-shapes of describes the YORP states of individual family fragments where they are reset or modified by minute changes in their shapes or surface features following the stochastic YORP model \citep[][]{Statler2009,Cotto-Figueroa2015}. When applied to asteroid families, the stochastic YORP model entails that the functional form of asteroid family V-shape described by Eq.~\ref{eqn.apvDvsCfinal} with $\alpha \; = \; 1.0$ becomes distorted or inwardly curved as asteroid families age, particularly for asteroid families older than 500 Myrs and for asteroids smaller than $\sim$1 km. This effect of stochastic YORP is similar to the effect of size-dependent $\Gamma$ on asteroid family V-shapes described by Eq.~\ref{eqn.apvDvsCfinal} with $\alpha \; < \; 1.0$. 

However, the lack of a clear trend with decreasing $\alpha_{normalized}$ with increasing age, $\ie$ asteroid families becoming more curved with age, suggests that smaller, $D$ 1$\sim$3 km asteroids may not be as affected by stochastic YORP cycles as predicted by \citep[][]{Bottke2015a} for asteroid families 500 Myrs to $\gtrsim$ 2 Gyrs old. In fact, the opposite trend seems to be the case because some Gyr-old families such as the Eos, Hygiea, Koronis, Ursula and Nemausa have less curvature, (\ie~$\alpha \sim 0.9$) compared to families with ages $<$500 Myrs such as the Astrid, Erigone, Massalia, Naema and Tamara with (0.7 $\lesssim$ $\alpha$ $\lesssim$ 0.8) suggesting that the overall trend between family ages and $\alpha$ seems to be inconclusive or unfavorable to the stochastic YORP model when applied to asteroid family V-shapes. More importantly, the lower bound in asteroid size used in the V-shape determination technique excludes asteroids affected by the stochastic YORP. For instance, in the case of the Eunomia, Hygiea and Koronis families the smallest asteroids used were $\sim$ 7 km, much larger than the 1-2 km size at which the stochastic YORP becomes apparent. However one possibility is that the time scale on which YORP becomes stochastic for asteroids is longer than predicted by \citet[][]{Bottke2015a} as indicated by recent simulations of the evolving shapes of certain asteroids due to rotational stress \citep[][]{McMahon2017}. The result of YORP being less stochastic is that the change in shape of a family's V-shape due to stochastic YORP may occur on longer time scales than predicted by \citet[][]{Bottke2015a}. Instead of stochastic YORP having an effect on an asteroid family's V-shape after 500 Myrs, it may take an effect on much longer time scales than can be recognized within the range of  ages of asteroid families studied in this paper.

An alternative explanation of curved family V-shapes is that family fragments are non-uniform in density, with smaller fragments having a higher density compared to larger objects, resulting in lower drift rates for smaller asteroid as determined by Eq.~\ref{eqn.yarkorate_final}. Additionally, re-accumulation of material following the disruption of parent body with uniform density may result in less dense larger fragments \citep[][]{Michel2001,Michel2015}. In fact, larger fragments have greater gravity and are able to re-accumulate more debris into a more loosely compact body than smaller fragments.

However, bulk density measurements of S and C-type 50 - 200 km asteroids are relatively homogenous with a slight increase in $\rho$ for larger objects past 200 km \citep[][]{Carry2012}. This increase in $\rho$ at larger asteroid sizes is possibly due to grain compaction \citep[][]{Consolmagno2008}, but it is beyond the size affected by the Yarkovsky effect \citep[][]{Vokrouhlicky2015}. Measured bulk densities of small km-scale asteroid bodies from spacecraft missions (such as the NEAR-Shoemaker mission to Eros \citep[][]{Yeomans2000}, Hayabusa spacecraft's mission to Itokawa \citep[][]{Fujiwara2006} and the Rosetta spacecraft's flyby of Lutetia \citep[][]{Drummond2010}), the YORP effect \citep[][]{Lowry2014} or observation of binaries \citep[][]{Hanus2017,Carry2015,Margot2015} are comparable to their larger counterparts suggesting that there is no size dependence on asteroid $\rho$ for asteroids in the sub-km to 10 km scale. 

The weak dependence of $\rho$ with size would cause asteroid family V-shapes to have no curvature or be slightly curved with $\alpha$ $\gtrsim$ 1.0 which is opposite to the $\alpha$ $<$ 1.0 measured for family V-shapes throughout the Main Belt. This implies that the inward curvature of asteroid family V-shapes is probably not caused by density inhomogeneities with asteroid size within a family. 

Although the main goal of this study is to not redetermine the ages of asteroid families, the ages of asteroid families calculated with Eqs.~\ref{eqn.familyagenoejectionpvg} and \ref{eqn.familyagenoejectionpvgsolar} with $\alpha$ V-shape measurements can be compared to  ages calculated assuming $\alpha \; = \; 1.0$ \citep[\eg][]{Broz2013b, Spoto2015}. The age of asteroid families revised with $\alpha$ measurements as described in Section~\ref{s.ages} are summarized in Tables~\ref{t.completefams}, \ref{t.clippedfams} and \ref{t.halfvfams}. The average relative difference between the revised age and the age determined with $\alpha \; = \; 1.0$ is -12.0 $\pm$ 26.0$\; \%$ implying that the ages of asteroid families are overestimated on average when $\alpha$ is assumed to be unity. The absolute relative difference between asteroid family ages calculated with V-shape technique determined $\alpha$ and $\alpha \; = \; 1.0$ is on average 22.0 $\pm$ 19.0$\; \%$.

\newpage
\appendix
\section{Appendix}

\begin{table}[]
\centering
\caption{Description of variables.}
\label{my-label}
\begin{tabular}{ll}
\hline
\multicolumn{1}{|l|}{Variable}      & \multicolumn{1}{l|}{Description}                                                      \\ \hline
\multicolumn{1}{|l|}{$D$}           & \multicolumn{1}{l|}{Asteroid diameter in km}                                                                 \\ \hline
\multicolumn{1}{|l|}{$a$}           & \multicolumn{1}{l|}{Semi-major axis in au.}                                           \\ \hline
\multicolumn{1}{|l|}{$e$}           & \multicolumn{1}{l|}{Eccentricity.}                                                    \\ \hline
\multicolumn{1}{|l|}{$i$}           & \multicolumn{1}{l|}{Inclination in degrees.}                                                     \\ \hline
\multicolumn{1}{|l|}{$D_r$}         & \multicolumn{1}{l|}{Reciprocal of the diameter, $\frac{1}{D}$ in km$^{-1}$.}          \\ \hline
\multicolumn{1}{|l|}{$a_c$}         & \multicolumn{1}{l|}{The location of the V-shape centere in au.}                        \\ \hline
\multicolumn{1}{|l|}{$n$}           & \multicolumn{1}{l|}{Mean motion in $\frac{\mathrm{rad}}{\mathrm{s}}$}             \\ \hline
\multicolumn{1}{|l|}{$V_{ev}$}      & \multicolumn{1}{l|}{Ejection velocity in $\frac{\mathrm{m}}{\mathrm{s}}$.}            \\ \hline
\multicolumn{1}{|l|}{$\alpha$} & \multicolumn{1}{l|}{The $\alpha$ of a V-shape according to  by Eq.~\ref{eqn.apvDvsCfinal}.}                               \\ \hline
\multicolumn{1}{|l|}{$\pv$}         & \multicolumn{1}{l|}{Visual albedo.}                                                   \\ \hline
\multicolumn{1}{|l|}{$C$}           & \multicolumn{1}{l|}{Total V-shape width in au.}                                       \\ \hline
\multicolumn{1}{|l|}{$\alpha_{EV}$}      & \multicolumn{1}{l|}{$\alpha$ defined for an ejection velocity V-shape defined by Eq.~\ref{eqn.deltaaejection_a_vs_D_r}.}           \\ \hline
\multicolumn{1}{|l|}{$N_{out}$}     & \multicolumn{1}{l|}{Number density of objects between the nominal and outer V-shapes.} \\ \hline
\multicolumn{1}{|l|}{$N_{in}$}      & \multicolumn{1}{l|}{Number density of objects between the nominal and inner V-shapes.} \\ \hline
\multicolumn{1}{|l|}{$dC$}          & \multicolumn{1}{l|}{Difference in $C$ between the nominal and outer/inner V-shapes.}  \\ \hline
\multicolumn{1}{|l|}{$H$}           & \multicolumn{1}{l|}{Absolute magnitude.}                                              \\ \hline
\multicolumn{1}{|l|}{$C_{YE}$}      & \multicolumn{1}{l|}{V-shape width due to Yarkovsky spreading of fragments in au.}     \\ \hline
\multicolumn{1}{|l|}{$C_{EV}$}      & \multicolumn{1}{l|}{V-shape width due to the initial ejection of fragments in au.}    \\ \hline
\multicolumn{1}{|l|}{$N$}           & \multicolumn{1}{l|}{Number of family members used with the V-shape technique.}        \\ \hline
\multicolumn{1}{|l|}{$\alpha_{YE}$} & \multicolumn{1}{l|}{The $\alpha$ of a Yarkovsky V-shape.}                                  \\ \hline
\multicolumn{1}{|l|}{$\rho$}        & \multicolumn{1}{l|}{Asteroid density in $\gpcmc$.}                                    \\ \hline
\multicolumn{1}{|l|}{$A$}           & \multicolumn{1}{l|}{Bond albedo.}                                                     \\ \hline
\multicolumn{1}{|l|}{$\theta$}      & \multicolumn{1}{l|}{Asteroid obliquity.}                                              \\ \hline
                                    &                                                                                      
\end{tabular}
\end{table}

\subsection{Complete V-shape families}
\label{s.appendixcompletevshape}
\subsubsection{Erigone}
\label{s.erigone}

The Erigone asteroid family located in the inner Main Belt was first identified by \citet[][]{Zappala1995} and consists of mostly C-type asteroids \citep[][]{Masiero2013,Spoto2015}. The V-shape identification technique was applied to 1,742 asteroids belonging to the Erigone asteroid family as defined by \citet[][]{Nesvorny2015a}. Eqs.~\ref{eq.border_method_N_outer} and \ref{eq.border_method_N_inner} are integrated with the interval [$0.04,0.73$] for the Dirac delta function $\delta(D_{r,j}-D_r )$. Eq.~\ref{eqn.apvDvsCfinal} is truncated to 0.04 km$^{-1}$ for $D_r$ $ <$ $0.04$ km$^{-1}$ and to 0.73 km$^{-1}$ for $D_r$ $>$ 0.73 km$^{-1}$. Asteroid $H$ values were converted to $D$ using Eq.~\ref{eq.HtoD} using the value of $\pv$ = 0.05 typical for members of the Erigone family \citep[][]{Masiero2013,Spoto2015}. 

\begin{figure}
\centering
\hspace*{-1.1cm}
\ifincludeplots
\includegraphics[scale=0.55]{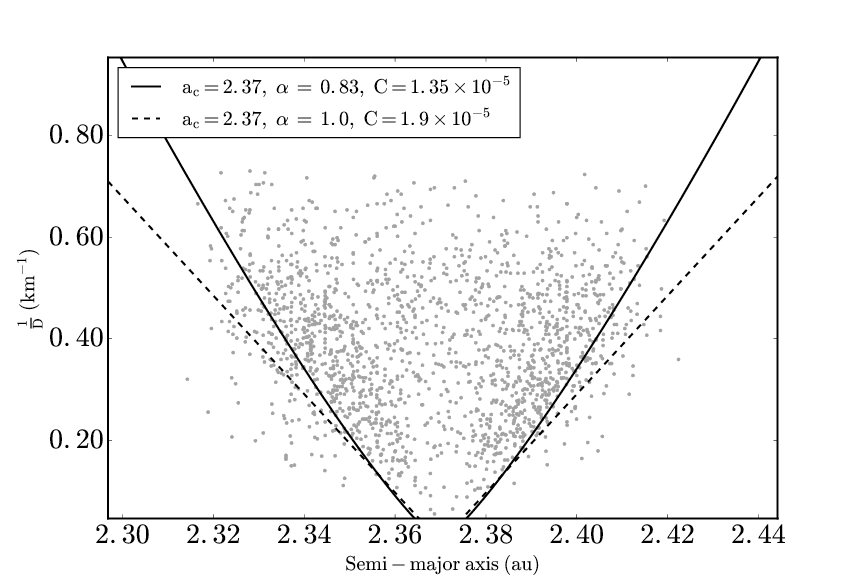}
\else
I am not enabling plots.
\fi
\caption{$a$ vs.$\frac{1}{D}$ plot for Erigone with V-shape borders that have $\alpha \; = \; 0.83$ and $\alpha \; = \; 1.0$.}
\label{fig.ErigoneTwoVs}
\end{figure}

$\frac{N_{in}^2}{N_{out}}$ at $(a_c, \; C, \; \alpha) \; = \; (2.37 \; \mathrm{au}, \; 1.34 \times 10^{-5} \; \mathrm{au}, \;  \sim0.85)$ as seen in the top panel of Fig.~\ref{fig.ErigoneBorderDensity1} and is $\sim$ 3 standard deviations above the mean value of $\frac{N_{in}^2}{N_{out}}$ The technique was repeated with the joint Erigone and Martes family defined by \citet[][]{Milani2014} resulting in similar results as seen in Fig.~\ref{fig.ErigoneBorderDensity1Milani}.We repeated the process in $\sim$2,000 Monte Carlo runs where the physical parameters of the family fragments were randomly varied in each run as described in Section~\ref{s.montecarlo}.  The $\pv$ of asteroids in the Monte Carlo trails was assumed to be the average value of $\pv$ for family fragments in the Erigone family fragments of 0.05 with an uncertainty of 0.01 \citep[][]{Spoto2015}. The Monte Carlo trial values of $\alpha$ in $\sim$ is $\sim$0.83 with a RMS uncertainty of 0.04 as seen in Fig.~\ref{fig.ErigoneMC}. The Erigone family V-shape is better fit with $\alpha \; = \; 0.83$  than the V-shape with $\alpha \; = \; 1.0$ as seen in Fig.~\ref{fig.ErigoneTwoVs}.

 The family age of 90 $\pm$ 40 Myrs is calculated using Eq.~\ref{eqn.familyagenoejectionpvg}, with $C_{YE}$ = $5.6 \; \times 10^{-6}$ au calculated from Eq.~\ref{eqn.Ccombo} where $C\; = \; 1.35 \; \times 10^{-5}$ au.  The value of $\mu_{\alpha} \; =$ 0.83 and $C_{EV} \; = \; 7.90 \; \times 10^{-6}$ au calculated using Eq.~\ref{eqn.VEVvsCalphaFinal} assuming $V_{EV}$ = 30 $\mps$ from \citet[][]{Vokrouhlicky2006b}. Note that the 90 Myr age from this estimate is the same minimum amount of time needed to maintain a steady state population of C-type asteroids in the z2 resonance that interacts with members of the Erigone family \citep[][]{Carruba2016h}.
 
\subsubsection{Massalia}
\label{s.Massalia}

The Massalia asteroid family located in the inner Main Belt was first identified by \citet[][]{Zappala1995} and consists of mostly S-type asteroids \citep[][]{Masiero2013,Spoto2015}. The V-shape identification technique was applied to 6,414 asteroids belonging to the Massalia asteroid family as defined by \citet[][]{Nesvorny2015a}. The interval [$0.09,2.2$] for the Dirac delta function $\delta(D_{r,j}-D_r )$ is used and Eq.~\ref{eqn.apvDvsCfinal} is truncated to 0.04 km$^{-1}$ for $D_r$ $ <$ $0.04$ km$^{-1}$ and to 0.73 km$^{-1}$ for $D_r$ $>$ 0.73 km$^{-1}$. Asteroid $H$ values were converted to $D$ using Eq.~\ref{eq.HtoD} and $\pv$ = 0.24 typical for members of the Massalia family \citep[][]{Masiero2013,Spoto2015}. The peak in $\frac{N_{in}^2}{N_{out}}$ at $(a_c, \; C, \; \alpha) \; = \; (2.41 \; \mathrm{au}, \; 1.95 \times 10^{-5} \; \mathrm{au}, \;  \sim0.76)$.The technique was repeated with the Massalia family defined by \citet[][]{Milani2014} resulting in similar results as seen in Fig.~\ref{fig.MassaliaBorderAlphaMilani}.

\begin{figure}
\centering
\hspace*{-0.7cm}
\ifincludeplots
\includegraphics[scale=0.455]{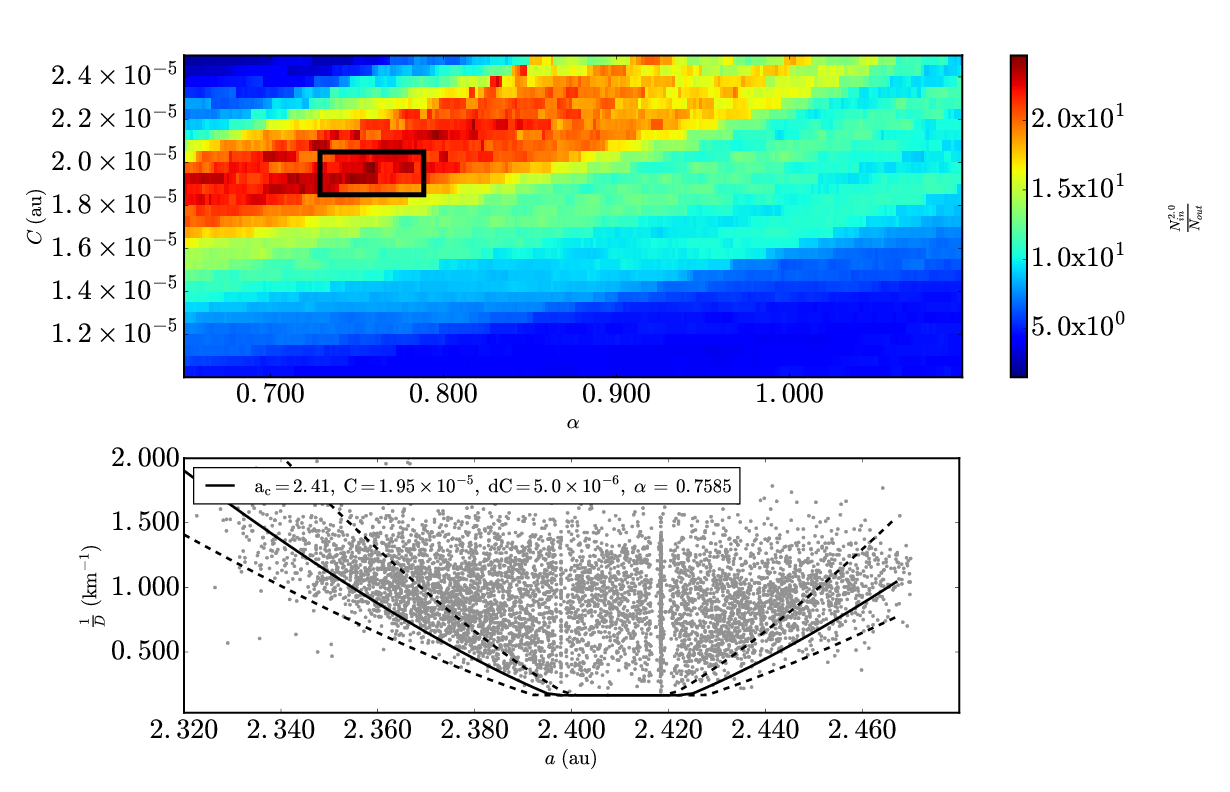}
\else
I am not enabling plots.
\fi
\caption{The same as Fig.~\ref{fig.synErig200Myrs} for Massalia asteroid family data from \citet[][]{Nesvorny2015a}. (Top panel) $\Delta \alpha$ is equal to $1.8 \times 10^{-3}$ au and $\Delta C$, is equal to $5.0 \times 10^{-7}$ au. (Bottom Panel) $D_r(a,a_c,C\pm dC,\pv,\alpha)$ is plotted with $\pv = 0.24$, $a_c$ = 2.41 au and $dC \; = \; 5.0 \x 10^{-6}$ au.}
\label{fig.MassaliaBorderAlpha}
\end{figure}

\begin{figure}
\centering
\hspace*{-0.7cm}
\ifincludeplots
\includegraphics[scale=0.455]{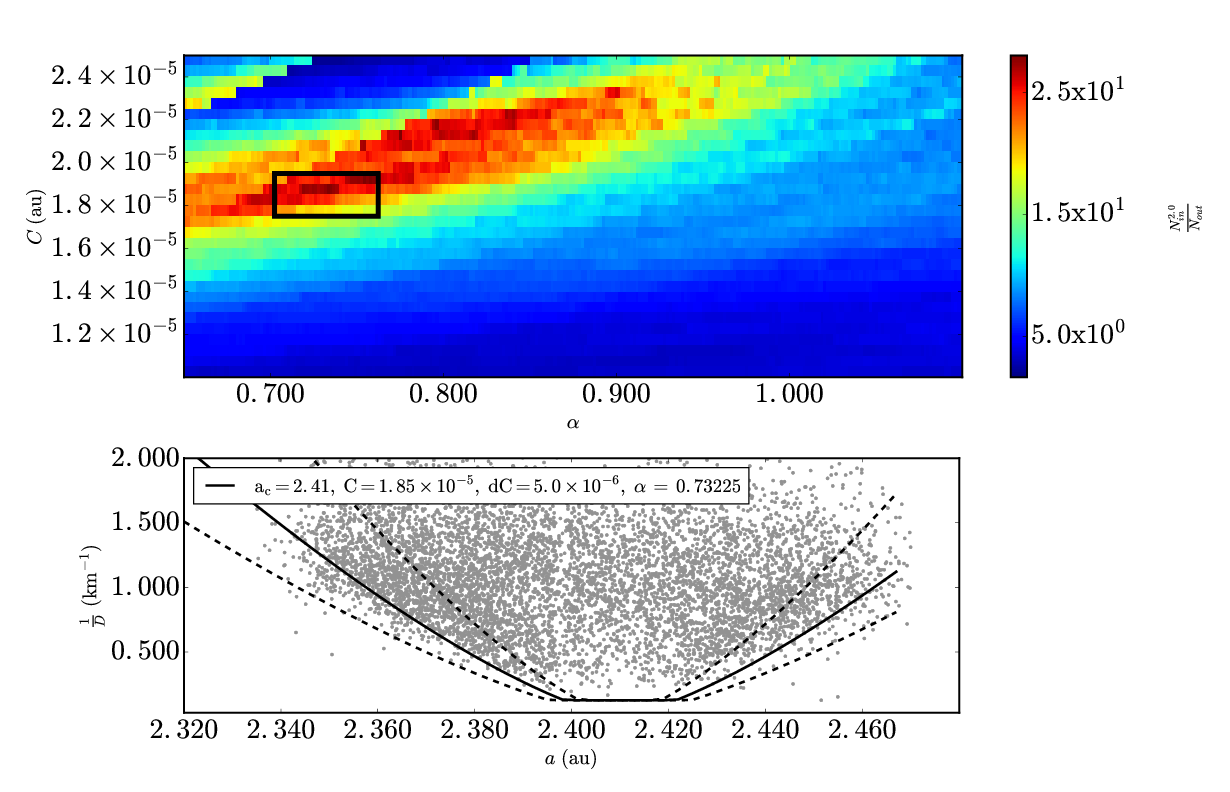}
\else
I am not enabling plots.
\fi
\caption{Same as Fig.~\ref{fig.MassaliaBorderAlpha}, but repeated for the Massalia family defined by \citet[][]{Milani2014}.}
\label{fig.MassaliaBorderAlphaMilani}
\end{figure} 

\begin{figure}
\centering
\hspace*{-0.9cm}
\ifincludeplots
\includegraphics[scale=0.3225]{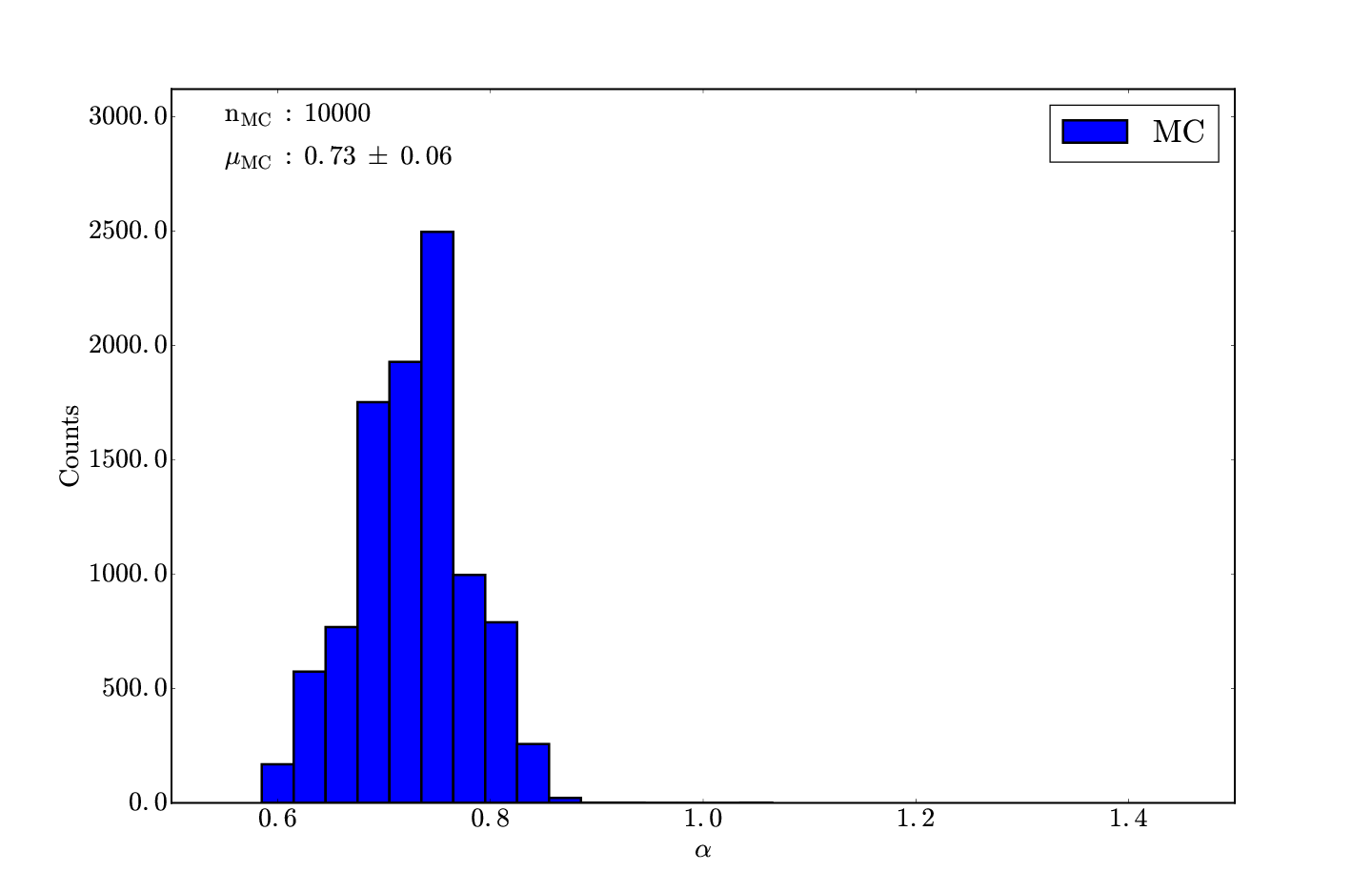}
\else
I am not enabling plots.
\fi
\caption{The same as Fig.~\ref{fig.ErigoneMC} with $\sim$10,000 trials repeating the V-shape technique for the Massalia family. The mean of the distribution is centered at $\alpha$ = 0.73 $\pm$ 0.06 and the bin size in the histogram is 0.03.}
\label{fig.MassaliaMC}
\end{figure}

\begin{figure}
\centering
\hspace*{-1.1cm}
\ifincludeplots
\includegraphics[scale=0.55]{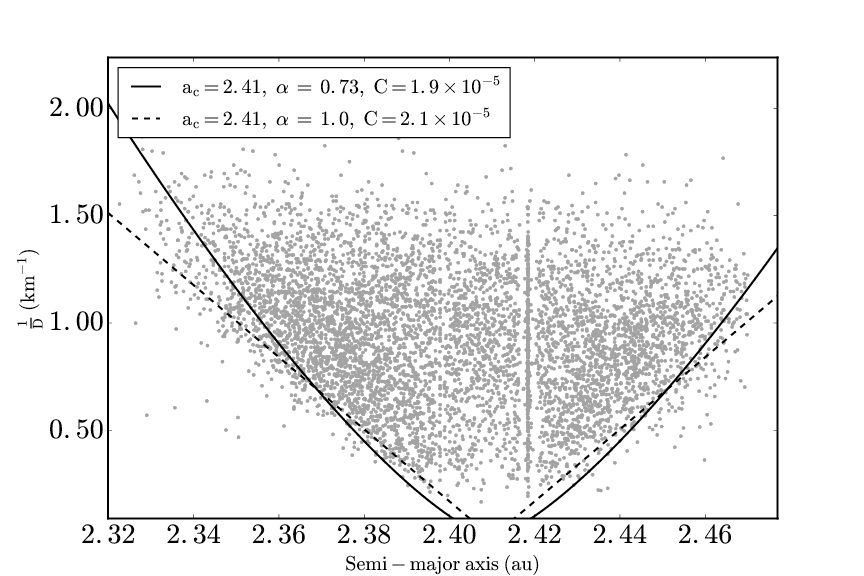}
\else
I am not enabling plots.
\fi
\caption{$a$ vs.$\frac{1}{D}$ plot for Massalia with V-shape borders that have $\alpha \; = \; 0.73$ and $\alpha \; = \; 1.0$.}
\label{fig.MassaliaTwoVs}
\end{figure}

$\sim$10,000 Monte Carlo runs were completed by randomizing $H$ magnitudes by  0.25 and $\pv$ values were assumed to be 0.24 with an uncertainty of 0.07 as described for the Massalia family \citep[][]{Spoto2015}. The mean value of $\alpha$ is $\sim$0.73 $\pm$ 0.06 as seen in Fig.~\ref{fig.MassaliaMC}. The Massalia family V-shape is better fit with $\alpha \; = \; 0.73$  than the V-shape with $\alpha \; = \; 1.0$ as seen in Fig.~\ref{fig.MassaliaTwoVs}.

 The family age of 150 $\pm$ 70 Myrs is calculated using Eq.~\ref{eqn.familyagenoejectionpvg}, with $C_{YE}$ = $1.1 \; \times 10^{-5}$ au calculated from Eq.~\ref{eqn.Ccombo} where $C\; = \; 1.95 \; \times 10^{-5}$ au. The value of $\mu_{\alpha} \; =$ 0.83 and $C_{EV} \; = \; 8.8 \; \times 10^{-6}$ au calculated using Eq.~\ref{eqn.VEVvsCalphaFinal} assuming $V_{EV}$ = 20 $\mps$ from \citet[][]{Vokrouhlicky2006b}, $a_c$  = 2.41 au, $e_{\mu}$ = 0.16, $\rho_{mu}$ = 2.3 $\gpcmc$, $\pv$ = 0.24 and $G_{\mu}$ = 0.24. 

\subsubsection{Misa(2)}
\label{s.Misa}

\begin{figure}
\centering
\hspace*{-0.7cm}
\ifincludeplots
\includegraphics[scale=0.455]{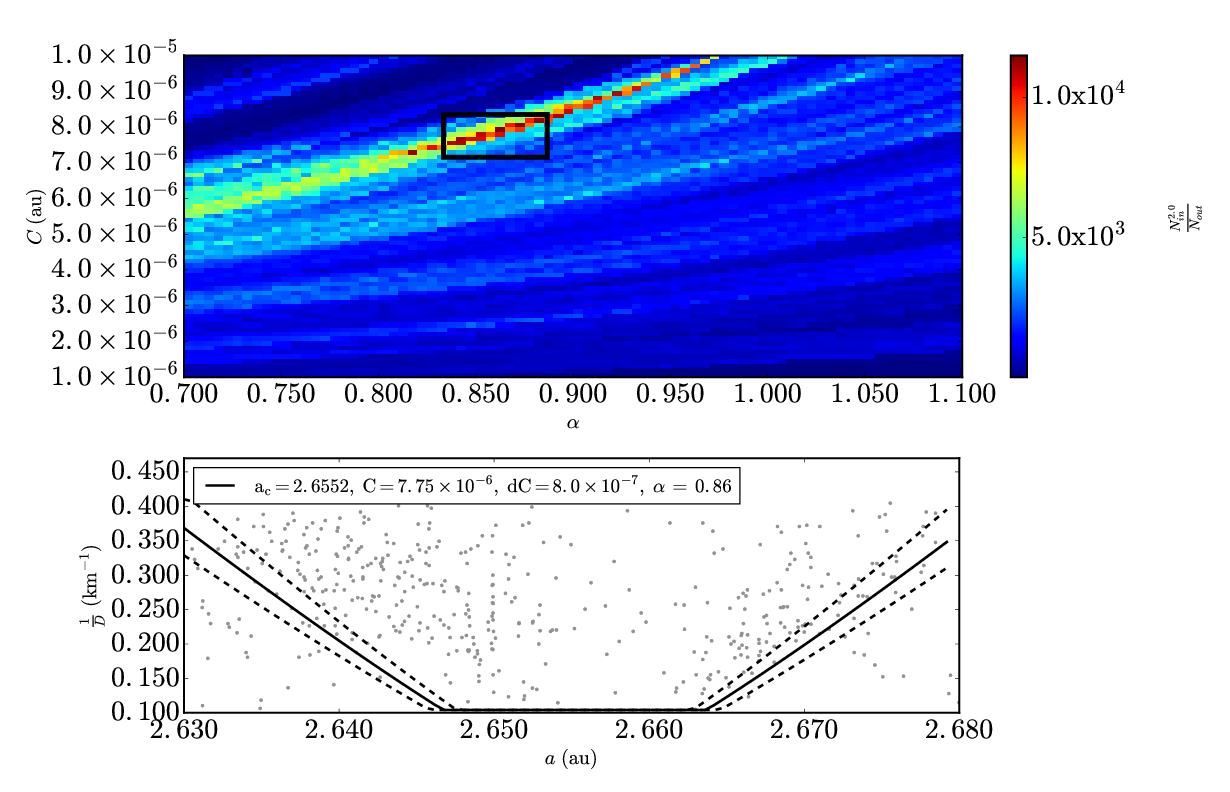}
\else
I am not enabling plots.
\fi
\caption{The same as Fig.~\ref{fig.synErig200Myrs} for the Misa sub-family with data from \citet[][]{Nesvorny2015a}. (Top panel) $\Delta \alpha$ is equal to $5.0 \times 10^{-3}$ au and $\Delta C$, is equal to $1.3 \times 10^{-6}$ au. (Bottom Panel) $D_r(a,a_c,C\pm dC,\pv,\alpha)$ is plotted with $\pv = 0.1$, $a_c$ = 2.655 au and $dC \; = \; 1.3 \x 10^{-7}$ au.}\label{fig.MisaSubAlpha}
\end{figure}

The C-type Misa family has been noted to have  sub asteroid family located inside of it \citep[][]{Milani2014,Nesvorny2015a} that we will call Misa(2).  The V-shape identification technique was applied to 427 asteroids belonging to the Misa(2) asteroid family as defined by \citet[][]{Nesvorny2015a}.  $\frac{N_{in}^2}{N_{out}}$ at $(a_c, \; C, \; \alpha) $ maximizes at $ \; (2.66 \; \mathrm{au}, \; 7.75 \times 10^{-6} \; \mathrm{au}, \;  \sim0.86)$ as seen in the top panel of Fig.~\ref{fig.MisaSubAlpha} and is $\sim$ 8 standard deviations above the mean value of $\frac{N_{in}^2}{N_{out}}$. 

The Monte Carlo tests have a mean value of $\alpha$ is $\sim$0.87 $\pm$ 0.11 with positive skew as seen in Fig.~\ref{fig.MisaSubMC}. The family age of 120 $\pm$ 60 Myrs is calculated using Eq.~\ref{eqn.familyagenoejectionpvg}, with $C_{YE}$ = $5.5 \; \times 10^{-6}$ au calculated from Eq.~\ref{eqn.Ccombo} where $C\; = \; 7.75 \; \times 10^{-6}$ au and  The value of $\mu_{\alpha} \; =$ 0.87 and $C_{EV} \; = \; 8.8 \; \times 10^{-6}$ au calculated using Eq.~\ref{eqn.VEVvsCalphaFinal} assuming $V_{EV}$ = 12 $\mps$ which is the escape speed of a 27 km diameter body with $\rho$ = 1.4 $\gpcmc$. We estimate the $D$ of the parent body of the Misa(2) family by using the technique of \citet[][]{Tanga1999}. The calculation was repeating using the same parameters except with $\alpha$ = 1.0 and $C\; = \; 9.5 \; \times 10^{-6}$ au obtaining a value of 130 $\pm$ 60 Myrs.  

\subsubsection{Tamara}
\label{s.Tamara}

\begin{figure}
\centering
\hspace*{-0.7cm}
\ifincludeplots
\includegraphics[scale=0.455]{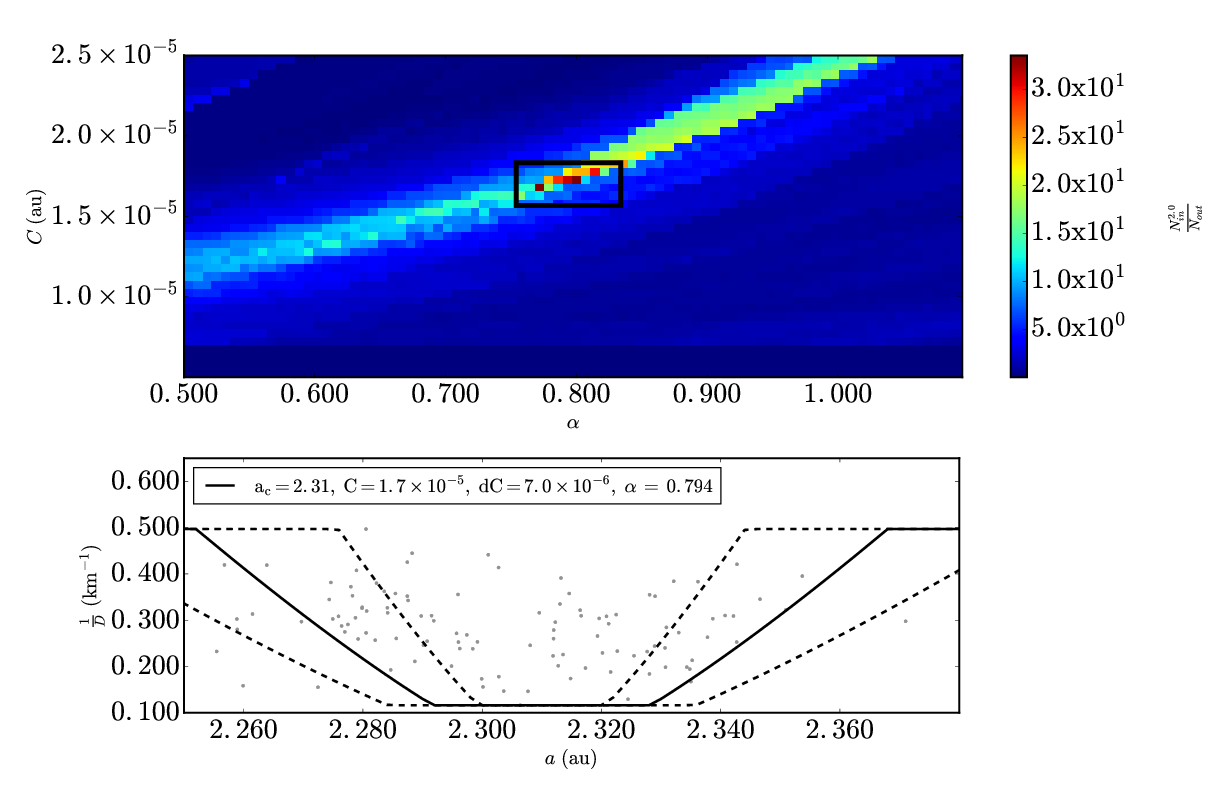}
\else
I am not enabling plots.
\fi
\caption{The same as Fig.~\ref{fig.synErig200Myrs} for the Tamara asteroid family data from \citet[][]{Novakovic2017}. (Top panel) $\Delta \alpha$ is equal to $7.0 \times 10^{-3}$ au and $\Delta C$, is equal to $5.0 \times 10^{-7}$ au. (Bottom Panel) $D_r(a,a_c,C\pm dC,\pv,\alpha)$ is plotted with $\pv = 0.06$, $a_c$ = 2.310 au and $dC \; = \; 7.0 \x 10^{-6}$ au.}
\label{fig.TamaraAlpha}
\end{figure}

\begin{figure}
\centering
\hspace*{-0.9cm}
\ifincludeplots
\includegraphics[scale=0.3225]{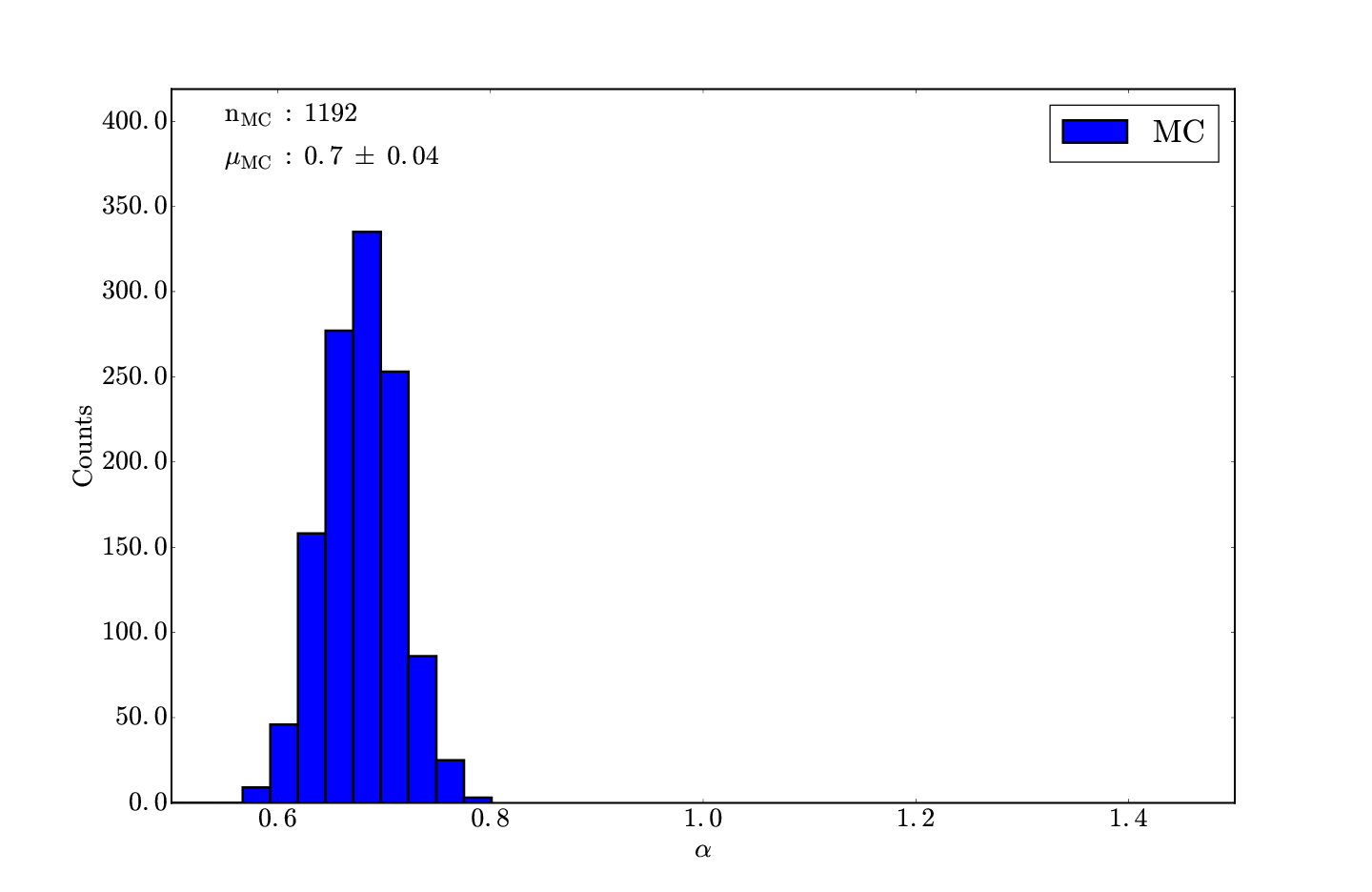}
\else
I am not enabling plots.
\fi
\caption{The same as Fig.~\ref{fig.ErigoneMC} with $\sim$1,200 trials repeating the V-shape technique for the Tamara family. The mean of the distribution is centered at $\alpha$ = 0.70 $\pm$ 0.04 and the bin size in the histogram is 0.03.}
\label{fig.TamaraMC}
\end{figure}

The Tamara is a dark family of C-type asteroids located near the high $i$ Phocaea region of the MB \citep[][]{Novakovic2017}.  The V-shape identification technique was applied to 111 asteroids belonging to the Tamara asteroid family as defined by \citet[][]{Novakovic2017} with $\pv \; <$ 0.1. Only asteroids with known $D$ measurements from \citet[][]{Masiero2011} were used. Asteroid $\pv$ values were calculated with $H$ values from \citet[][]{Veres2015} and $D$ from \citet[][]{Masiero2011} according to
\begin{equation}
\label{eq.DHtoPV}
\pv = 8.94 \x 10^{16} \; \frac{10^{0.4 \; (m_\odot  \; - \; H)}}{D^2}
 \end{equation}
 from \citet[][]{Harris2002}. The interval [$0.10,0.58$] for the Dirac delta function $\delta(D_{r,j}-D_r )$ is used and Eq.~\ref{eqn.apvDvsCfinal} is truncated to 0.10 km$^{-1}$ for $D_r$ $ <$ $0.10$ km$^{-1}$ and to 0.58 km$^{-1}$ for $D_r$ $>$ 0.58 km$^{-1}$. The peak in $\frac{N_{in}^2}{N_{out}}$ at $(a_c, \; C, \; \alpha) \; = \; (2.31 \; \mathrm{au}, \; 1.7 \times 10^{-5} \; \mathrm{au}, \;  \sim0.79)$ as seen in the top panel of Fig.~\ref{fig.TamaraAlpha} and is $\sim$5 standard deviations above the mean. $\sim$1,200 with a mean value of $\alpha$ is $\sim$0.70 $\pm$ 0.04 as seen in Fig.~\ref{fig.TamaraMC}.

 The family age of 120 $\pm$ 60 Myrs is calculated using Eq.~\ref{eqn.familyagenoejectionpvg}, with $C_{YE}$ = $5.7 \; \times 10^{-6}$ au calculated from Eq.~\ref{eqn.Ccombo} where $C\; = \; 1.5 \; \times 10^{-5}$ au. The value of $\mu_{\alpha} \; =$ 0.87 and $C_{EV} \; = \; 9.3 \; \times 10^{-6}$ au calculated using Eq.~\ref{eqn.VEVvsCalphaFinal} assuming $V_{EV}$ = 46 $\mps$ which is the escape speed of a 53 km diameter body with $\rho$ = 1.4 $\gpcmc$. We estimate the $D$ of the parent body of the Tamara family by using the technique of \citet[][]{Tanga1999}. The other parameters in Eq.~\ref{eqn.familyagenoejectionpvg} used to calculate the family age for Tamara are $a_c$  = 2.31 au, $e_{\mu}$ = 0.2, $\rho_{mu}$ = 1.4 $\gpcmc$, $\pv$ = 0.06 and $G_{\mu}$ = 0.15. The calculation was repeating using the same parameters except with $\alpha$ = 1.0 and $C\; = \; 2.3 \; \times 10^{-5}$ au obtaining a value of 180 $\pm$ 90 Myrs.  

\subsection{Clipped V-shape families}
\label{s.appendixclippedvshape}
\subsubsection{Agnia}
\label{s.Agnia}

\begin{figure}
\centering
\hspace*{-0.7cm}
\ifincludeplots
\includegraphics[scale=0.455]{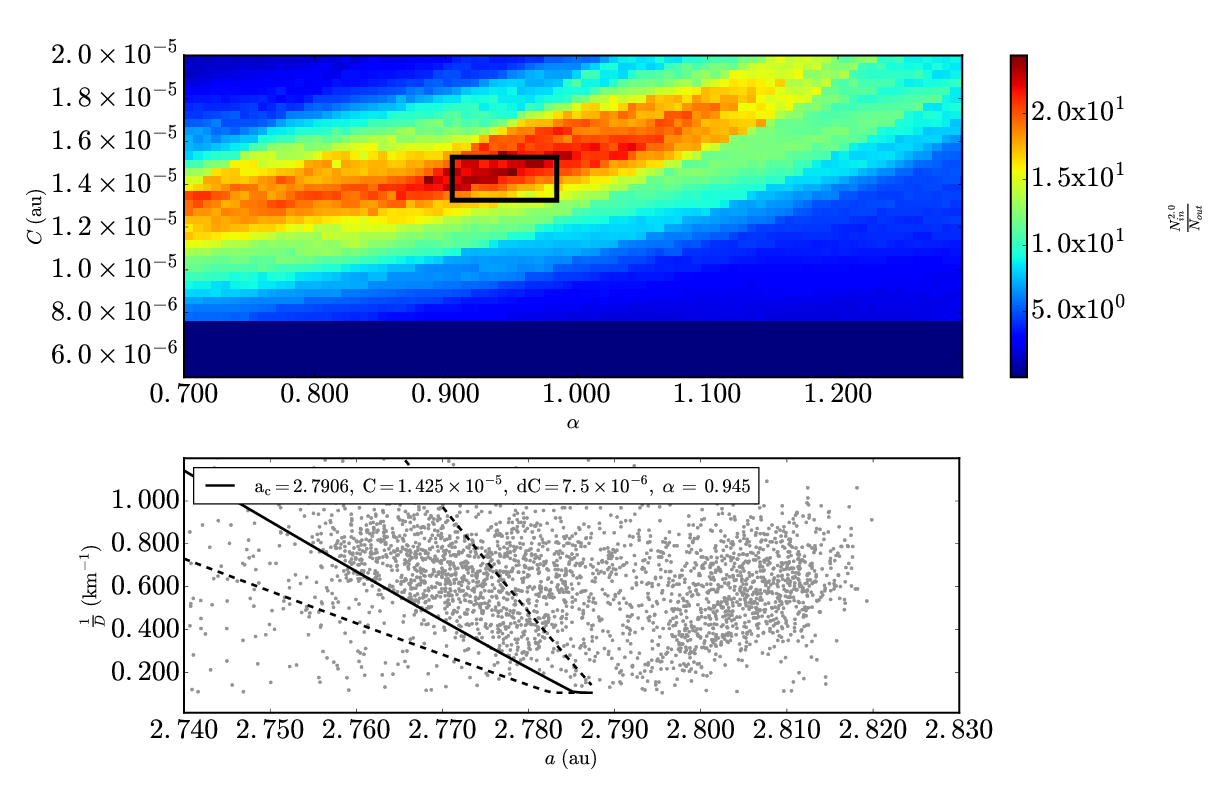}
\else
I am not enabling plots.
\fi
\caption{Same as Fig.~\ref{fig.AgniaAlpha}, but repeated for the Agnia family defined by \citet[][]{Milani2014}.}
\label{fig.AgniaBorderAlphaMilani}
\end{figure}

\begin{figure}
\centering
\hspace*{-0.9cm}
\ifincludeplots
\includegraphics[scale=0.3225]{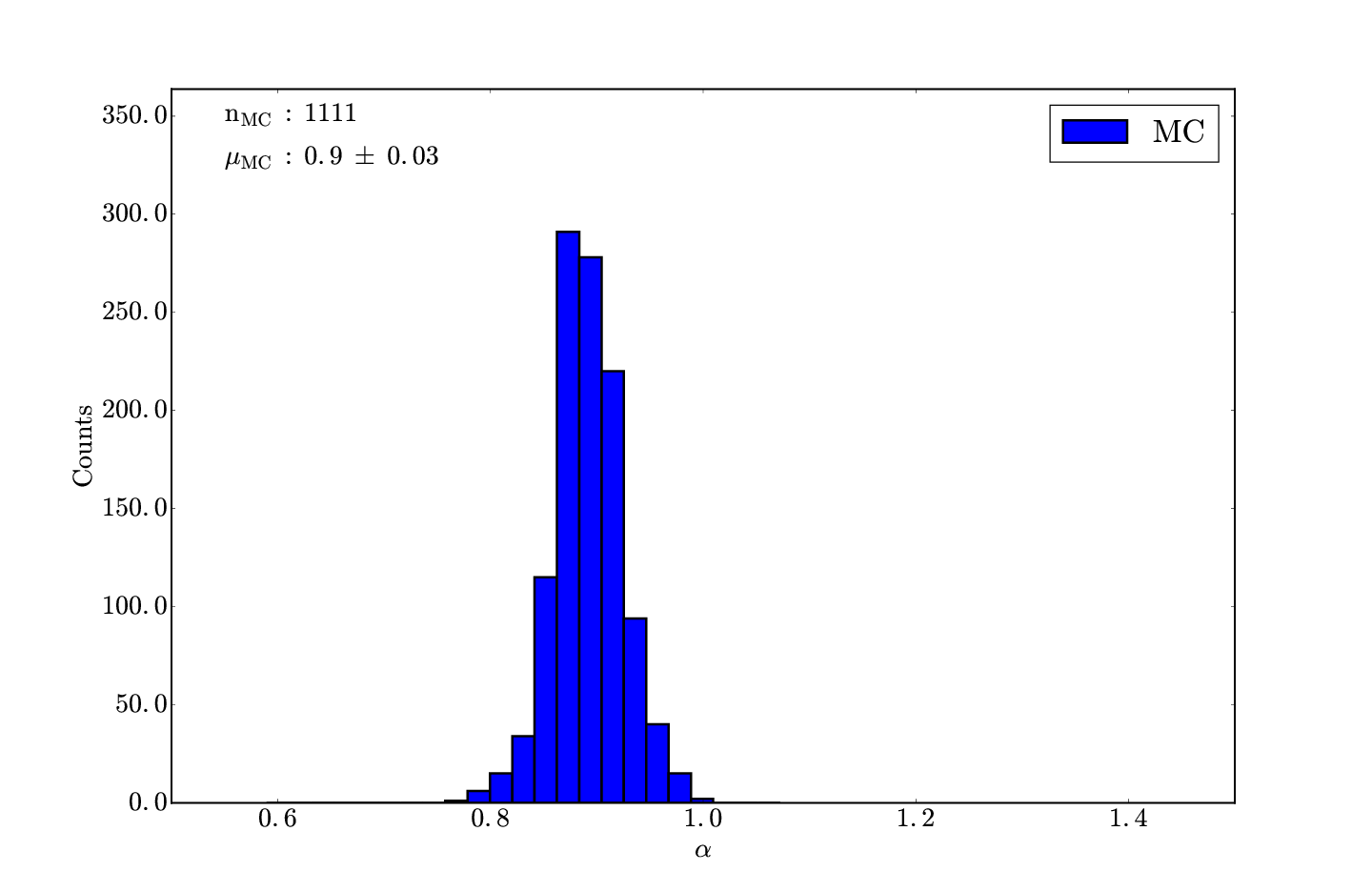}
\else
I am not enabling plots.
\fi
\caption{The same as Fig.~\ref{fig.ErigoneMC} with $\sim$1,100 trials repeating the V-shape technique for the Agnia family. The mean of the distribution is centered at $\alpha$ = 0.90 $\pm$ 0.03 and the bin size in the histogram is 0.02.}
\label{fig.AgniaMC}
\end{figure}

The S-type Agnia family is located in the central region of the MB bordering the 5:2 MMR with Jupiter \citep[][]{Zappala1995} and contains sub family Jitka \cite[][]{Milani2014}. The V-shape identification technique was applied to 2,123 asteroids belonging to the Agnia asteroid family as defined by \citet[][]{Nesvorny2015a}. The interval [$0.10,1.32$] for the Dirac delta function $\delta(D_{r,j}-D_r )$ is used and Eq.~\ref{eqn.apvDvsCfinal} is truncated to 0.10 km$^{-1}$ for $D_r$ $ <$ $0.10$ km$^{-1}$ and to 1.32 km$^{-1}$ for $D_r$ $>$ 1.32 km$^{-1}$. Asteroid $H$ values were converted to $D$ using Eq.~\ref{eq.HtoD} and $\pv$ = 0.18 typical for members of the Agnia family \citep[][]{Masiero2013,Spoto2015}. The peak in $\frac{N_{in}^2}{N_{out}}$ at $(a_c, \; C, \; \alpha) \; = \; (2.79 \; \mathrm{au}, \; 1.54 \times 10^{-5} \; \mathrm{au}, \;  \sim0.91)$ as seen in the top panel of Fig.~\ref{fig.AgniaAlpha} and is $\sim$3 standard deviations above the mean. The technique was repeated with the Agnia family defined by \citet[][]{Milani2014} resulting in similar results as seen in Fig.~\ref{fig.AgniaBorderAlphaMilani}. The Monte Carlo mean value of $\alpha$ is $\sim$0.90 $\pm$ 0.03 as seen in Fig.~\ref{fig.AgniaMC}.

 The family age of 120 $\pm$ 60 Myrs is calculated using Eq.~\ref{eqn.familyagenoejectionpvg}, with $C_{YE}$ = $7.9 \; \times 10^{-6}$ au calculated from Eq.~\ref{eqn.Ccombo} where $C\; = \; 1.5 \; \times 10^{-5}$ au and is similar to the 130 Myr age calcluated by \citet[][]{Vokrouhlicky2006b}. The value of $\mu_{\alpha} \; =$ 0.9 and $C_{EV} \; = \; 7.5 \; \times 10^{-6}$ au calculated using Eq.~\ref{eqn.VEVvsCalphaFinal} assuming $V_{EV}$ = 15 $\mps$ from \citep[][]{Vokrouhlicky2006b}. The calculation was repeating using the same parameters except with $\alpha$ = 1.0 and $C\; = \; 1.8 \; \times 10^{-5}$ obtaining a value of 100 $\pm$ 50 Myrs.  

\subsubsection{Astrid}
\label{s.Astrid}
\begin{figure}
\centering
\hspace*{-0.7cm}
\ifincludeplots
\includegraphics[scale=0.455]{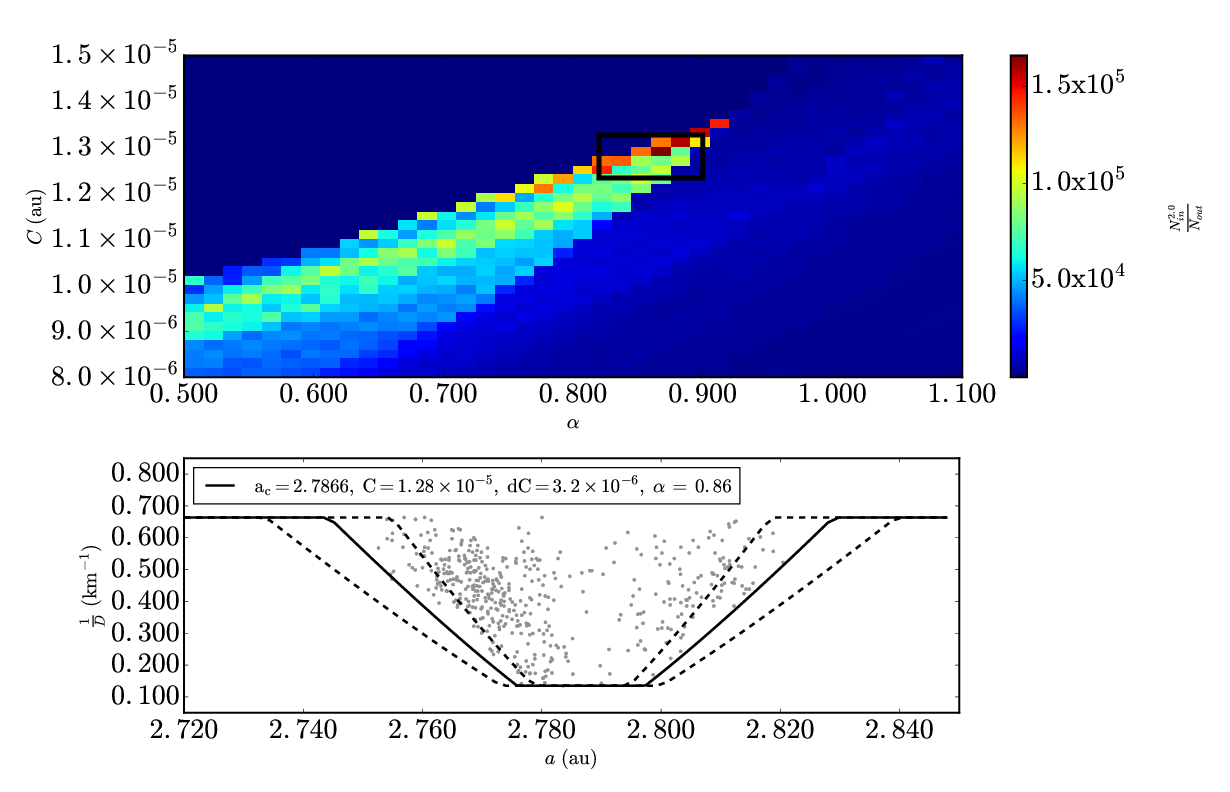}
\else
I am not enabling plots.
\fi
\caption{The same as Fig.~\ref{fig.synErig200Myrs} for Astrid asteroid family data from \citet[][]{Nesvorny2015a}. (Top panel) $\Delta \alpha$ is equal to $1.5 \times 10^{-2}$ au and $\Delta C$, is equal to $2.0 \times 10^{-7}$ au. (Bottom Panel) $D_r(a,a_c,C\pm dC,\pv,\alpha)$ is plotted with $\pv = 0.08$, $a_c$ = 2.787 au and $dC \; = \; 3.2 \x 10^{-6}$ au.}
\label{fig.AstridAlpha}
\end{figure}

The C-type Astrid family is located in the central region of the MB and borders the 5:2 MMR with Jupiter \citep[][]{Zappala1995}. Members of the family interact with the $s$ - $s_C$ nodal resonances with the asteroid Ceres affecting the distribution of its family members in $a$ vs $sini$ space \citep[][]{Carruba2016i}. The interval [$0.13,0.67$] for the Dirac delta function $\delta(D_{r,j}-D_r )$ is used and Eq.~\ref{eqn.apvDvsCfinal} is truncated to 0.13 km$^{-1}$ for $D_r$ $ <$ $0.13$ km$^{-1}$ and to 0.67 km$^{-1}$ for $D_r$ $>$ 0.67 km$^{-1}$. Asteroid $H$ values were converted to $D$ using Eq.~\ref{eq.HtoD} and $\pv$ = 0.18 typical for members of the Astrid family \citep[][]{Masiero2013,Spoto2015}. The peak in $\frac{N_{in}^2}{N_{out}}$ at $(a_c, \; C, \; \alpha) \; = \; (2.79 \; \mathrm{au}, \; 1.28 \times 10^{-5} \; \mathrm{au}, \;  \sim0.86)$ as seen in the top panel of Fig.~\ref{fig.AstridAlpha} and is $~$6 standard deviations above the mean value. The mean value of $\alpha$ from the Monte Carlo test is $\sim$0.81 $\pm$ 0.07 as seen in Fig.~\ref{fig.AstridAlpha}.

\begin{figure}
\centering
\hspace*{-0.9cm}
\ifincludeplots
\includegraphics[scale=0.3225]{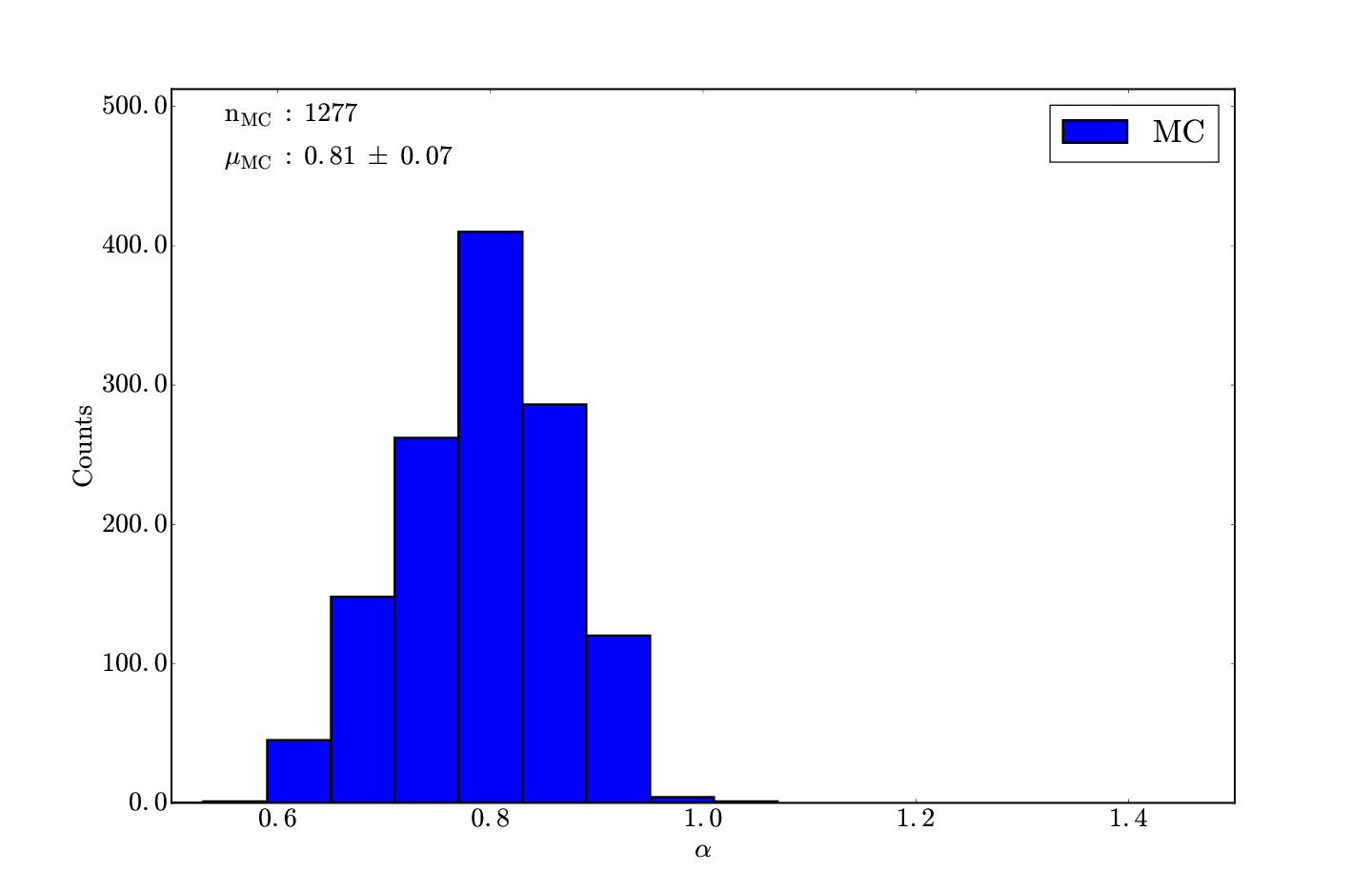}
\else
I am not enabling plots.
\fi
\caption{The same as Fig.~\ref{fig.ErigoneMC} with $\sim$1,300 trials repeating the V-shape technique for the Astrid family. The mean of the distribution is centered at $\alpha$ = 0.81 $\pm$ 0.07 and the bin size in the histogram is 0.06.}
\label{fig.PaduaMC}
\end{figure}

 The family age of 110 $\pm$ 60 Myrs is calculated using Eq.~\ref{eqn.familyagenoejectionpvg}, with $C_{YE}$ = $3.9 \; \times 10^{-6}$ au calculated from Eq.~\ref{eqn.Ccombo} where $C\; = \; 1.2 \; \times 10^{-5}$ au. This age is in agreement with the $\sim$140 Myr age for the astrid family by \citep[][]{Carruba2016i}. The value of $\mu_{\alpha} \; =$ 0.81 and $C_{EV} \; = \; 8.1 \; \times 10^{-6}$ au calculated using Eq.~\ref{eqn.VEVvsCalphaFinal} assuming $V_{EV}$ = 15 $\mps$ from \citep[][]{Vokrouhlicky2006b}. 
 
\subsubsection{Baptistina}
\label{s.Baptistina}
\begin{figure}
\centering
\hspace*{-0.7cm}
\ifincludeplots
\includegraphics[scale=0.455]{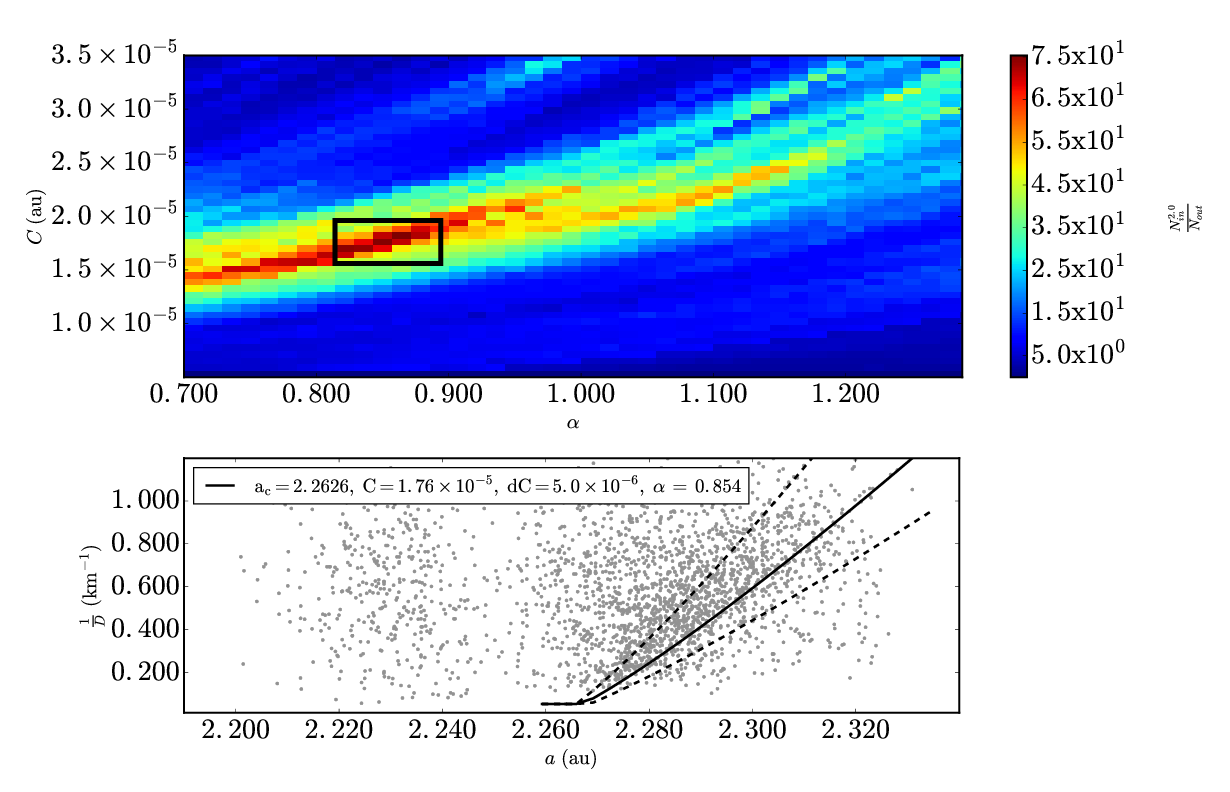}
\else
I am not enabling plots.
\fi
\caption{The same as Fig.~\ref{fig.synErig200Myrs} for Baptistina asteroid family data from \citet[][]{Nesvorny2015a}. (Top panel) $\Delta \alpha$ is equal to $1.4 \times 10^{-2}$ au and $\Delta C$, is equal to $6.0 \times 10^{-7}$ au. (Bottom Panel) $D_r(a,a_c,C\pm dC,\pv,\alpha)$ is plotted with $\pv = 0.16$, $a_c$ = 2.262 au and $dC \; = \; 5.0 \x 10^{-6}$ au.}
\label{fig.BaptistinaAlpha}
\end{figure}

\begin{figure}
\centering
\hspace*{-0.9cm}
\ifincludeplots
\includegraphics[scale=0.3225]{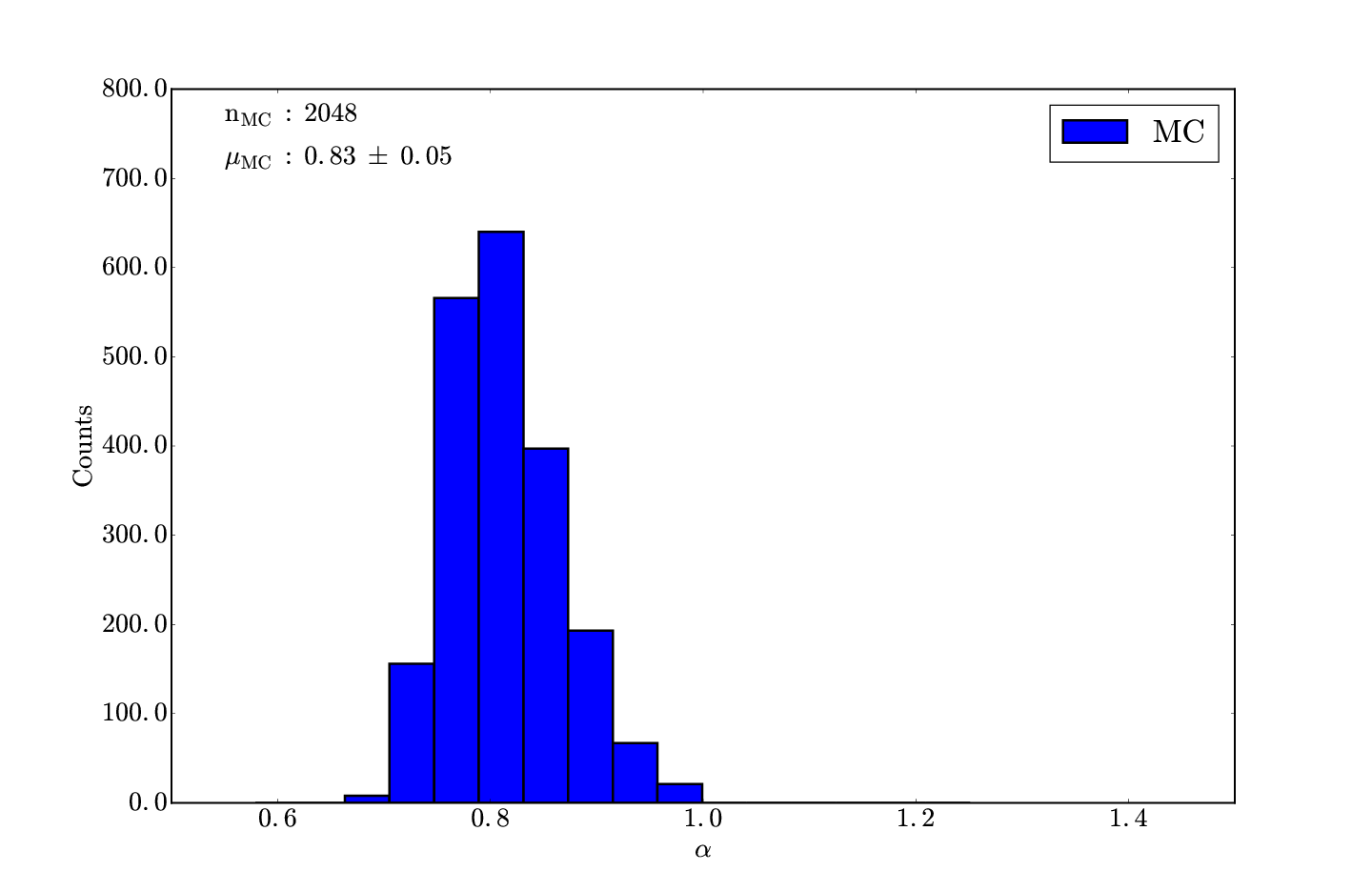}
\else
I am not enabling plots.
\fi
\caption{The same as Fig.~\ref{fig.ErigoneMC} with $\sim$2,000 trials repeating the V-shape technique for the Baptistina family. The mean of the distribution is centered at $\alpha$ = 0.83 $\pm$ 0.05 and the bin size in the histogram is 0.04.}
\label{fig.BaptistinaMC}
\end{figure}

The X-type Baptistina family is located in the inner region of the MB and borders the 7:2 /5:9 MMR with Jupiter/Mars \citep[][]{Knezevic2003,Mothe-Diniz2005,Bottke2007}. The taxonomy of the Baptistina families may also be closer to S-types \citep[][]{Reddy2009, Reddy2011}.The V-shape identification technique was applied to 2,450 asteroids belonging to the Baptistina asteroid family as defined by \citet[][]{Nesvorny2015a}. The peak in $\frac{N_{in}^2}{N_{out}}$ at $(a_c, \; C, \; \alpha) \; = \; (2.26 \; \mathrm{au}, \; 1.76 \times 10^{-5} \; \mathrm{au}, \;  \sim0.85)$ as seen in the top panel of Fig.~\ref{fig.BaptistinaAlpha} and is $\sim$5 standard deviations above the mean value. $\sim$2,000 Monte Carlo runs where the mean value of $\alpha$ is $\sim$0.83 $\pm$ 0.05 as seen in Fig.~\ref{fig.BaptistinaMC}.

 The family age of 200 $\pm$ 100 Myrs is calculated using Eq.~\ref{eqn.familyagenoejectionpvg}, with $C_{YE}$ = $1.0 \; \times 10^{-5}$ au calculated from Eq.~\ref{eqn.Ccombo} where $C\; = \; 1.76 \; \times 10^{-5}$ au. The value of $\mu_{\alpha} \; =$ 0.83 and $C_{EV} \; = \; 7.6 \; \times 10^{-6}$ au calculated using Eq.~\ref{eqn.VEVvsCalphaFinal} assuming $V_{EV}$ = 21 $\mps$ from \citep[][]{Broz2013a}.
  
\subsubsection{Dora(2)}
\label{s.Dora2}
\begin{figure}
\centering
\hspace*{-0.7cm}
\ifincludeplots
\includegraphics[scale=0.455]{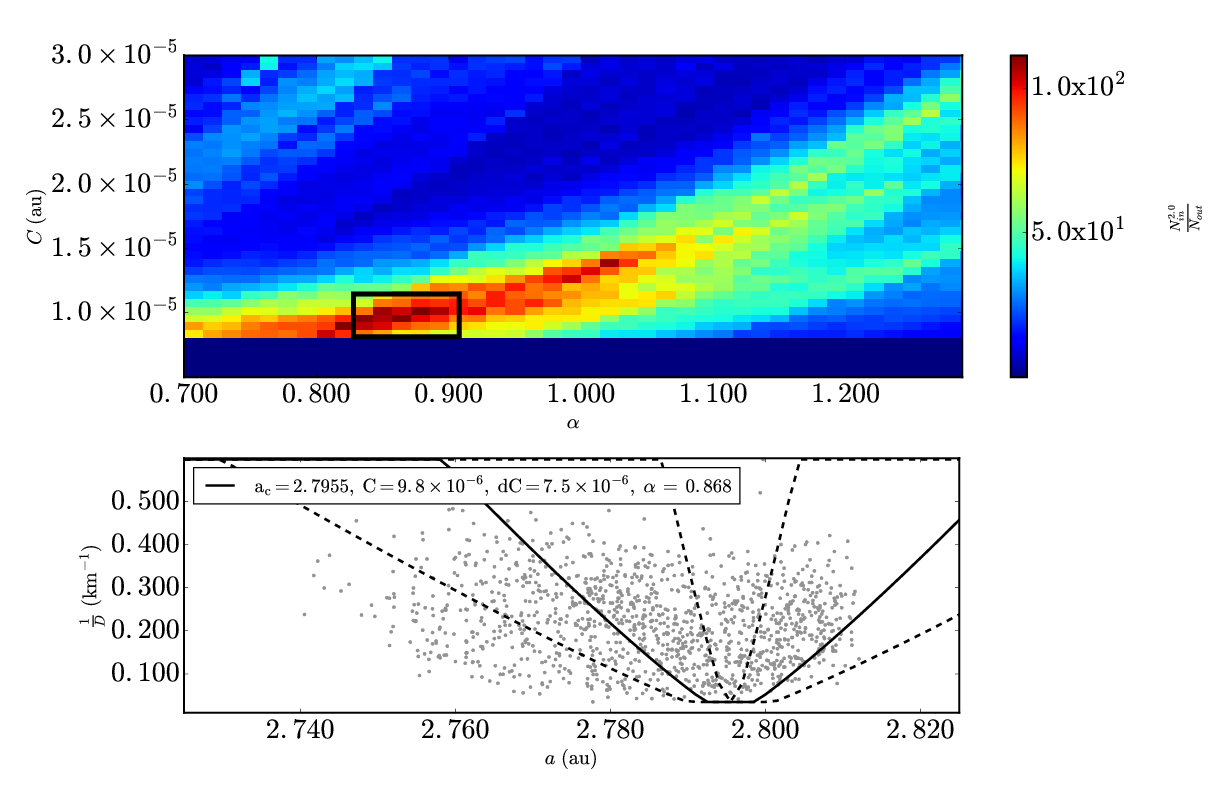}
\else
I am not enabling plots.
\fi
\caption{The same as Fig.~\ref{fig.synErig200Myrs} for the Dora asteroid family data from \citet[][]{Nesvorny2015a}. (Top panel) $\Delta \alpha$ is equal to $1.4 \times 10^{-2}$ au and $\Delta C$, is equal to $6.0 \times 10^{-7}$ au. (Bottom Panel) $D_r(a,a_c,C\pm dC,\pv,\alpha)$ is plotted with $\pv = 0.05$, $a_c$ = 2.796 au and $dC \; = \; 7.5 \x 10^{-6}$ au.}
\label{fig.DoraAlpha}
\end{figure}

\begin{figure}
\centering
\hspace*{-0.9cm}
\ifincludeplots
\includegraphics[scale=0.3225]{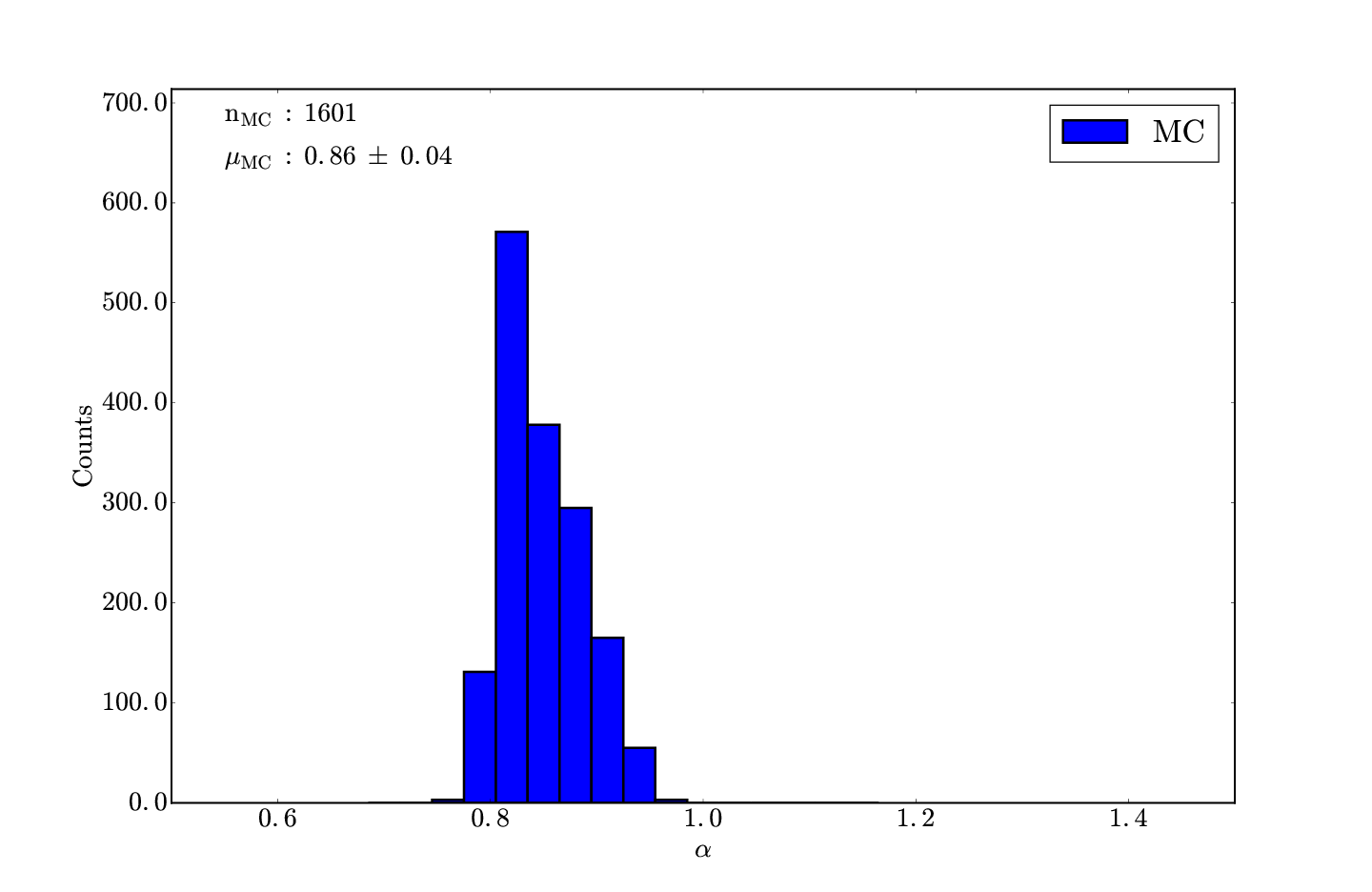}
\else
I am not enabling plots.
\fi
\caption{The same as Fig.~\ref{fig.ErigoneMC} with $\sim$1,600 trials repeating the V-shape technique for the Dora(2) family. The mean of the distribution is centered at $\alpha$ = 0.86 $\pm$ 0.04 and the bin size in the histogram is 0.03.}
\label{fig.DoraMC}
\end{figure}

The C-type Dora located in the central region of the MB contains a sub-family with a clipped V-shape \citep[][]{Nesvorny2015a} that we will call Dora(2), and borders the 5:2 MMR with Jupiter. The V-shape identification technique was applied to 1,223 asteroids belonging to the Dora asteroid family as defined by \citet[][]{Nesvorny2015a}. The peak in $\frac{N_{in}^2}{N_{out}}$ at $(a_c, \; C, \; \alpha) \; = \; (2.8 \; \mathrm{au}, \; 9.8 \times 10^{-5} \; \mathrm{au}, \;  \sim0.87)$ for the Dora(2) sub-family as seen in the top panel of Fig.~\ref{fig.DoraAlpha} and is $\sim$4 standard deviations above the mean value. The mean value of $\alpha$ in the Monte Carlo trials is $\sim$0.86 $\pm$ 0.04 as seen in Fig.~\ref{fig.DoraMC}.

\begin{figure}
\centering
\hspace*{-1.1cm}
\ifincludeplots
\includegraphics[scale=0.55]{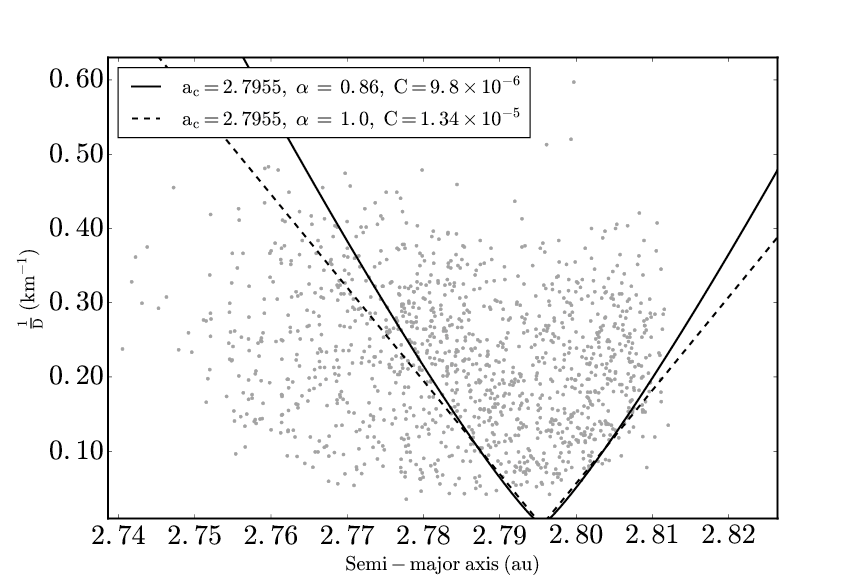}
\else
I am not enabling plots.
\fi
\caption{$a$ vs.$\frac{1}{D}$ plot for Dora(2) with V-shape borders that have $\alpha \; = \; 0.86$ and $\alpha \; = \; 1.0$}
\label{fig.DoraTwoVs}
\end{figure}

V-shapes with $(a_c, \; C, \; \alpha) \; = \; (2.8 \; \mathrm{au}, \; 9.8 \times 10^{-6} \; \mathrm{au}, \; 0.86)$ and  $(a_c, \; C, \; \alpha) \; = \; (2.8 \; \mathrm{au}, \; 1.3 \times 10^{-5} \; \mathrm{au}, \; 1.0)$ according to Eq.~\ref{eqn.apvDvsCfinal} are over plot on the the V-shape with $\alpha \; = \; 1.0$ was obtained by repeating the V-shape technique with the fixed value of $\alpha$ = 1.0. The Dora(2) family V-shape is better fit with $\alpha \; = \; 0.86$  than the V-shape with $\alpha \; = \; 1.0$ as seen in Fig.~\ref{fig.DoraTwoVs}.

 The family age of 100 $\pm$ 50 Myrs is calculated using Eq.~\ref{eqn.familyagenoejectionpvg}, with $C_{YE}$ = $5.8 \; \times 10^{-6}$ au calculated from Eq.~\ref{eqn.Ccombo} where $C\; = \; 9.8 \; \times 10^{-6}$ au. The value of $\mu_{\alpha} \; =$ 0.86 and $C_{EV} \; = \; 4.0 \; \times 10^{-6}$ au calculated using Eq.~\ref{eqn.VEVvsCalphaFinal} assuming $V_{EV}$ = 15 $\mps$ which is the escape speed of a 27 km diameter body with $\rho$ = 1.4 $\gpcmc$. 
  
\subsubsection{Eos}
\label{s.Eos}
The K-type Eos family is located in the outer region of the MB, and is bracketed by the 7:3 and 11:5 MMR and the z1 resonance, and bisected by the 9:4 MMR with Jupiter, respectively \citep[][]{Hirayama1918,Zappala1990, Carruba2007,Broz2013a}. The V-shape identification technique was applied to 6,897 asteroids belonging to the Eos asteroid family as defined by \citet[][]{Nesvorny2015a}. The interval [$0.04, 0.34$] for the Dirac delta function $\delta(D_{r,j}-D_r )$ is used and Eq.~\ref{eqn.apvDvsCfinal} is truncated to 0.05 km$^{-1}$ for $D_r$ $ <$ $0.05$ km$^{-1}$ and to 0.34 km$^{-1}$ for $D_r$ $>$ 0.34 km$^{-1}$. The lower bound on including objects with $D_r$ $<$ 0.05 excludes objects that have not had their original spin axes modified by the YORP effect over the age of the Eos family \citep[][]{Hanus2018}. Asteroid $H$ values were converted to $D$ using Eq.~\ref{eq.HtoD} and $\pv$ = 0.13 typical for members of the Eos family \citep[][]{Masiero2013,Spoto2015}.
  
\begin{figure}
\centering
\hspace*{-0.7cm}
\ifincludeplots
\includegraphics[scale=0.455]{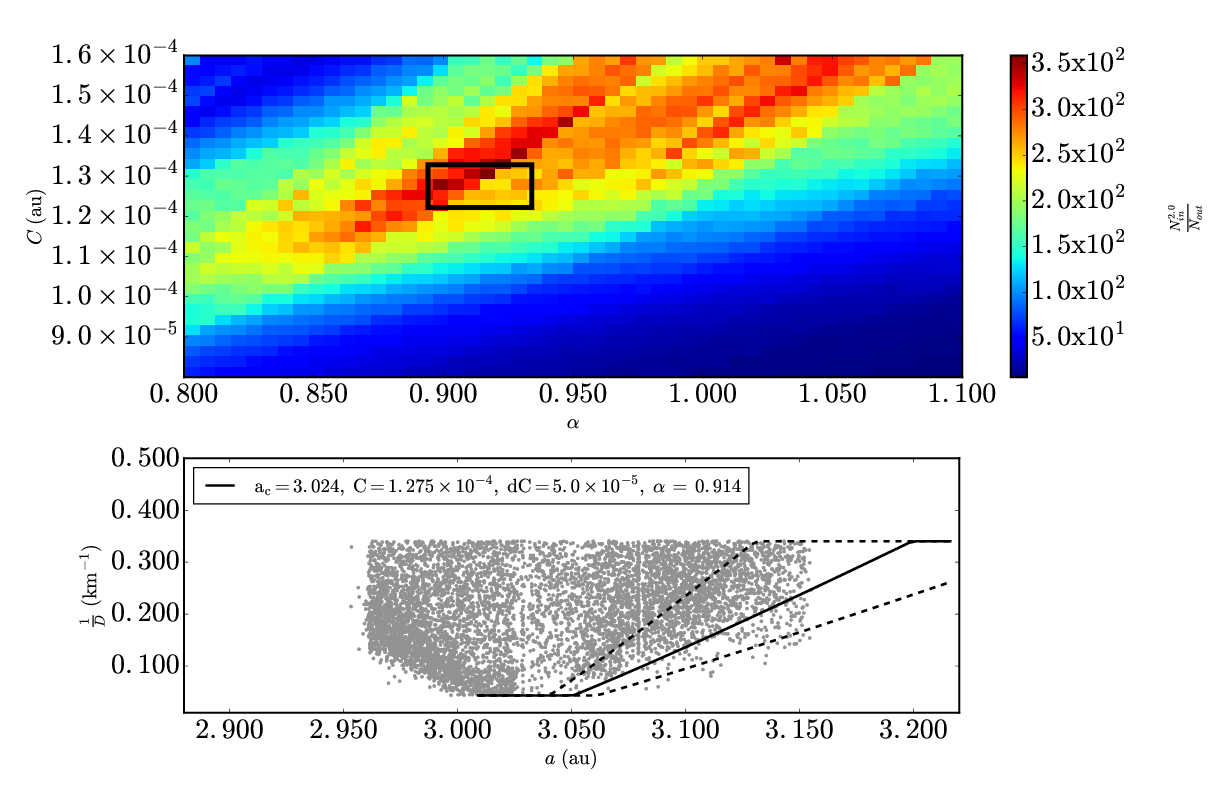}
\else
I am not enabling plots.
\fi
\caption{The same as Fig.~\ref{fig.synErig200Myrs} for Eos asteroid family data from \citet[][]{Nesvorny2015a}. (Top panel) $\Delta \alpha$ is equal to $6.0 \times 10^{-3}$ au and $\Delta C$, is equal to $2.5 \times 10^{-6}$ au. (Bottom Panel) $D_r(a,a_c,C\pm dC,\pv,\alpha)$ is plotted with $\pv = 0.13$, $a_c$ = 3.024 au and $dC \; = \; 5.0 \x 10^{-5}$ au.}
\label{fig.EosAlpha}
\end{figure}

The peak in $\frac{N_{in}^2}{N_{out}}$ at $(a_c, \; C, \; \alpha) \; = \; (3.02 \; \mathrm{au}, \; 1.28 \times 10^{-4} \; \mathrm{au}, \;  \sim0.91)$ as seen in the top panel of Fig.~\ref{fig.EosAlpha} and is $\sim$3 standard deviations above the mean value. The mean value of $\alpha$ in the Monte Carlo trials is $\sim$0.92 $\pm$ 0.02 as seen in Fig.~\ref{fig.EosAlpha}.

\begin{figure}
\centering
\hspace*{-0.9cm}
\ifincludeplots
\includegraphics[scale=0.3225]{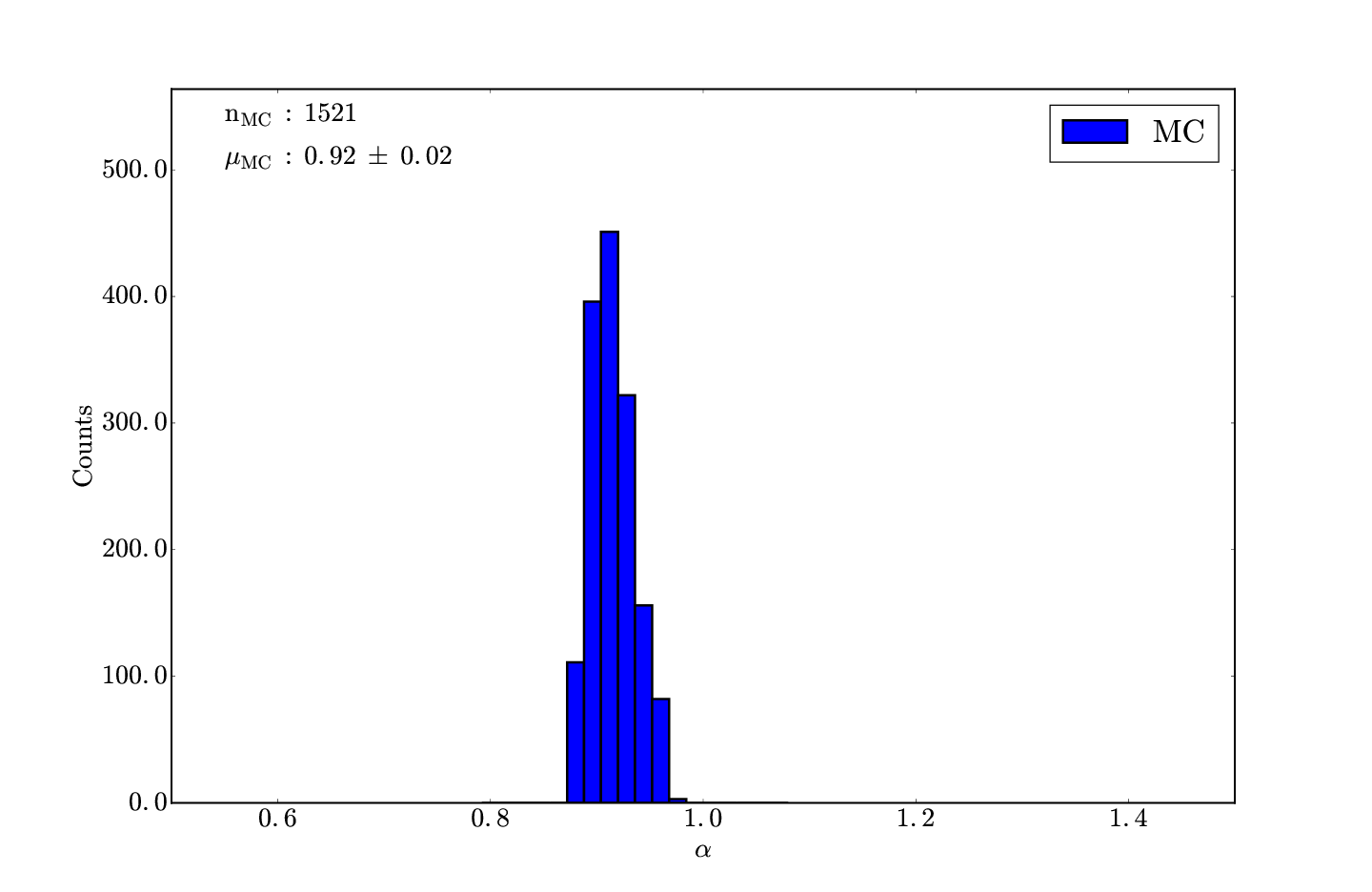}
\else
I am not enabling plots.
\fi
\caption{The same as Fig.~\ref{fig.ErigoneMC} with $\sim$1,500 trials repeating the V-shape technique for the Eos family. The mean of the distribution is centered at $\alpha$ = 0.92 $\pm$ 0.02 and the bin size in the histogram is 0.02.}
\label{fig.EosMC}
\end{figure}

 The family age of 1.08 $\pm$ 0.54 Gyrs is calculated using Eq.~\ref{eqn.familyagenoejectionpvgsolar}, with $C_{YE}$ = $9.3 \; \times 10^{-5}$ au calculated from Eq.~\ref{eqn.Ccombo} where $C\; = \; 1.3 \; \times 10^{-4}$ au similar to $\sim$1.3 Gyr age given by \citep[][]{Vokrouhlicky2006a}. The value of $\mu_{\alpha} \; =$ 0.92 and $C_{EV} \; = \; 3.4 \; \times 10^{-5}$ au calculated using Eq.~\ref{eqn.VEVvsCalphaFinal} assuming $V_{EV}$ = 70 $\mps$ from \citep[][]{Nesvorny2015a}. The other parameters in Eq.~\ref{eqn.familyagenoejectionpvg} used to calculate the family age for Eos are $a_c$  = 3.024 au, $e_{\mu}$ = 0.07, $\rho_{mu}$ = 2.3 $\gpcmc$, $\pv$ = 0.13 and $G_{\mu}$ = 0.24. The calculation was repeating using the same parameters except with $\alpha$ = 1.0 and $C\; = \; 1.5  \; \times 10^{-4}$ obtaining a value of 1.13 $\pm$ 0.56 Gyrs.

\subsubsection{Eunomia}
\label{s.Eunomia}
The S-type Eunomia family is located in the central region of the MB  and is bracketed by the 3:1 and 8:3 MMRs with Jupiter \citep[][]{Zappala1990}. The V-shape identification technique was applied to 1,311 asteroids belonging to the Eunomia asteroid family as defined by \citet[][]{Nesvorny2015a}. The interval [$0.05,0.21$] for the Dirac delta function $\delta(D_{r,j}-D_r )$ is used and Eq.~\ref{eqn.apvDvsCfinal} is truncated to 0.05 km$^{-1}$ for $D_r$ $ <$ $0.05$ km$^{-1}$ and to 0.21 km$^{-1}$ for $D_r$ $>$ 0.21 km$^{-1}$. Asteroid $H$ values were converted to $D$ using Eq.~\ref{eq.HtoD} and $\pv$ = 0.19 typical for members of the Eunomia family \citep[][]{Masiero2013,Spoto2015}.
  
\begin{figure}
\centering
\hspace*{-0.7cm}
\ifincludeplots
\includegraphics[scale=0.455]{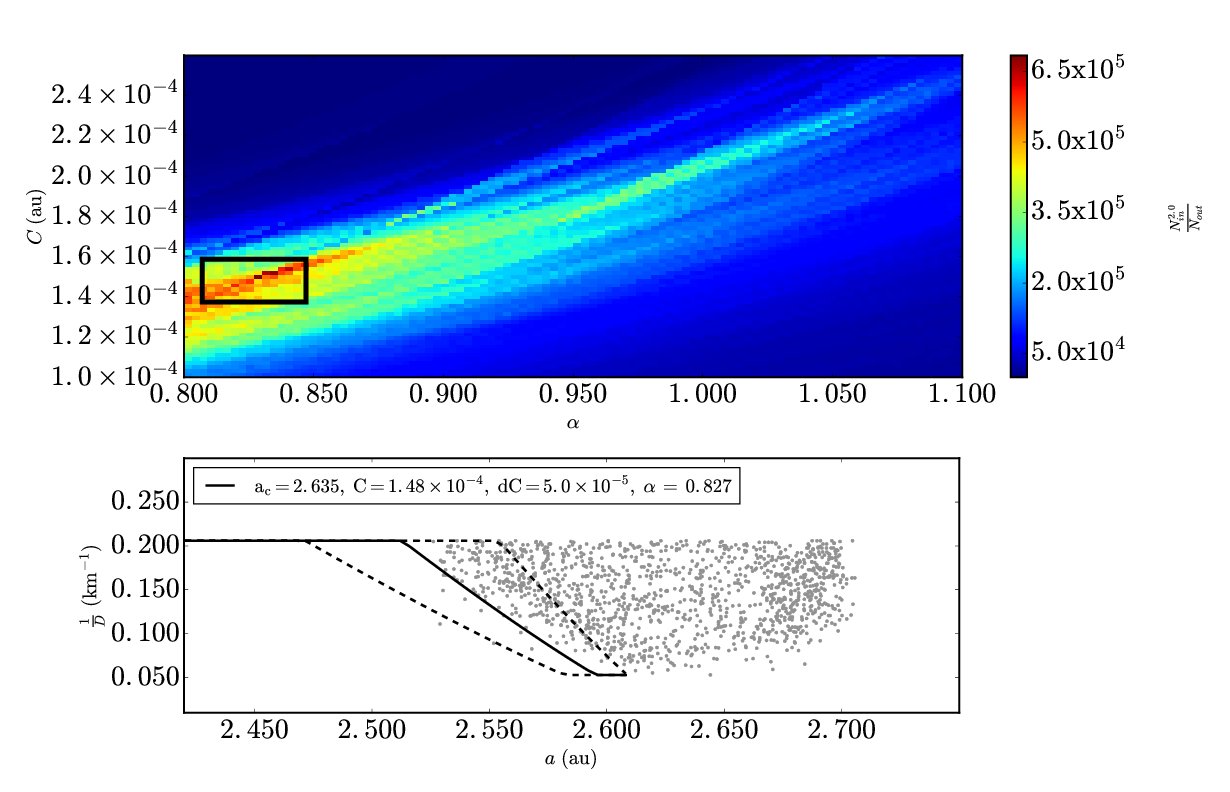}
\else
I am not enabling plots.
\fi
\caption{The same as Fig.~\ref{fig.synErig200Myrs} for Eunomia asteroid family data from \citet[][]{Nesvorny2015a}. (Top panel) $\Delta \alpha$ is equal to $3.0 \times 10^{-3}$ au and $\Delta C$, is equal to $2.0 \times 10^{-6}$ au. (Bottom Panel) $D_r(a,a_c,C\pm dC,\pv,\alpha)$ is plotted with $\pv = 0.19$, $a_c$ = 2.635 au and $dC \; = \; 5.0 \x 10^{-5}$ au.}
\label{fig.EunomiaAlpha}
\end{figure}

\begin{figure}
\centering
\hspace*{-0.9cm}
\ifincludeplots
\includegraphics[scale=0.3225]{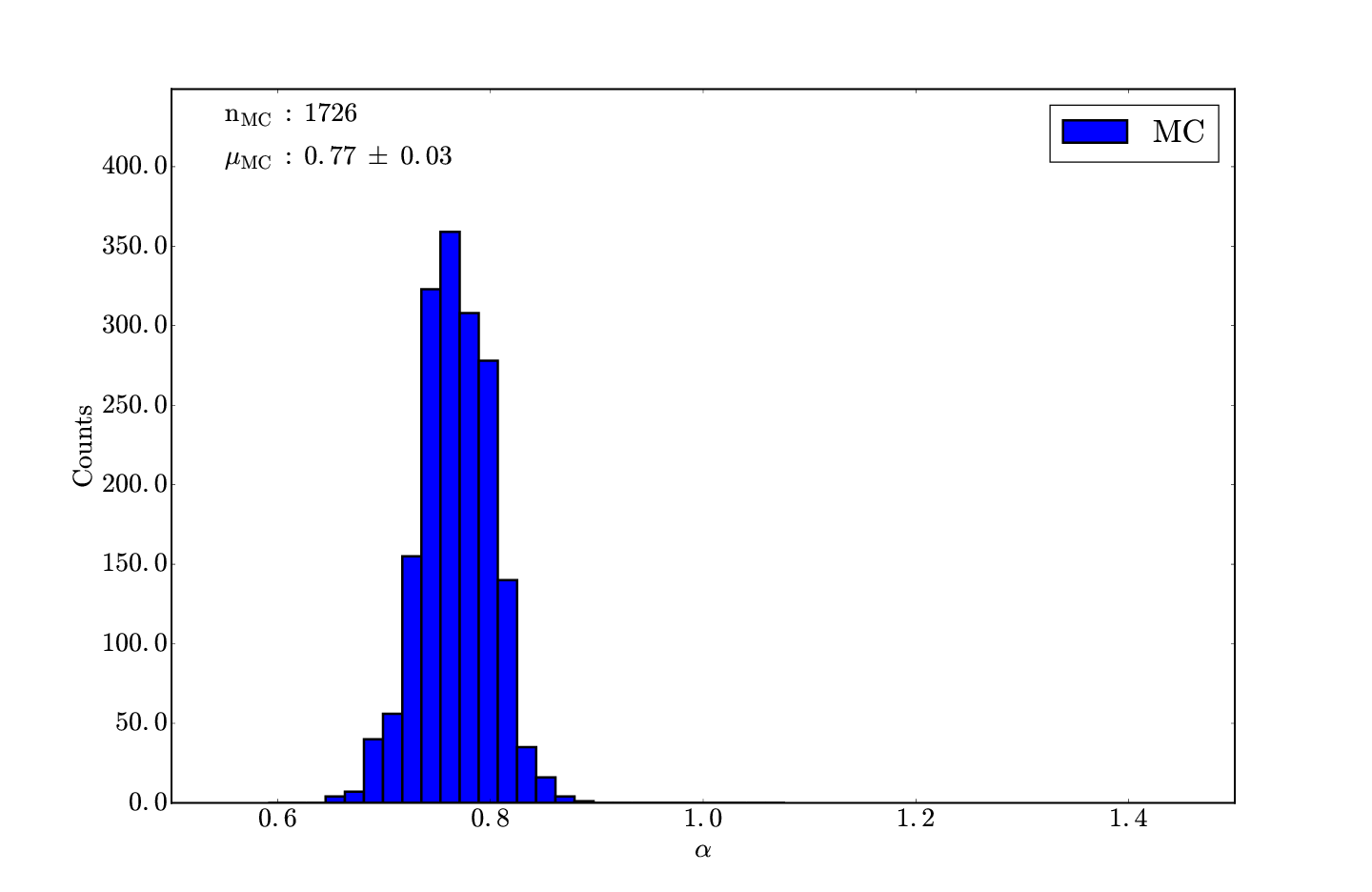}
\else
I am not enabling plots.
\fi
\caption{The same as Fig.~\ref{fig.ErigoneMC} with $\sim$1,700 trials repeating the V-shape technique for the Eunomia family. The mean of the distribution is centered at $\alpha$ = 0.77 $\pm$ 0.03 and the bin size in the histogram is 0.02.}
\label{fig.EunomiaMC}
\end{figure}

The peak in $\frac{N_{in}^2}{N_{out}}$ at $(a_c, \; C, \; \alpha) \; = \; (2.64 \; \mathrm{au}, \; 1.48 \times 10^{-4} \; \mathrm{au}, \;  \sim0.83)$ as seen in the top panel of Fig.~\ref{fig.EunomiaAlpha} and is $\sim$5 standard deviations above the mean value. $\sim$1,700 Monte Carlo runs where the the mean value of $\alpha$ is $\sim$0.77 $\pm$ 0.03 as seen in Fig.~\ref{fig.EunomiaMC}.

 The family age of 1.1 $\pm$ 0.83 Gyrs is calculated using Eq.~\ref{eqn.familyagenoejectionpvgsolar}, with $C_{YE}$ = $5.9 \; \times 10^{-5}$ au calculated from Eq.~\ref{eqn.Ccombo} where $C\; = \; 1.32 \; \times 10^{-4}$ au  and overlaps with the 1.6-2.7 Gyr age found by \citep[][]{Carruba2016c}. The value of $\mu_{\alpha} \; =$ 0.77 and $C_{EV} \; = \; 7.3 \; \times 10^{-5}$ au calculated using Eq.~\ref{eqn.VEVvsCalphaFinal} assuming $V_{EV}$ = 15 $\mps$ which is the escape speed of a 276 km diameter body with $\rho$ = 2.3 $\gpcmc$. 
 
\subsubsection{Hoffmeister}
\label{s.Hoffmeister}
\begin{figure}
\centering
\hspace*{-0.7cm}
\ifincludeplots
\includegraphics[scale=0.455]{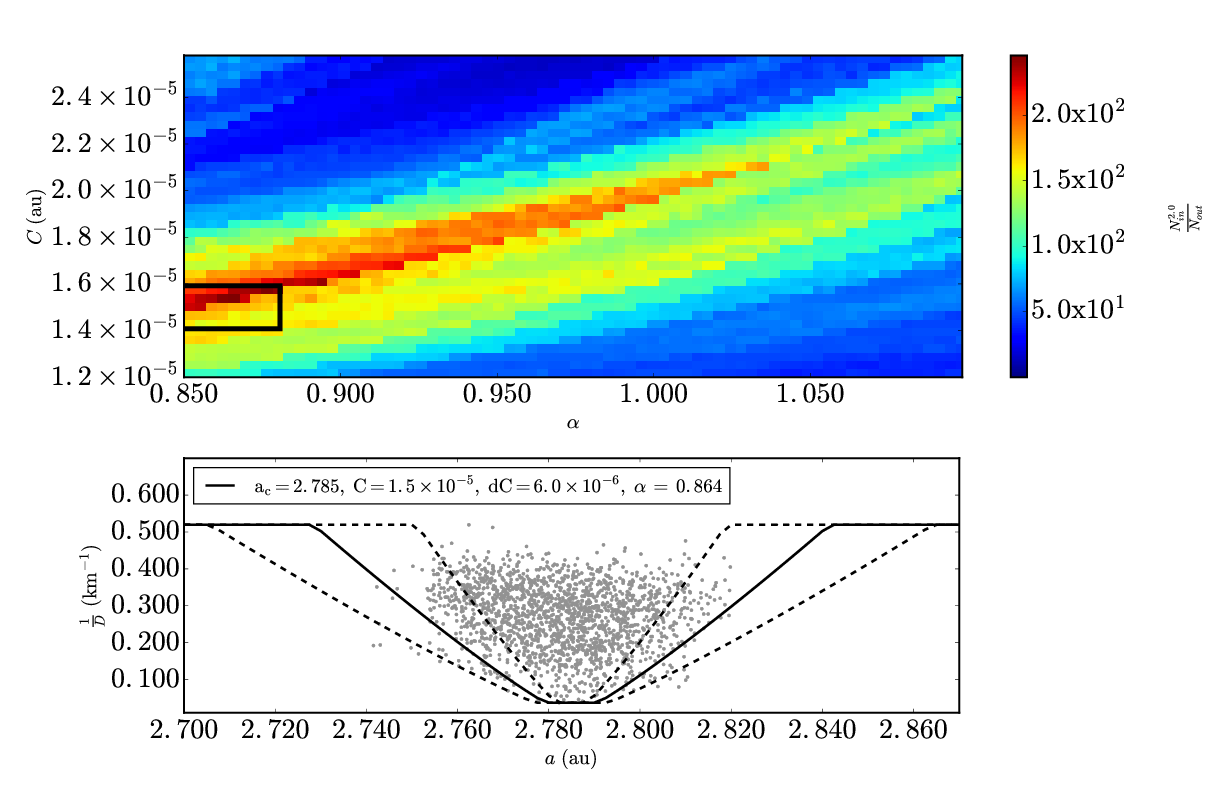}
\else
I am not enabling plots.
\fi
\caption{The same as Fig.~\ref{fig.synErig200Myrs} for Hoffmeister asteroid family data from \citet[][]{Nesvorny2015a}. (Top panel) $\Delta \alpha$ is equal to $3.5 \times 10^{-3}$ au and $\Delta C$, is equal to $3.5 \times 10^{-7}$ au. (Bottom Panel) $D_r(a,a_c,C\pm dC,\pv,\alpha)$ is plotted with $\pv = 0.04$, $a_c$ = 2.785 au and $dC \; = \; 6.0 \x 10^{-6}$ au.}
\label{fig.HoffmeisterAlpha}
\end{figure}

\begin{figure}
\centering
\hspace*{-0.9cm}
\ifincludeplots
\includegraphics[scale=0.3225]{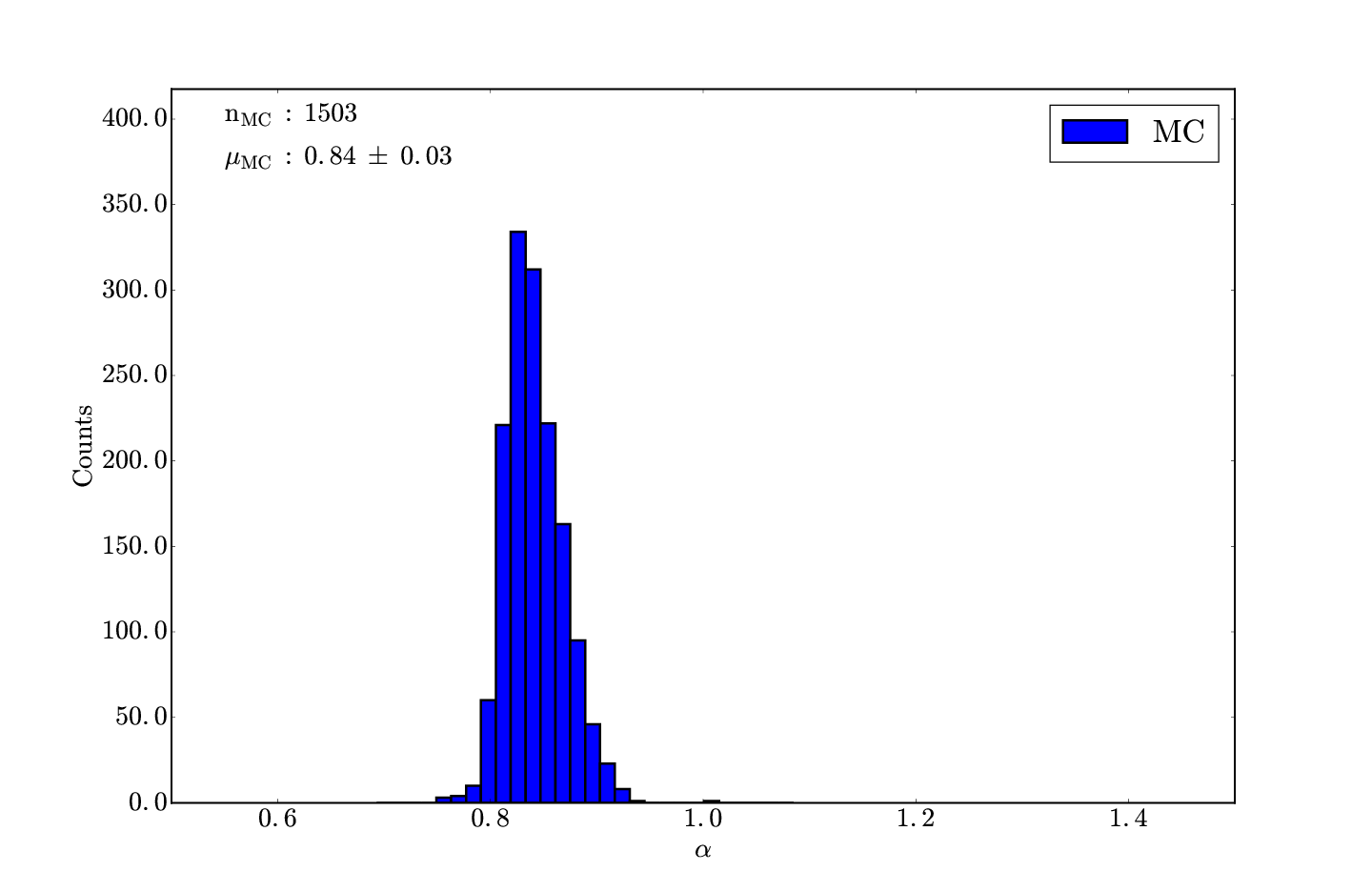}
\else
I am not enabling plots.
\fi
\caption{The same as Fig.~\ref{fig.ErigoneMC} with $\sim$1,500 trials repeating the V-shape technique for the Hoffmeister family. The mean of the distribution is centered at $\alpha$ = 0.84 $\pm$ 0.03 and the bin size in the histogram is 0.01.}
\label{fig.HoffmeisterMC}
\end{figure}

The C-type Hoffmeister family is located in the central region of the MB bracketed between the 3:1:1 three body resonance with Jupiter and Saturn at 2.752 au and 5:2 MMR with Jupiter at 2.82 au and interacts with the $s$-$s_C$ nodal resonance with Ceres \citep[][]{Zappala1995,Novakovic2015,Carruba2016g}. The V-shape identification technique was applied to 1,773 asteroids belonging to the Hoffmeister asteroid family as defined by \citet[][]{Nesvorny2015a}. The peak in $\frac{N_{in}^2}{N_{out}}$ at $(a_c, \; C, \; \alpha) \; = \; (2.79 \; \mathrm{au}, \; 1.92 \times 10^{-5} \; \mathrm{au}, \;  \sim0.86)$ as seen in the top panel of Fig.~\ref{fig.HoffmeisterAlpha} and is $\sim$4 standard deviations above the mean value. $\sim$1,500 Monte Carlo runs were completed by randomizing $H$ magnitudes by  0.25 and $\pv$ values were assumed to be 0.04 with an uncertainty of 0.01 with a mean value of $\alpha$ is $\sim$0.84 $\pm$ 0.03 as seen in Fig.~\ref{fig.HoffmeisterMC}.

 The family age of 220 $\pm$ 110 Myrs is calculated using Eq.~\ref{eqn.familyagenoejectionpvg}, with $C_{YE}$ = $1.0 \; \times 10^{-5}$ au calculated from Eq.~\ref{eqn.Ccombo} where $C\; = \; 1.5 \; \times 10^{-5}$ au and is in agreement with the age of $\sim$220 Myrs for the Hoffmeister given by \citep[][]{Carruba2016g}. The value of $\mu_{\alpha} \; =$ 0.84 and $C_{EV} \; = \; 4.7 \; \times 10^{-6}$ au calculated using Eq.~\ref{eqn.VEVvsCalphaFinal} assuming $V_{EV}$ = 20 $\mps$ from \citep[][]{Carruba2016g}. 
 
 \subsubsection{Hungaria}
\label{s.Hungaria}

The E-type Hungaria family is located interior to the inner region of the MB and is bracketed by numerous secular resonances within 1.87 au and the 4:1 MMR with Jupiter at 2.06 au \citep[][]{Warner2009a, Milani2010a}. The V-shape identification technique was applied to 2,337 asteroids belonging to the Hungaria asteroid family as defined by \citet[][]{Nesvorny2015a}. The interval [$0.28,1.22$] for the Dirac delta function $\delta(D_{r,j}-D_r )$ is used and Eq.~\ref{eqn.apvDvsCfinal} is truncated to 0.28 km$^{-1}$ for $D_r$ $ <$ $0.28$ km$^{-1}$ and to 1.23 km$^{-1}$ for $D_r$ $>$ 1.23 km$^{-1}$. Asteroid $H$ values were converted to $D$ using Eq.~\ref{eq.HtoD} and $\pv$ = 0.35 typical for members of the Hungaria family \citep[][]{Shepard2008,Spoto2015}.
  
\begin{figure}
\centering
\hspace*{-0.7cm}
\ifincludeplots
\includegraphics[scale=0.455]{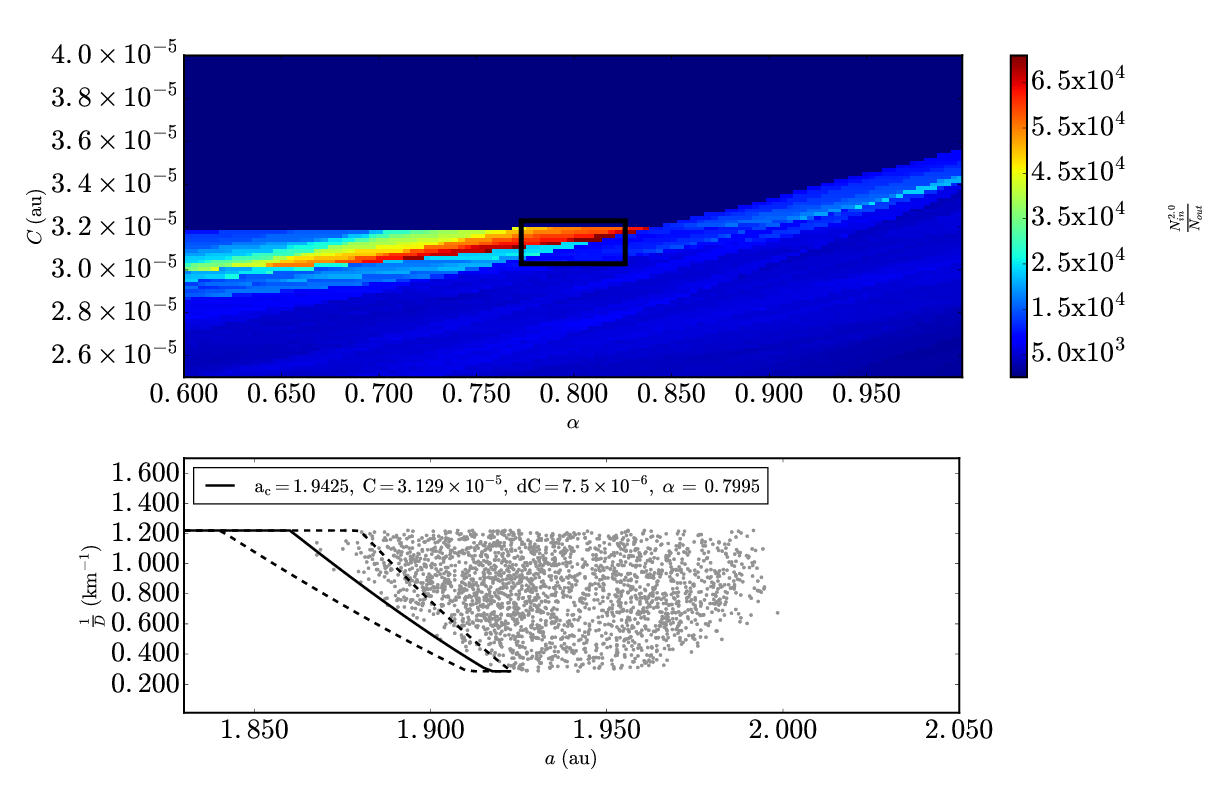}
\else
I am not enabling plots.
\fi
\caption{The same as Fig.~\ref{fig.synErig200Myrs} for Hungaria asteroid family data from \citet[][]{Nesvorny2015a}. (Top panel) $\Delta \alpha$ is equal to $3.5 \times 10^{-3}$ au and $\Delta C$, is equal to $1.7 \times 10^{-7}$ au. (Bottom Panel) $D_r(a,a_c,C\pm dC,\pv,\alpha)$ is plotted with $\pv = 0.35$, $a_c$ = 1.9425 au and $dC \; = \; 7.5 \x 10^{-6}$ au.}
\label{fig.HungariaAlpha}
\end{figure}

\begin{figure}
\centering
\hspace*{-0.9cm}
\ifincludeplots
\includegraphics[scale=0.3225]{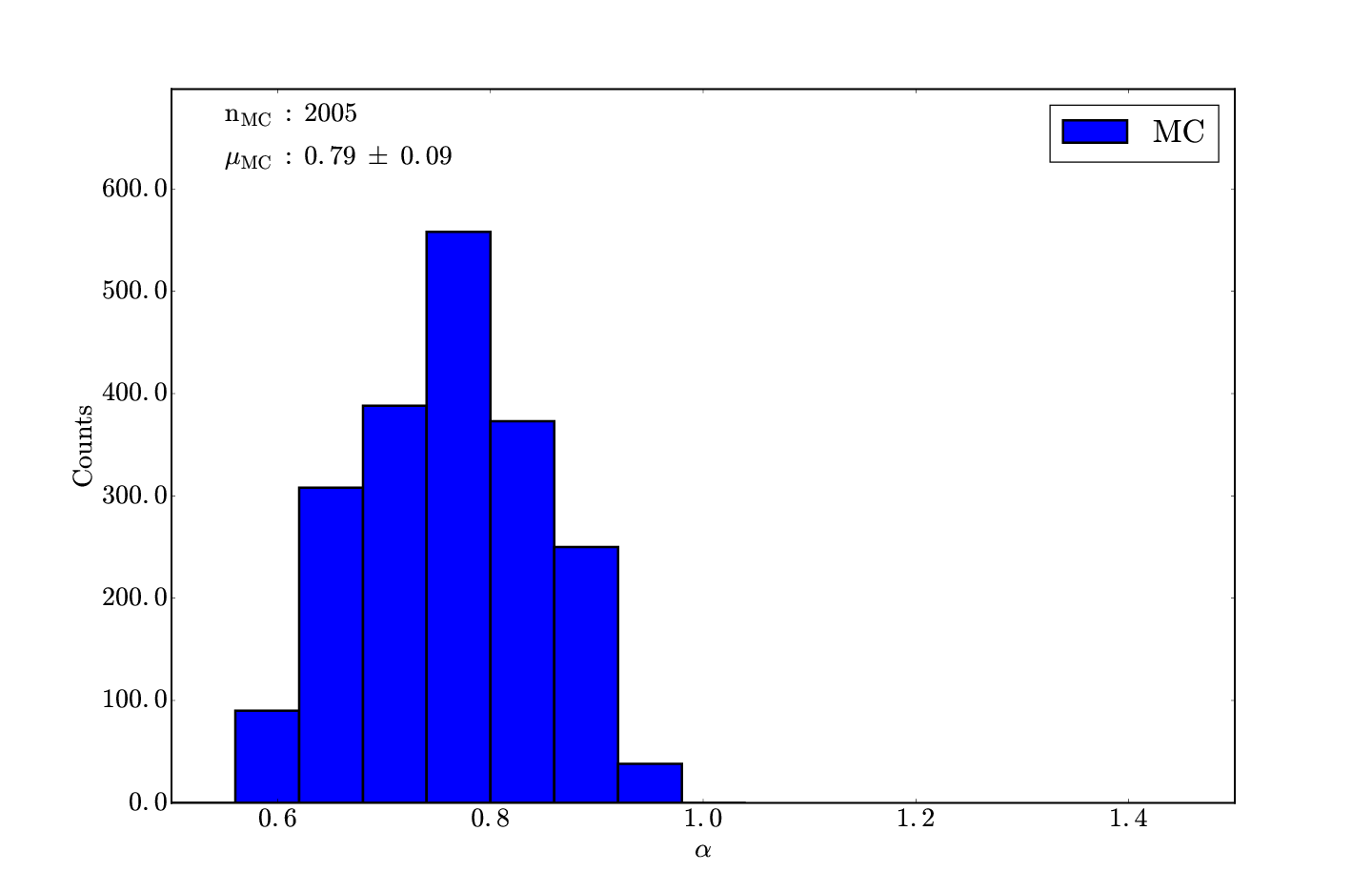}
\else
I am not enabling plots.
\fi
\caption{The same as Fig.~\ref{fig.ErigoneMC} with $\sim$2,000 trials repeating the V-shape technique for the Hungaria family. The mean of the distribution is centered at $\alpha$ = 0.79 $\pm$ 0.09 and the bin size in the histogram is 0.06.}
\label{fig.HungariaMC}
\end{figure}

The peak in $\frac{N_{in}^2}{N_{out}}$ at $(a_c, \; C, \; \alpha) \; = \; (1.94 \; \mathrm{au}, \; 3.1 \times 10^{-5} \; \mathrm{au}, \;  \sim0.8)$ as seen in the top panel of Fig.~\ref{fig.HungariaAlpha} and is $\sim$7 standard deviations above the mean value. $\sim$2,000 Monte Carlo runs were completed bwith a mean value of $\alpha$ is $\sim$0.90 $\pm$ 0.03 as seen in Fig.~\ref{fig.HungariaMC}.

 The family age of 300 $\pm$ 150 Myrs is calculated using Eq.~\ref{eqn.familyagenoejectionpvg}, with $C_{YE}$ = $2.1 \; \times 10^{-5}$ au calculated from Eq.~\ref{eqn.Ccombo} where $C\; = \; 3.1 \; \times 10^{-5}$ au. The value of $\mu_{\alpha} \; =$ 0.79 and $C_{EV} \; = \; 1.0 \; \times 10^{-5}$ au calculated using Eq.~\ref{eqn.VEVvsCalphaFinal} assuming $V_{EV}$ = 25 $\mps$ which is the escape speed of a 41.4 km diameter body with $\rho$ = 2.7 $\gpcmc$.
  
\subsubsection{Hygiea}
\label{s.Hygiea}

\begin{figure}
\centering
\hspace*{-0.7cm}
\ifincludeplots
\includegraphics[scale=0.455]{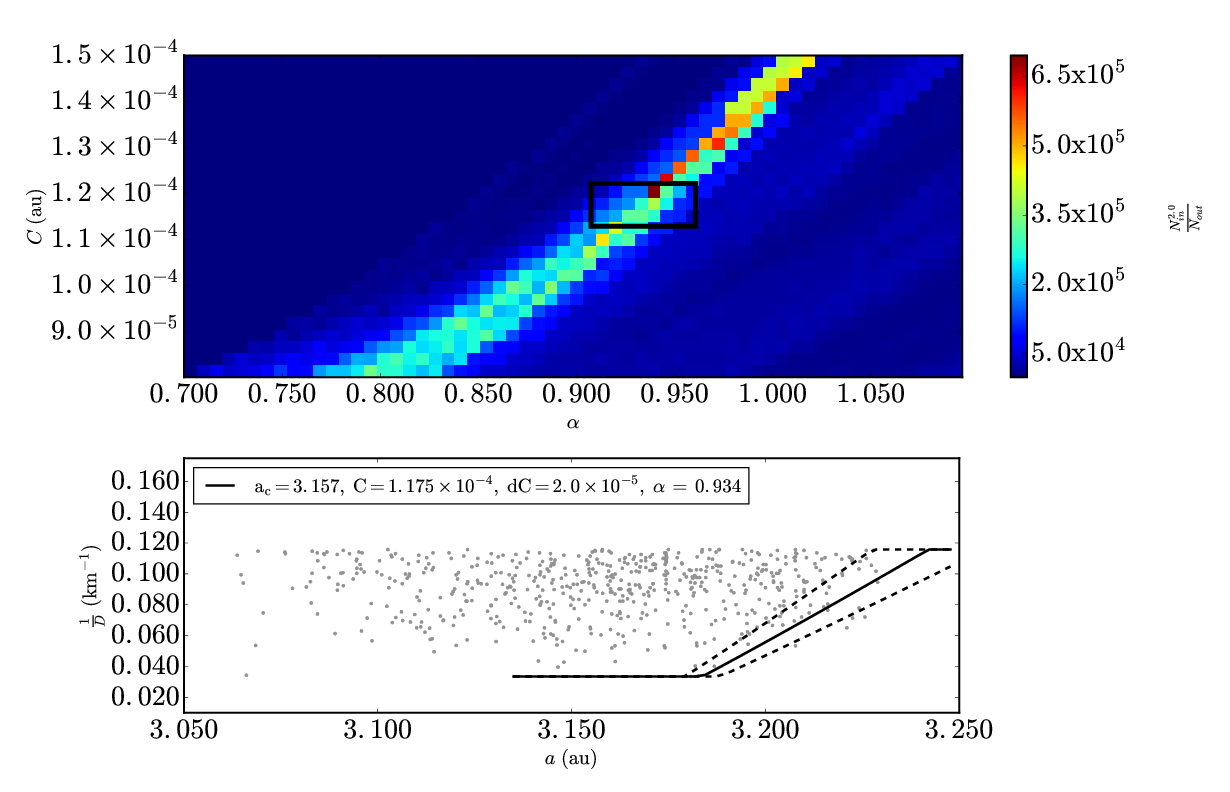}
\else
I am not enabling plots.
\fi
\caption{The same as Fig.~\ref{fig.synErig200Myrs} for Hygiea asteroid family data from \citet[][]{Nesvorny2015a}. (Top panel) $\Delta \alpha$ is equal to $6.5 \times 10^{-3}$ au and $\Delta C$, is equal to $2.5 \times 10^{-6}$ au. (Bottom Panel) $D_r(a,a_c,C\pm dC,\pv,\alpha)$ is plotted with $\pv = 0.06$, $a_c$ = 3.157 au and $dC \; = \; 2.0 \x 10^{-5}$ au.}
\label{fig.HygieaAlpha}
\end{figure}

\begin{figure}
\centering
\hspace*{-0.9cm}
\ifincludeplots
\includegraphics[scale=0.3225]{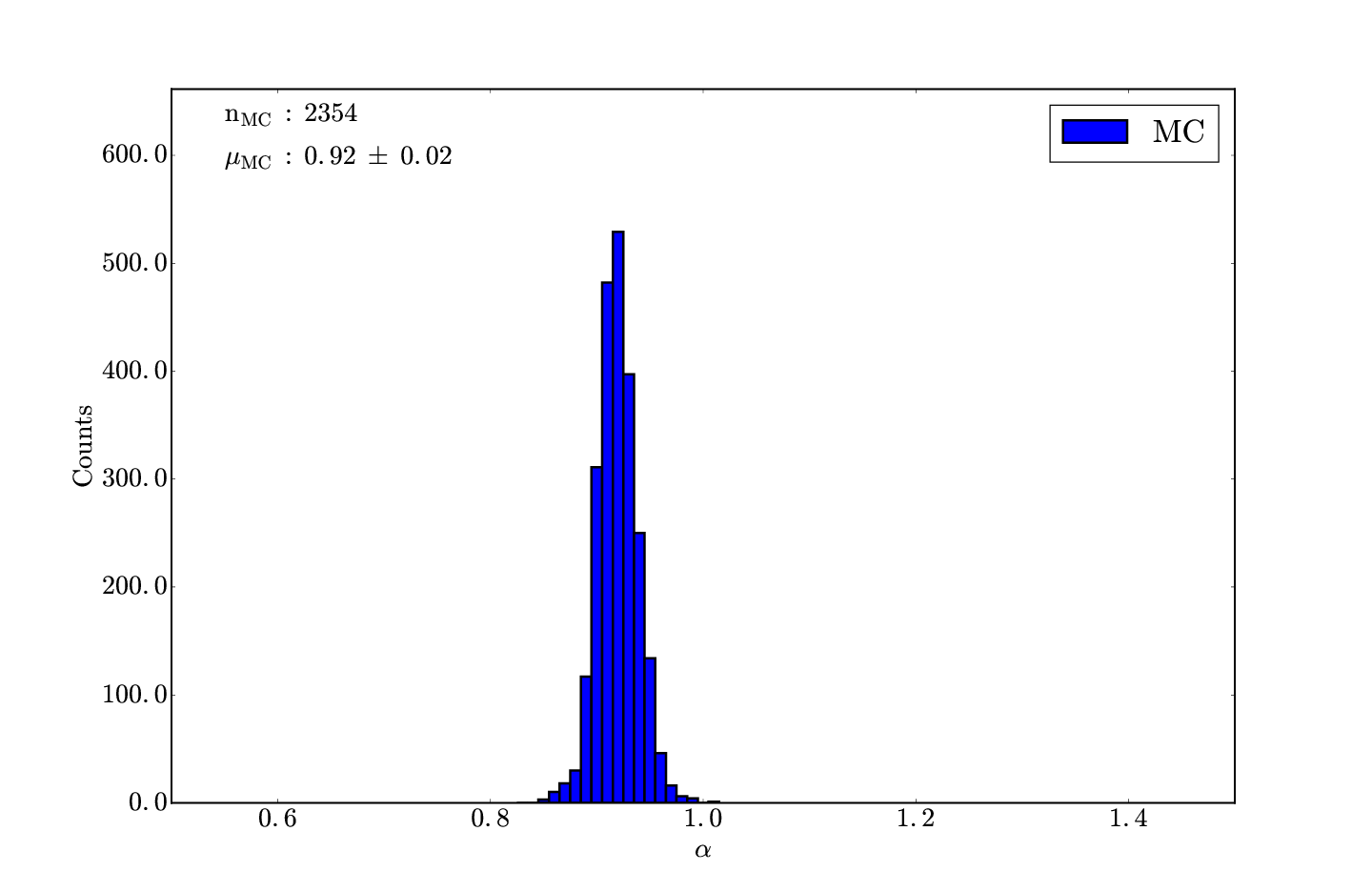}
\else
I am not enabling plots.
\fi
\caption{The same as Fig.~\ref{fig.ErigoneMC} with $\sim$2,400 trials repeating the V-shape technique for the Hygiea family. The mean of the distribution is centered at $\alpha$ = 0.92 $\pm$ 0.02 and the bin size in the histogram is 0.01.}
\label{fig.HygieaMC}
\end{figure}

The C-type Hygiea family is located in the outer region of the MB and borders the 9:4 MMR with Jupiter \citep[][]{Zappala1995,Carruba2014}. The V-shape identification technique was applied to 553 asteroids belonging to the Hygiea asteroid family as defined by \citet[][]{Nesvorny2015a}. The peak in $\frac{N_{in}^2}{N_{out}}$ at $(a_c, \; C, \; \alpha) \; = \; (3.16 \; \mathrm{au}, \; 1.175 \times 10^{-4} \; \mathrm{au}, \;  \sim0.93)$ as seen in the top panel of Fig.~\ref{fig.HygieaAlpha} and is $\sim$8 standard deviations above the mean value. $\sim$2,350 Monte Carlo runs with a mean value of $\alpha$ is $\sim$0.92 $\pm$ 0.02 as seen in Fig.~\ref{fig.HygieaMC}.

 The family age of 0.92 $\pm$ 0.46 Gyrs is calculated using Eq.~\ref{eqn.familyagenoejectionpvgsolar}, with $C_{YE}$ = $5.2 \; \times 10^{-5}$ au calculated from Eq.~\ref{eqn.Ccombo} where $C\; = \; 1.2 \; \times 10^{-5}$ au,  however as discussed by \citet[][]{Carruba2014}, this age may be an upper limit as Hygiea can perturb its members affecting their $a$. The value of $\mu_{\alpha} \; =$ 0.92 and $C_{EV} \; = \; 6.5 \; \times 10^{-5}$ au calculated using Eq.~\ref{eqn.VEVvsCalphaFinal} assuming $V_{EV}$ = 190 $\mps$ from \citep[][]{Vokrouhlicky2006b}. The calculation was repeating using the same parameters except with $\alpha$ = 1.0 and $C\; = \; 1.4 \; \times 10^{-4}$ obtaining a value of 1.2 $\pm$ 0.58 Gyrs.

\subsubsection{Koronis}
\label{s.Koronis}
\begin{figure}
\centering
\hspace*{-0.7cm}
\ifincludeplots
\includegraphics[scale=0.455]{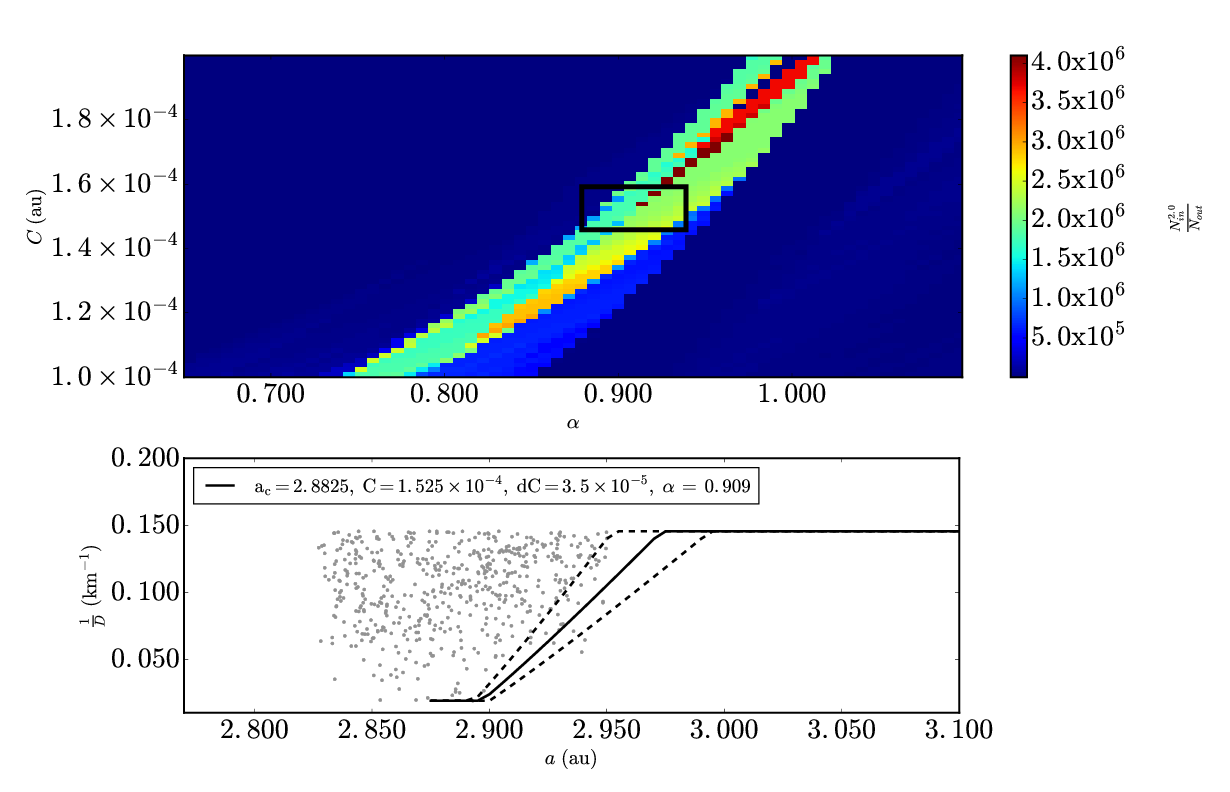}
\else
I am not enabling plots.
\fi
\caption{The same as Fig.~\ref{fig.synErig200Myrs} for Koronis asteroid family data from \citet[][]{Nesvorny2015a}. (Top panel) $\Delta \alpha$ is equal to $3.5 \times 10^{-3}$ au and $\Delta C$, is equal to $1.5 \times 10^{-6}$ au. (Bottom Panel) $D_r(a,a_c,C\pm dC,\pv,\alpha)$ is plotted with $\pv = 0.15$, $a_c$ = 2.883 au and $dC \; = \; 3.5 \x 10^{-5}$ au.}
\label{fig.KoronisAlpha}
\end{figure}

\begin{figure}
\centering
\hspace*{-0.9cm}
\ifincludeplots
\includegraphics[scale=0.3225]{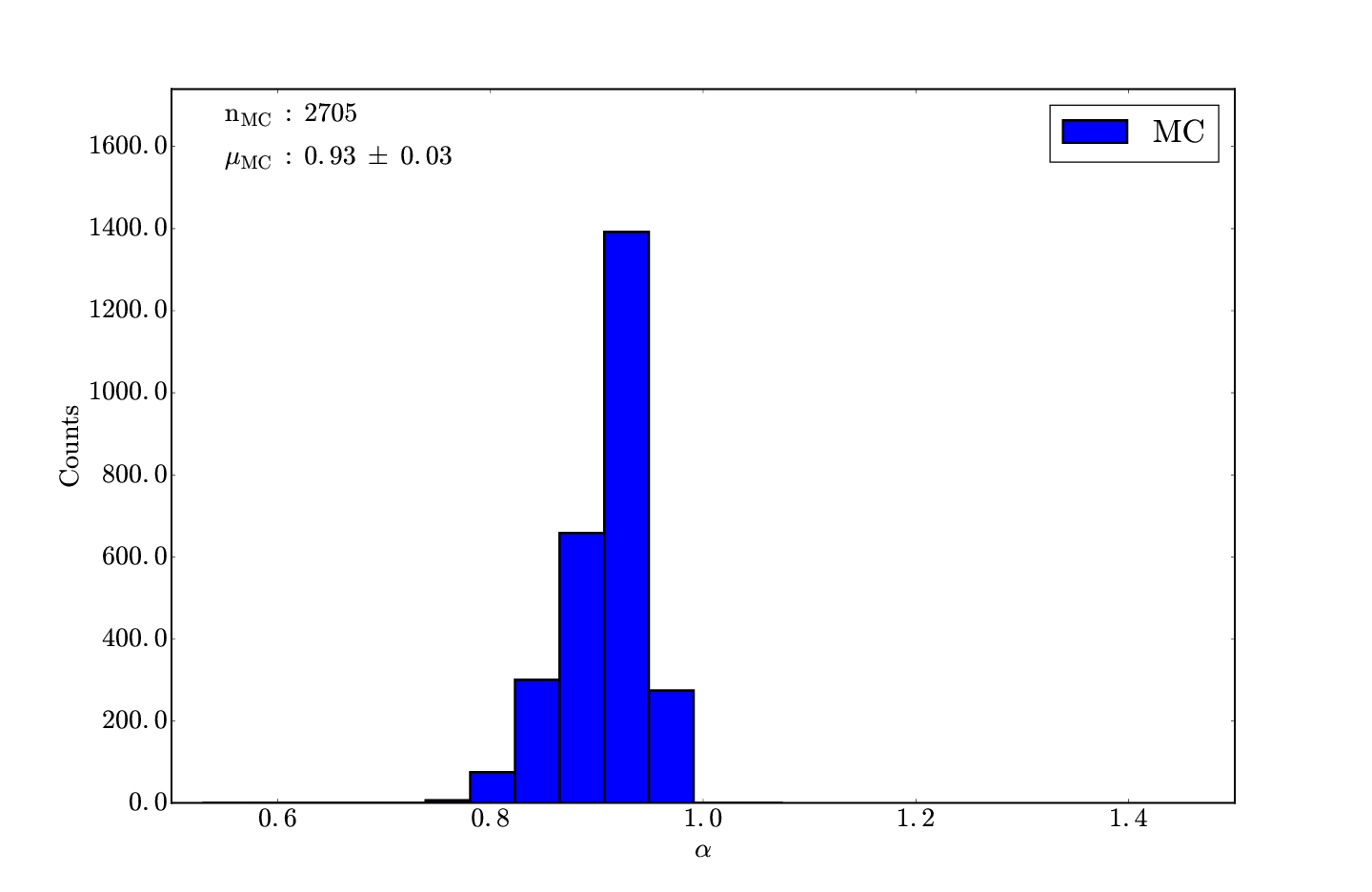}
\else
I am not enabling plots.
\fi
\caption{The same as Fig.~\ref{fig.ErigoneMC} with $\sim$2,700 trials repeating the V-shape technique for the Koronis family. The mean of the distribution is centered at $\alpha$ = 0.93 $\pm$ 0.03 and the bin size in the histogram is 0.04.}
\label{fig.KoronisMC}
\end{figure}

The S-type Koronis family is located in the outer region of the MB and is bracketed by the 5:2 and 7:3 MMRs with Jupiter \citep[][]{Hirayama1918, Zappala1995, Bottke2001}. The V-shape identification technique was applied to 516 asteroids belonging to the Koronis asteroid family as defined by \citet[][]{Nesvorny2015a}. The peak in $\frac{N_{in}^2}{N_{out}}$ at $(a_c, \; C, \; \alpha) \; = \; (2.88 \; \mathrm{au}, \; 1.53 \times 10^{-4} \; \mathrm{au}, \;  \sim0.91)$ as seen in the top panel of Fig.~\ref{fig.KoronisAlpha} and is $\sim$3 standard deviations above the mean value. $\sim$2,700 Monte Carlo runs were completed with a mean value of $\alpha$ is $\sim$0.93 $\pm$ 0.03 as seen in Fig.~\ref{fig.KoronisMC}.

 The family age of 1.94 $\pm$ 0.97 Gyrs is calculated using Eq.~\ref{eqn.familyagenoejectionpvgsolar}, with $C_{YE}$ = $1.1 \; \times 10^{-4}$ au calculated from Eq.~\ref{eqn.Ccombo} where $C\; = \; 1.52 \; \times 10^{-4}$ au  and overlaps with the age estimate of $\sim$ 2.4 Gyrs for Koronis family given by \citep[][]{Carruba2016c}. The value of $\mu_{\alpha} \; =$ 0.93 and $C_{EV} \; = \; 4.30 \; \times 10^{-5}$ au calculated using Eq.~\ref{eqn.VEVvsCalphaFinal} assuming $V_{EV}$ = 90 $\mps$ which is the escape speed of a 160 km diameter body with $\rho$ = 2.3 $\gpcmc$. This is similar to the estimate of $V_{EV}$ = 80 $\mps$ by \citet[][]{Carruba2016e} based on the $e$ and $i$ distribution of its family members.
 
\subsubsection{Naema}
\label{s.Naema}

\begin{figure}
\centering
\hspace*{-0.7cm}
\ifincludeplots
\includegraphics[scale=0.455]{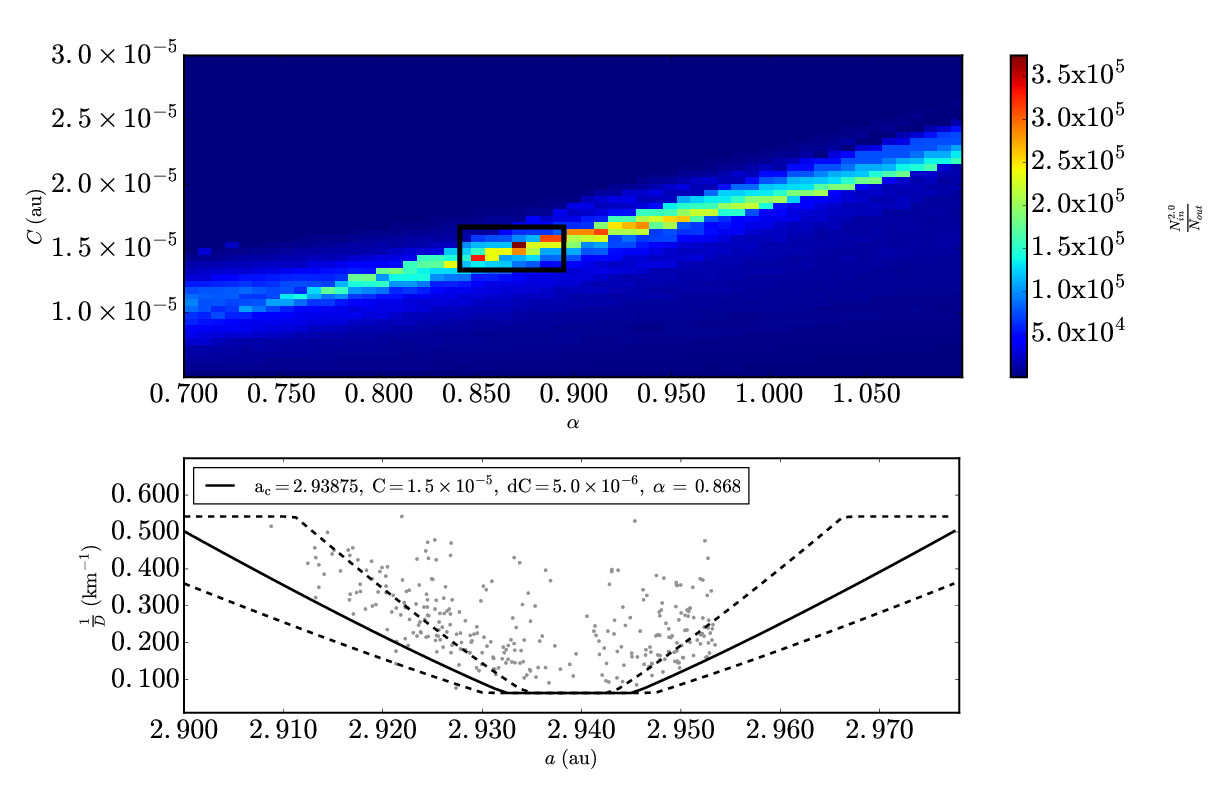}
\else
I am not enabling plots.
\fi
\caption{The same as Fig.~\ref{fig.synErig200Myrs} for Naema asteroid family data from \citet[][]{Nesvorny2015a}. (Top panel) $\Delta \alpha$ is equal to $7.0 \times 10^{-3}$ au and $\Delta C$, is equal to $5.0 \times 10^{-7}$ au. (Bottom Panel) $D_r(a,a_c,C\pm dC,\pv,\alpha)$ is plotted with $\pv = 0.08$, $a_c$ = 2.939 au and $dC \; = \; 5.0 \x 10^{-6}$ au.}
\label{fig.NaemaAlpha}
\end{figure}

\begin{figure}
\centering
\hspace*{-0.9cm}
\ifincludeplots
\includegraphics[scale=0.3225]{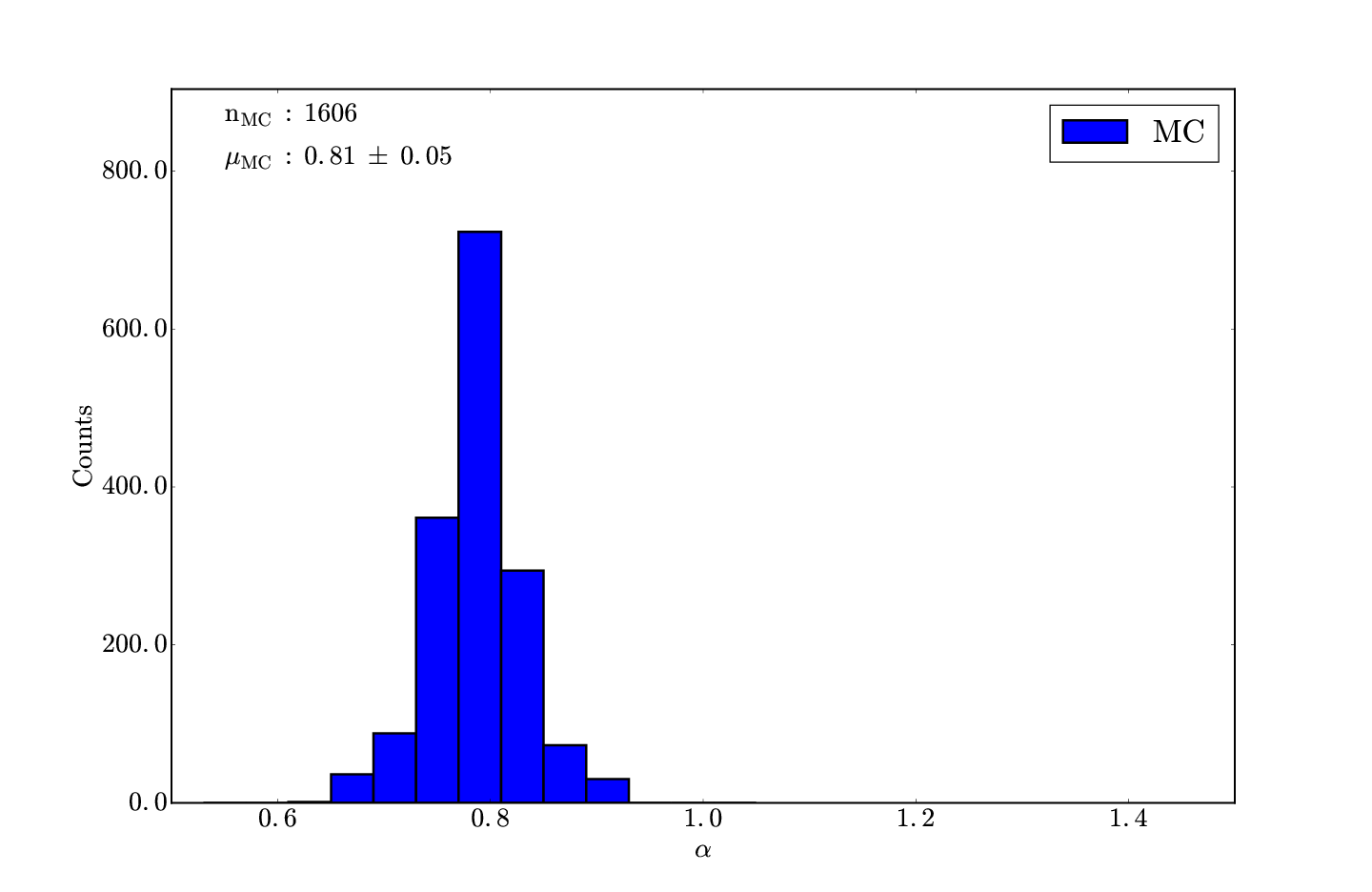}
\else
I am not enabling plots.
\fi
\caption{The same as Fig.~\ref{fig.ErigoneMC} with $\sim$1,600 trials repeating the V-shape technique for the Naema family. The mean of the distribution is centered at $\alpha$ = 0.81 $\pm$ 0.05 and the bin size in the histogram is 0.04.}
\label{fig.NaemaMC}
\end{figure}

The C-type Naema family is located in the outer region of the MB \citep[][]{Zappala1995}. The V-shape identification technique was applied to 281 asteroids belonging to the Naema asteroid family as defined by \citet[][]{Nesvorny2015a}. The peak in $\frac{N_{in}^2}{N_{out}}$ at $(a_c, \; C, \; \alpha) \; = \; (2.94 \; \mathrm{au}, \; 1.5 \times 10^{-5} \; \mathrm{au}, \;  \sim0.87)$ as seen in the top panel of Fig.~\ref{fig.NaemaAlpha} and is $\sim$7 standard deviations above the mean value. $\sim$1,600 Monte Carlo runs were completed with a mean value of $\alpha$ is $\sim$0.81 $\pm$ 0.05 as seen in Fig.~\ref{fig.NaemaMC}.

The family age of 40 $\pm$ 20 Myrs is calculated using Eq.~\ref{eqn.familyagenoejectionpvg}, with $C_{YE}$ = $2.4 \; \times 10^{-6}$ au calculated from Eq.~\ref{eqn.Ccombo} where $C\; = \; 1.5 \; \times 10^{-5}$ au. 
 
\subsubsection{Padua}
\label{s.Padua}

\begin{figure}
\centering
\hspace*{-0.7cm}
\ifincludeplots
\includegraphics[scale=0.455]{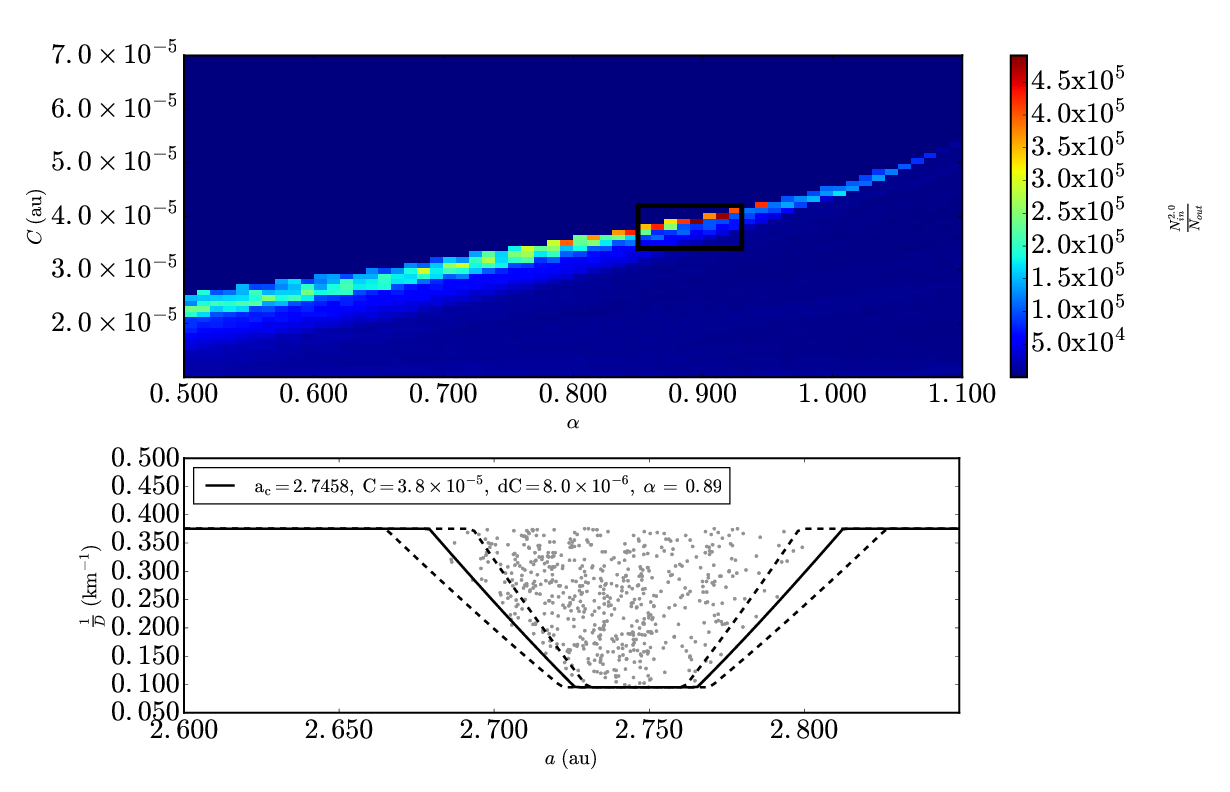}
\else
I am not enabling plots.
\fi
\caption{The same as Fig.~\ref{fig.synErig200Myrs} for Padua asteroid family data from \citet[][]{Nesvorny2015a}. (Top panel) $\Delta \alpha$ is equal to $1.0 \times 10^{-2}$ au and $\Delta C$, is equal to $1.0 \times 10^{-6}$ au. (Bottom Panel) $D_r(a,a_c,C\pm dC,\pv,\alpha)$ is plotted with $\pv = 0.24$, $a_c$ = 2.744 au and $dC \; = \; 8.0 \x 10^{-6}$ au.}
\label{fig.PaduaAlpha}
\end{figure}

The C-type Padua family is located in the central region of the MB and is bracketed by the 8:3 MMR with Jupiter at 2.705 au and the 3:1:1 three body resonance with Jupiter and Saturn at 2.752 au with Jupiter and Saturn \citep[][]{Carruba2009}. The V-shape identification technique was applied to 558 asteroids belonging to the Padua asteroid family as defined by \citet[][]{Nesvorny2015a}. The peak in $\frac{N_{in}^2}{N_{out}}$ at $(a_c, \; C, \; \alpha) \; = \; (2.746 \; \mathrm{au}, \; 3.8 \times 10^{-5} \; \mathrm{au}, \;  \sim0.89)$ as seen in the top panel of Fig.~\ref{fig.PaduaAlpha} and is $\sim$10 standard deviations above the mean value. $\sim$1,000 Monte Carlo runs with a mean value of $\alpha$ is $\sim$0.90 $\pm$ 0.11 and is positively skewed as seen in Fig.~\ref{fig.PaduaMC}.

\begin{figure}
\centering
\hspace*{-0.9cm}
\ifincludeplots
\includegraphics[scale=0.3225]{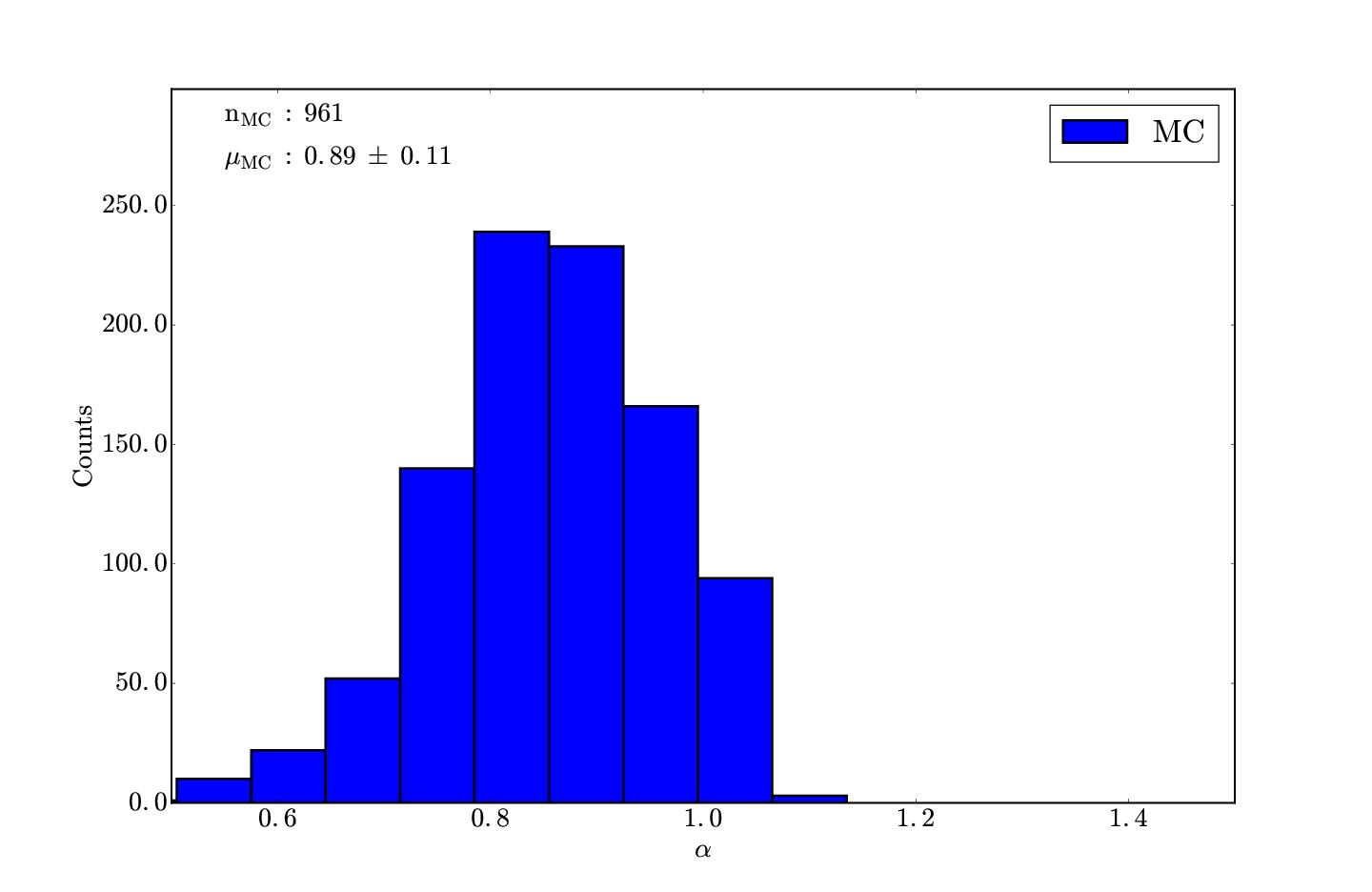}
\else
I am not enabling plots.
\fi
\caption{The same as Fig.~\ref{fig.ErigoneMC} with $\sim$1,000 trials repeating the V-shape technique for the Padua family. The mean of the distribution is centered at $\alpha$ = 0.89 $\pm$ 0.11 and the bin size in the histogram is 0.07.}
\label{fig.PaduaMC}
\end{figure}

The family age of 310 $\pm$ 160 Myrs is calculated using Eq.~\ref{eqn.familyagenoejectionpvg}, with $C_{YE}$ = $2.5 \; \times 10^{-5}$ au calculated from Eq.~\ref{eqn.Ccombo} where $C\; = \; 3.8 \; \times 10^{-5}$ au. The value of $\mu_{\alpha} \; =$ 0.89 and $C_{EV} \; = \; 1.3 \; \times 10^{-5}$ au calculated using Eq.~\ref{eqn.VEVvsCalphaFinal} assuming $V_{EV}$ = 15 $\mps$ from \citep[][]{Vokrouhlicky2006d}. This value is considerable less than the $V_{EV}$ = 30 $\mps$ described b\citet[][]{Carruba2009} and could be due to lack of information about the size of the original parent body of the Padua family causing $V_{EV}$ to be underestimated.

\subsection{Half V-shape families}
\label{s.appendixhalfvshape}
\subsubsection{Adeona}
\label{s.Adeona}

\begin{figure}
\centering
\hspace*{-0.7cm}
\ifincludeplots
\includegraphics[scale=0.455]{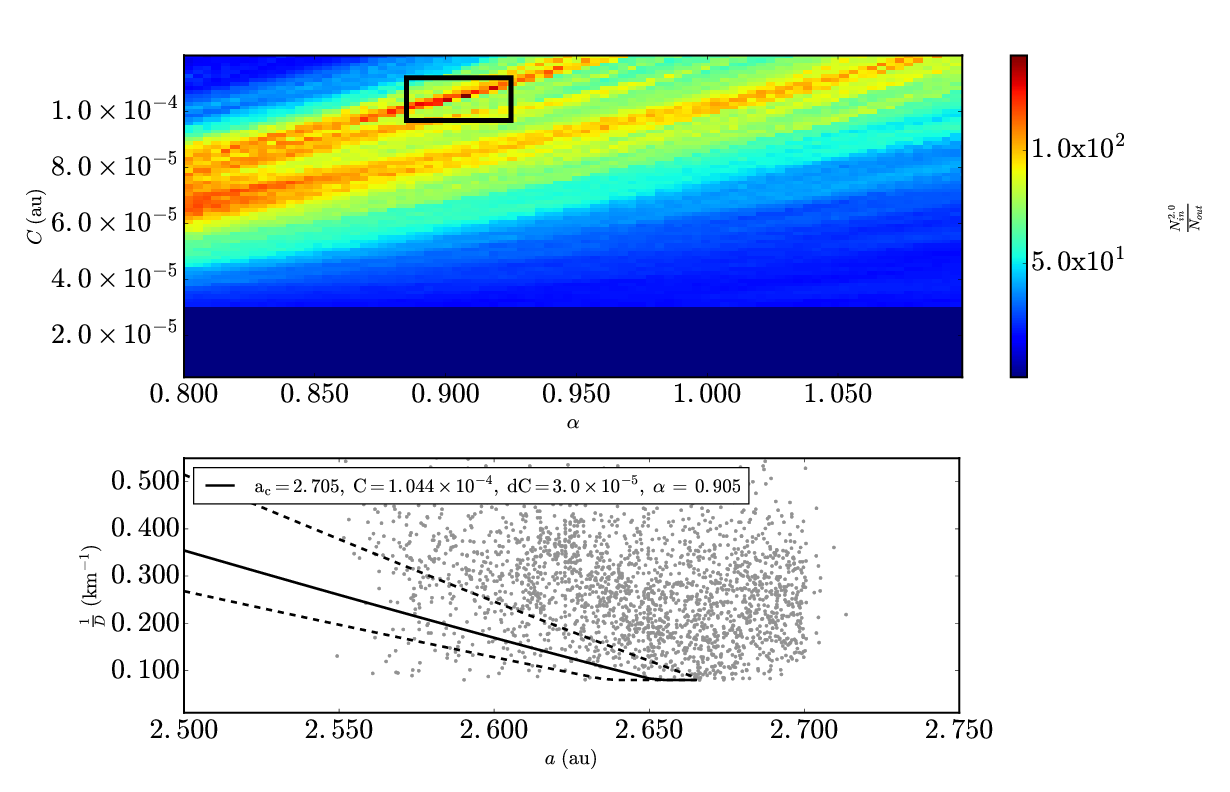}
\else
I am not enabling plots.
\fi
\caption{The same as Fig.~\ref{fig.synErig200Myrs} for Adeona asteroid family data from \citet[][]{Nesvorny2015a}. (Top panel) $\Delta \alpha$ is equal to $3.5 \times 10^{-3}$ au and $\Delta C$, is equal to $1.4 \times 10^{-6}$ au. (Bottom Panel) $D_r(a,a_c,C\pm dC,\pv,\alpha)$ is plotted with $\pv = 0.07$, $a_c$ = 2.705 au and $dC \; = \; 3.0 \x 10^{-5}$ au.}
\label{fig.AdeonaAlpha}
\end{figure}

\begin{figure}
\centering
\hspace*{-0.9cm}
\ifincludeplots
\includegraphics[scale=0.3225]{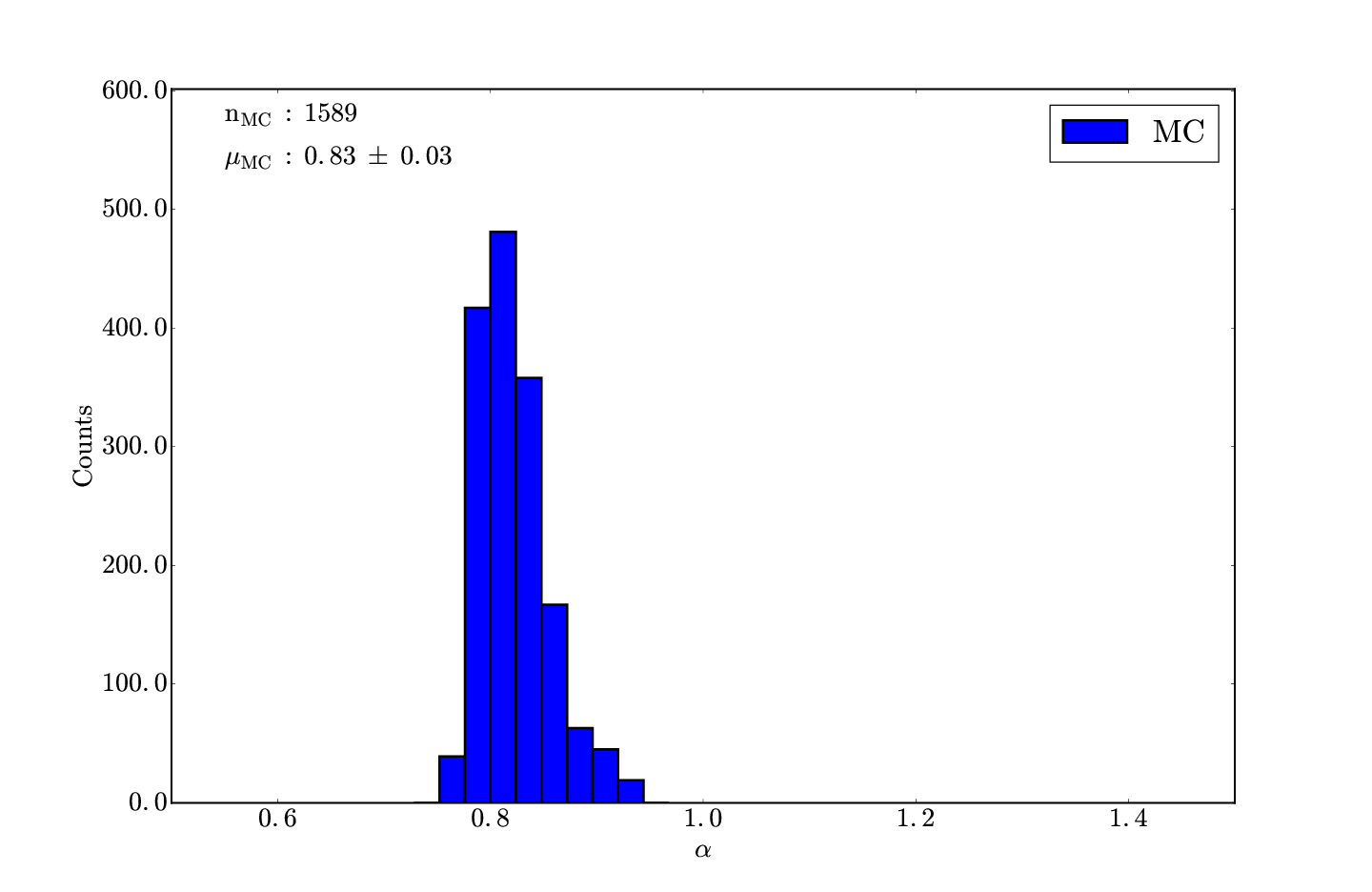}
\else
I am not enabling plots.
\fi
\caption{The same as Fig.~\ref{fig.ErigoneMC} with $\sim$1,600 trials repeating the V-shape technique for the Adeona family. The mean of the distribution is centered at $\alpha$ = 0.83 $\pm$ 0.03 and the bin size in the histogram is 0.01.}
\label{fig.AdeonaMC}
\end{figure}

The C-type Adeona family is located in the central region of the MB and borders the 3:8 resonance with Jupiter \citep[][]{Zappala1995,Carruba2003}. The V-shape identification technique was applied to 2,152 asteroids belonging to the Adeona asteroid family as defined by \citet[][]{Nesvorny2015a}. 
The peak in $\frac{N_{in}^2}{N_{out}}$ at $(a_c, \; C, \; \alpha) \; = \; (2.71 \; \mathrm{au}, \; 1.0 \times 10^{-5} \; \mathrm{au}, \;  \sim0.91)$ as seen in the top panel of Fig.~\ref{fig.AdeonaAlpha} and is $\sim$4 standard deviations above the mean value in the range 2.5 au $< \; a \;<$ 2.75 au, 0.5 $\times \; 10^{-5}$ au $< \; C \;<$ 12.0 $\times \; 10^{-4}$ au and 0.8 $< \; \alpha \;<$ 1.1.  A $dC \; = \; 3.0 \; \times 10^{-5}$ au was used. $\sim$1,600 with a the mean value of $\alpha$ is $\sim$0.83 $\pm$ 0.03 as seen in Fig.~\ref{fig.AdeonaMC}.

The family age of 1.4 $\pm$ 0.7 Gyrs is calculated using Eq.~\ref{eqn.familyagenoejectionpvgsolar}, with $C_{YE}$ = $7.2 \; \times 10^{-5}$ au calculated from Eq.~\ref{eqn.Ccombo} where $C\; = \; 1.0 \; \times 10^{-5}$ au. The value of $\mu_{\alpha} \; =$ 0.83 and $C_{EV} \; = \; 2.3 \; \times 10^{-5}$ au calculated using Eq.~\ref{eqn.VEVvsCalphaFinal} assuming $V_{EV}$ = 78 $\mps$ from \citep[][]{Vokrouhlicky2006b}. 
 
\subsubsection{Eulalia}
\label{s.Eulalia}

\begin{figure}
\centering
\hspace*{-0.9cm}
\ifincludeplots
\includegraphics[scale=0.3225]{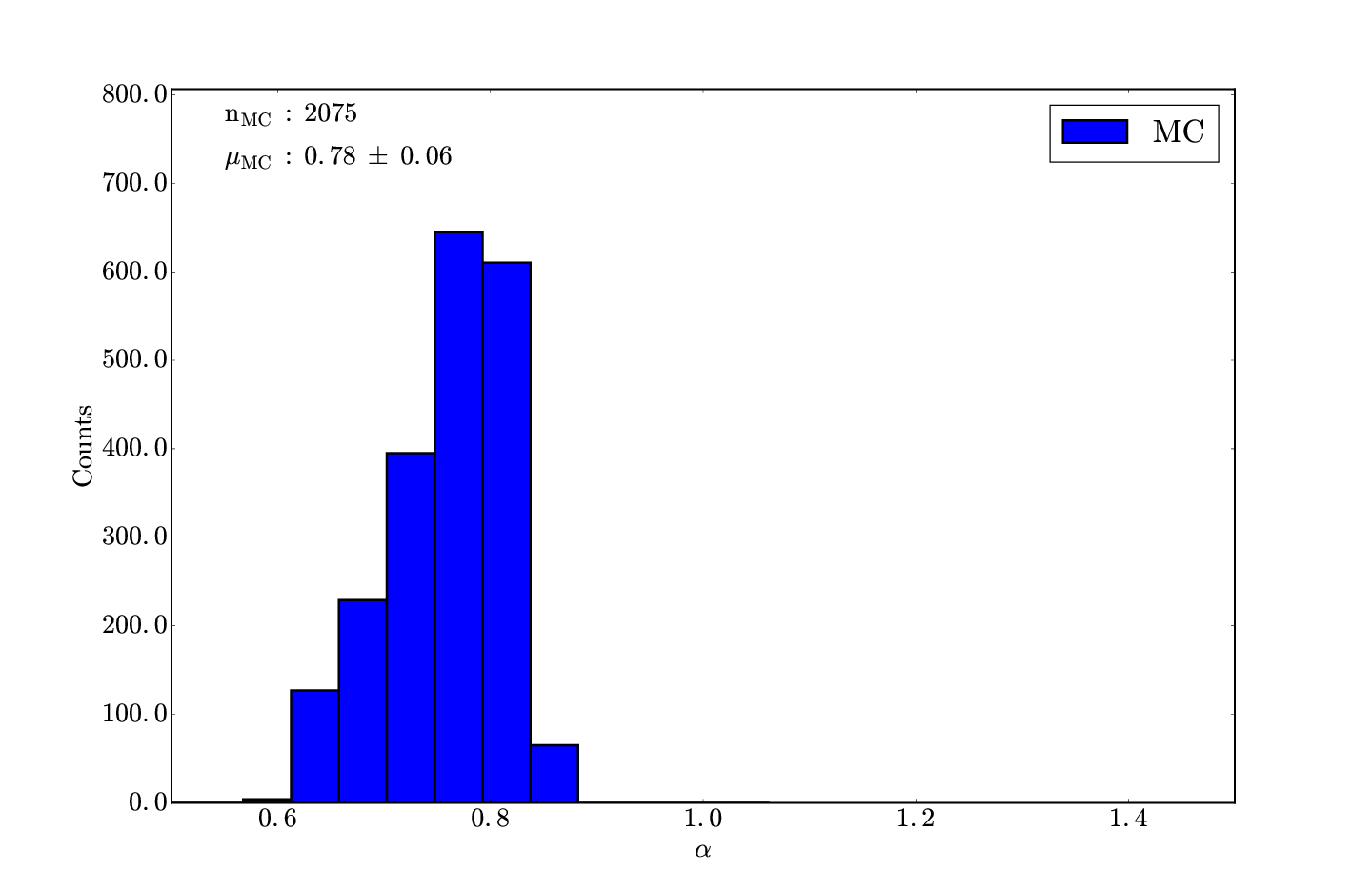}
\else
I am not enabling plots.
\fi
\caption{The same as Fig.~\ref{fig.ErigoneMC} with $\sim$2,100 trials repeating the V-shape technique for the Eulalia family. The mean of the distribution is centered at $\alpha$ = 0.78 $\pm$ 0.06 and the bin size in the histogram is 0.04.}
\label{fig.EulaliaMC}
\end{figure}

\begin{figure}
\centering
\hspace*{-1.1cm}
\ifincludeplots
\includegraphics[scale=0.425]{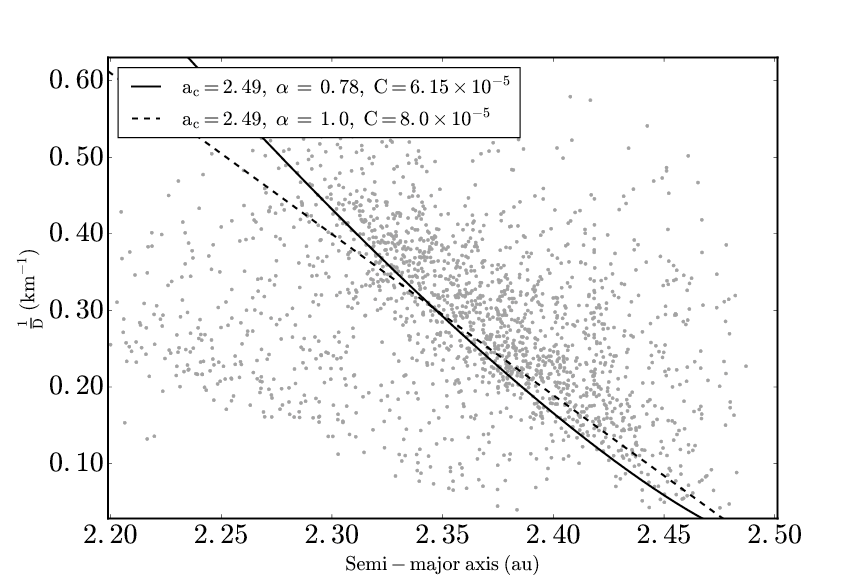}
\else
I am not enabling plots.
\fi
\caption{$a$ vs.$\frac{1}{D}$ plot for Eulalia with V-shape borders that have $\alpha \; = \; 0.78$ and $\alpha \; = \; 1.0$}
\label{fig.EulaliaTwoVs}
\end{figure}

The C-type Eulalia family is located in the inner region of the MB and borders the 3:1 resonance with Jupiter \citep[][]{Walsh2013}. The V-shape identification technique was applied to 2,123 asteroids belonging to the Nysa-Polana asteroid family as defined by \citet[][]{Nesvorny2015a}. The interval [$0.04,0.49$] for the Dirac delta function $\delta(D_{r,j}-D_r )$ is used and Eq.~\ref{eqn.apvDvsCfinal} is truncated to 0.04 km$^{-1}$ for $D_r$ $ <$ $0.04$ km$^{-1}$ and to 0.49 km$^{-1}$ for $D_r$ $>$ 0.49 km$^{-1}$. Asteroid $H$ values were converted to $D$ using Eq.~\ref{eq.HtoD} and $\pv$ = 0.06 typical for members of the Eulalia family \citep[][]{Walsh2013}. The peak in $\frac{N_{in}^2}{N_{out}}$ at $(a_c, \; C, \; \alpha) \; = \; (2.49 \; \mathrm{au}, \; 6.15 \times 10^{-5} \; \mathrm{au}, \;  \sim0.78)$ is $\sim$3 standard deviations above the mean value. $\sim$2,100 Monte Carlo runs were completed with a mean value of $\alpha$ is $\sim$0.78 $\pm$ 0.06 and is positively skewed as seen in Fig.~\ref{fig.EulaliaMC}. The Eulalia family V-shape is better fit with $\alpha \; = \; 0.78$  than the V-shape with $\alpha \; = \; 1.0$ as seen in Fig.~\ref{fig.EulaliaTwoVs}.

 The family age of 840 $\pm$ 420 Myrs is calculated using Eq.~\ref{eqn.familyagenoejectionpvg}, with $C_{YE}$ = $4.8 \; \times 10^{-6}$ au calculated from Eq.~\ref{eqn.Ccombo} where $C\; = \; 6.5 \; \times 10^{-5}$ au. The value of $\mu_{\alpha} \; =$ 0.78 and $C_{EV} \; = \; 1.4 \; \times 10^{-6}$ au calculated using Eq.~\ref{eqn.VEVvsCalphaFinal} assuming $V_{EV}$ = 58 $\mps$ which is the escape speed of a 130 km diameter body with $\rho$ = 1.4 $\gpcmc$. 
 
\subsubsection{Flora}
\label{s.Flora}
\begin{figure}
\centering
\hspace*{-0.7cm}
\ifincludeplots
\includegraphics[scale=0.455]{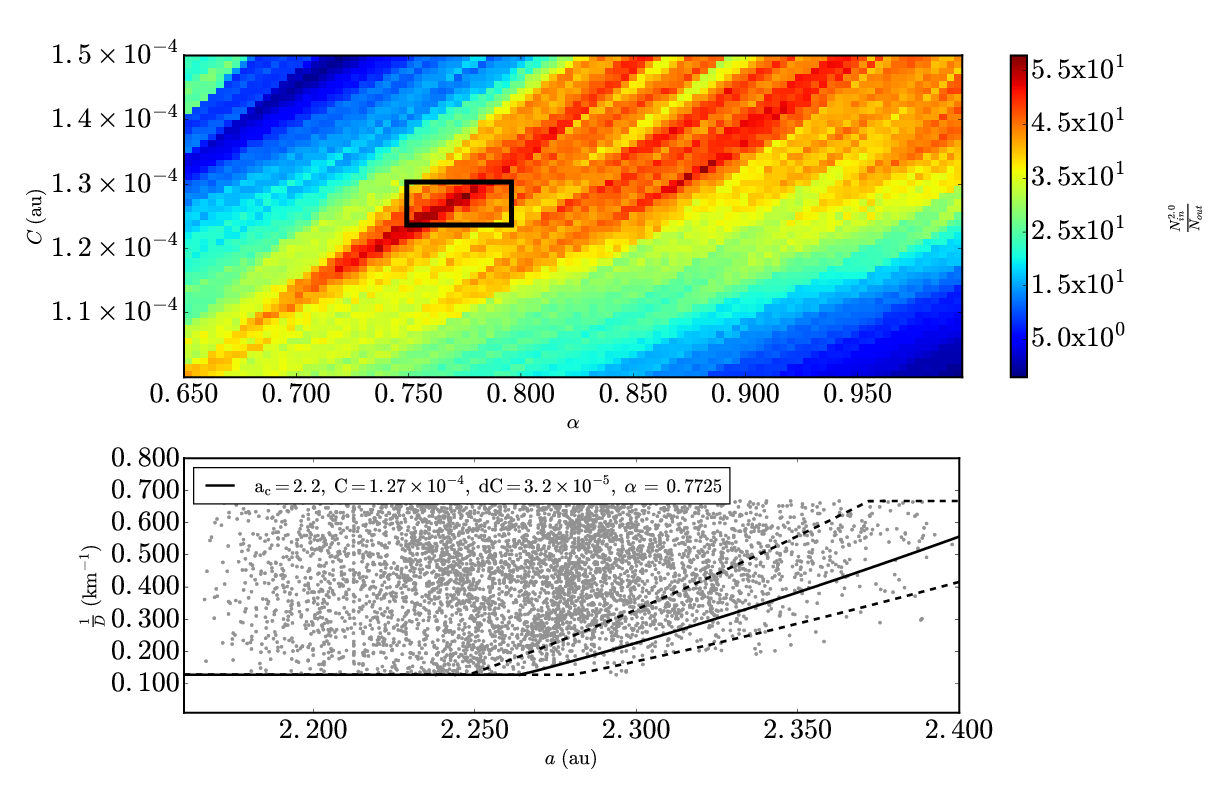}
\else
I am not enabling plots.
\fi
\caption{The same as Fig.~\ref{fig.synErig200Myrs} for Flora asteroid family data from \citet[][]{Nesvorny2015a}. (Top panel) $\Delta \alpha$ is equal to $7.0 \times 10^{-3}$ au and $\Delta C$, is equal to $4.0 \times 10^{-6}$ au. (Bottom Panel) $D_r(a,a_c,C\pm dC,\pv,\alpha)$ is plotted with $\pv = 0.29$, $a_c$ = 2.20 au and $dC \; = \; 3.2 \x 10^{-5}$ au.}
\label{fig.FloraAlpha}
\end{figure}

\begin{figure}
\centering
\hspace*{-0.9cm}
\ifincludeplots
\includegraphics[scale=0.3225]{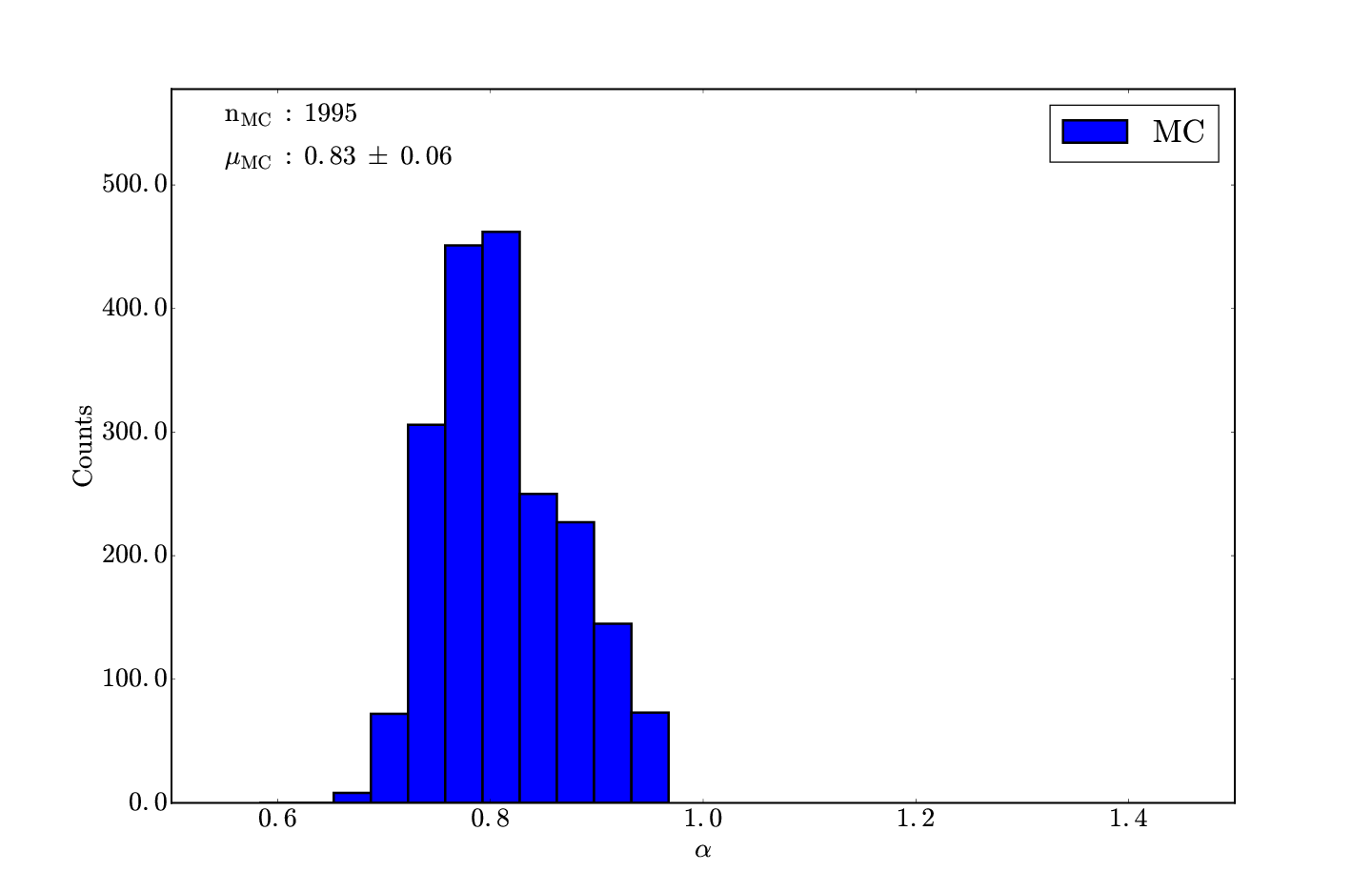}
\else
I am not enabling plots.
\fi
\caption{The same as Fig.~\ref{fig.ErigoneMC} with $\sim$2,000 trials repeating the V-shape technique for the Flora family. The mean of the distribution is centered at $\alpha$ = 0.83 $\pm$ 0.06 and the bin size in the histogram is 0.04.}
\label{fig.FloraMC}
\end{figure}

The S-type Flora family is located in the inner region of the MB and borders the $\nu_6$ resonance with Saturn at $\sim$ 2.16 au and is bisected by the 7:2 /5:9 MMR with Jupiter/Mars   \citep[][]{Hirayama1922,Zappala1990,Dykhuis2014}. The V-shape identification technique was applied to 5,362 asteroids belonging to the Flora asteroid family as defined by \citet[][]{Nesvorny2015a}. The peak in $\frac{N_{in}^2}{N_{out}}$ at $(a_c, \; C, \; \alpha) \; = \; (2.2 \; \mathrm{au}, \; 1.27 \times 10^{-4} \; \mathrm{au}, \;  \sim0.77)$ as seen in the top panel of Fig.~\ref{fig.FloraAlpha} and is $\sim$5 standard deviations above the mean value. $\sim$2,000 Monte Carlo runs were completed with the mean value of $\alpha$ is $\sim$0.83 $\pm$ 0.06 as seen in Fig.~\ref{fig.FloraMC}.

 The family age of 1.16 $\pm$ 0.58 Gyrs is calculated using Eq.~\ref{eqn.familyagenoejectionpvgsolar}, with $C_{YE}$ = $8.75 \; \times 10^{-5}$ au calculated from Eq.~\ref{eqn.Ccombo} where $C\; = \; 1.27 \; \times 10^{-4}$ au. This is in agreement with the lower bound on age of 1-1.4 Gyr family by \citet[][]{Vokrouhlicky2017b}, but as noted in their paper, the upper bound on the family's age is more compatible with the surface age of the asteroid Gaspara. We will have to consider our family age estimate on the lower bound of the acceptable ages constrained by the surface age of Gaspara. The value of $\mu_{\alpha} \; =$ 0.83 and $C_{EV} \; = \; 3.9 \; \times 10^{-5}$ au calculated using Eq.~\ref{eqn.VEVvsCalphaFinal} assuming $V_{EV}$ = 87 $\mps$ which is the escape speed of a 155 km diameter body with $\rho$ = 2.3 $\gpcmc$. 
 
\subsubsection{Maria}
\label{s.Maria}
\begin{figure}
\centering
\hspace*{-0.7cm}
\ifincludeplots
\includegraphics[scale=0.455]{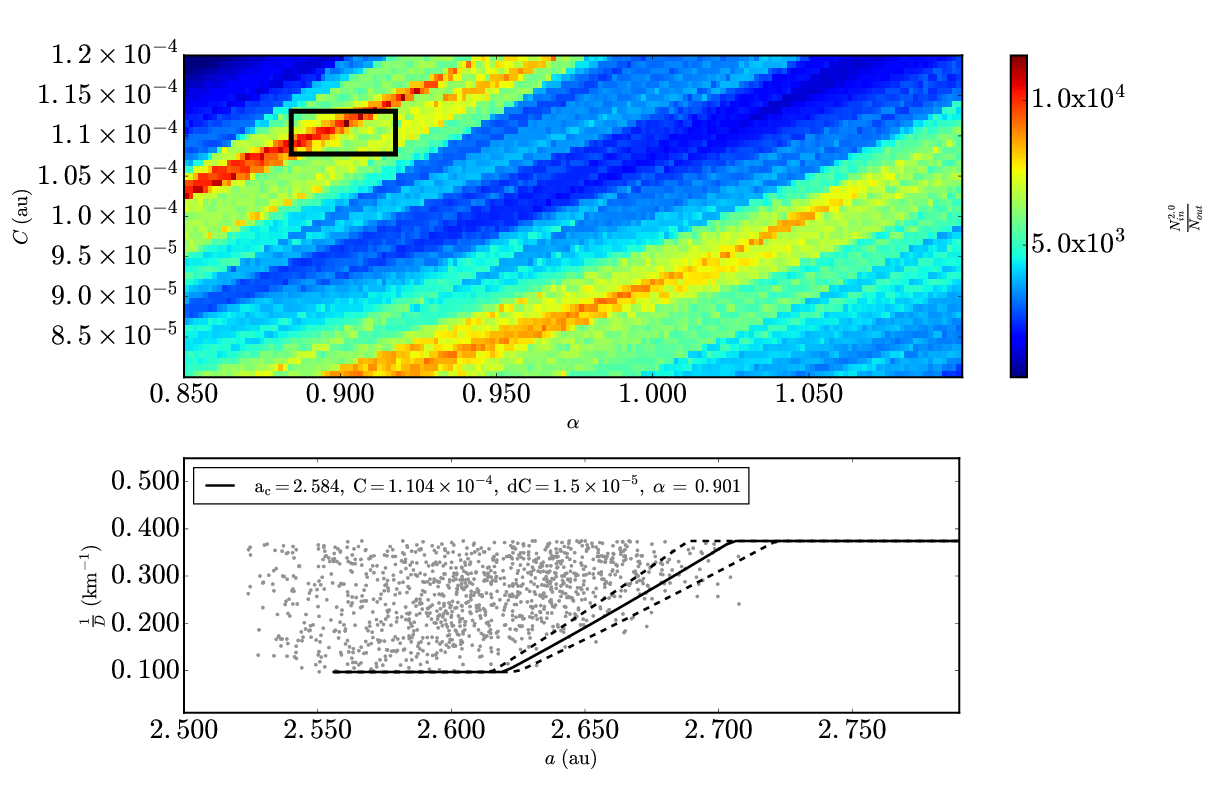}
\else
I am not enabling plots.
\fi
\caption{The same as Fig.~\ref{fig.synErig200Myrs} for Maria asteroid family data from \citet[][]{Nesvorny2015a}. (Top panel) $\Delta \alpha$ is equal to $1.5 \times 10^{-3}$ au and $\Delta C$, is equal to $8.0 \times 10^{-7}$ au. (Bottom Panel) $D_r(a,a_c,C\pm dC,\pv,\alpha)$ is plotted with $\pv = 0.25$, $a_c$ = 2.584 au and $dC \; = \; 1.5 \x 10^{-5}$ au.}
\label{fig.MariaAlpha}
\end{figure}

\begin{figure}
\centering
\hspace*{-0.7cm}
\ifincludeplots
\includegraphics[scale=0.455]{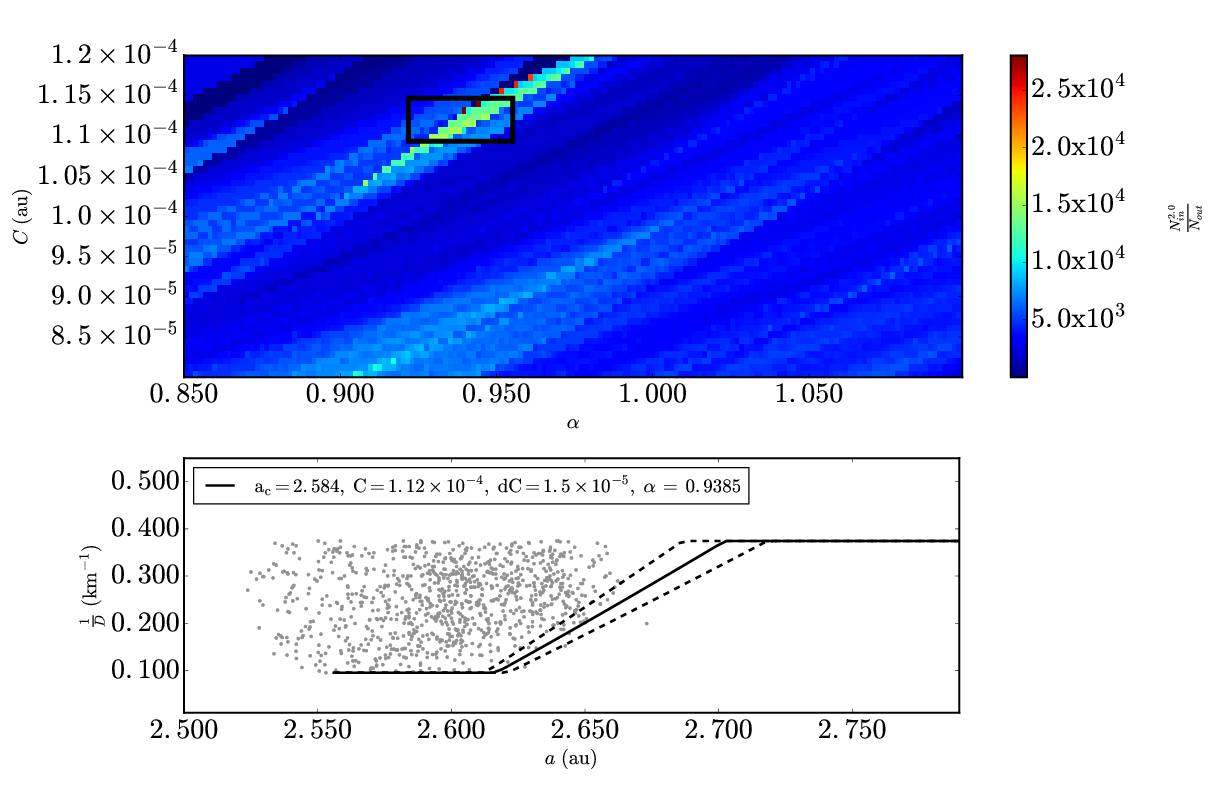}
\else
I am not enabling plots.
\fi
\caption{Milani Maria family.}
\label{fig.MariaBorderAlphaMilani}
\end{figure}

\begin{figure}
\centering
\hspace*{-0.9cm}
\ifincludeplots
\includegraphics[scale=0.3225]{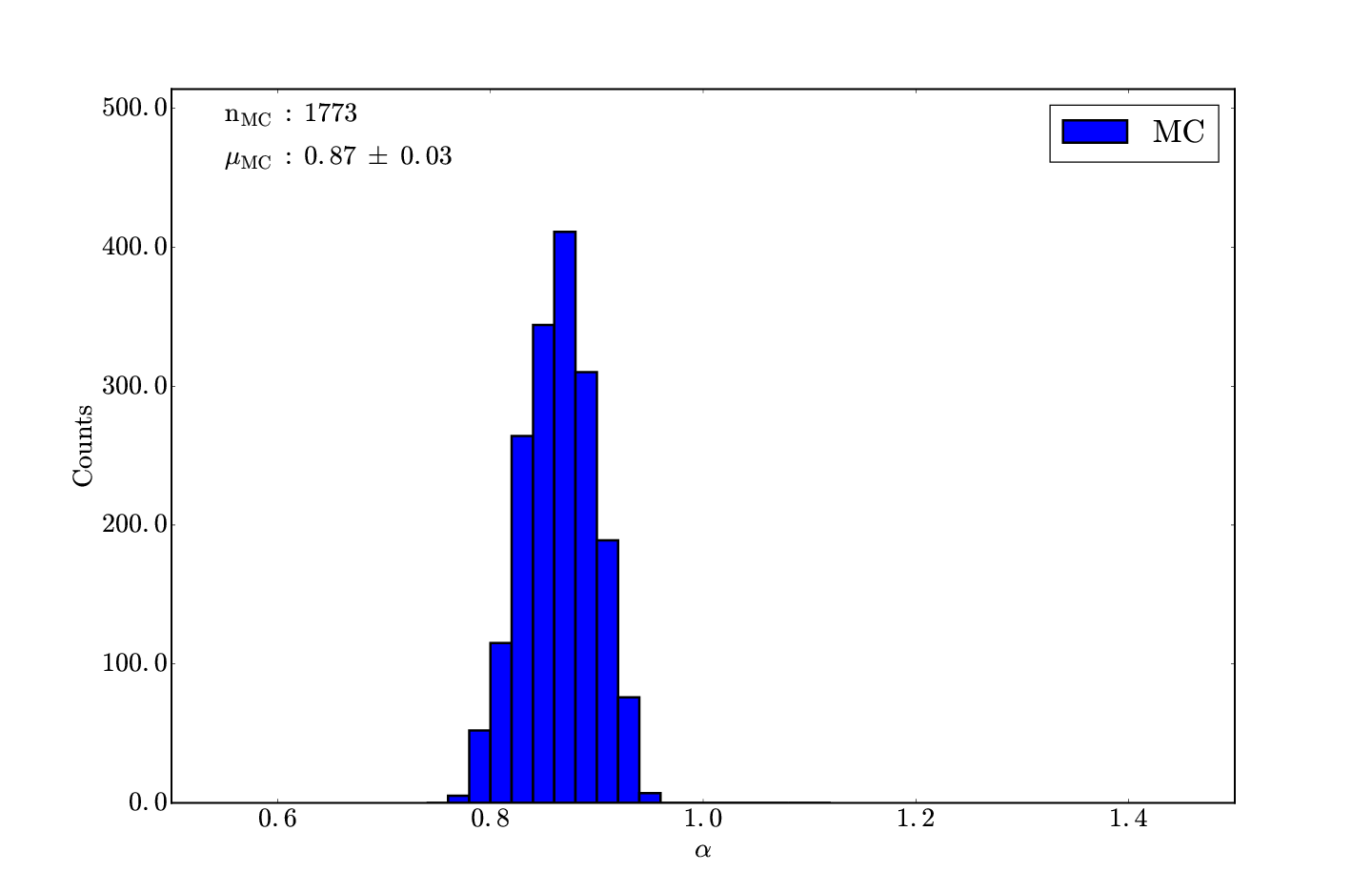}
\else
I am not enabling plots.
\fi
\caption{The same as Fig.~\ref{fig.ErigoneMC} with $\sim$1,800 trials repeating the V-shape technique for the Maria family. The mean of the distribution is centered at $\alpha$ = 0.87 $\pm$ 0.03 and the bin size in the histogram is 0.02.}
\label{fig.MariaMC}
\end{figure}

\begin{figure}
\centering
\hspace*{-1.1cm}
\ifincludeplots
\includegraphics[scale=0.55]{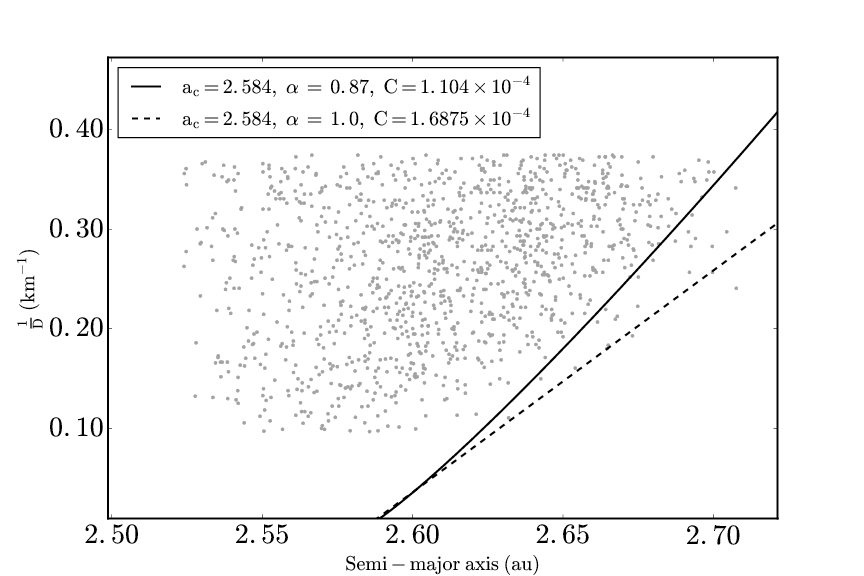}
\else
I am not enabling plots.
\fi
\caption{$a$ vs.$\frac{1}{D}$ plot for Maria with V-shape borders that have $\alpha \; = \; 0.87$ and $\alpha \; = \; 1.0$}
\label{fig.MariaTwoVs}
\end{figure}

The S-type Maria family is located in the central region of the MB and borders the 3:1 MMR with jupiter \citep[][]{Hirayama1922,Zappala1990}. The V-shape identification technique was applied to 1,144 asteroids belonging to the Maria asteroid family as defined by \citet[][]{Nesvorny2015a}. The peak in $\frac{N_{in}^2}{N_{out}}$ at $(a_c, \; C, \; \alpha) \; = \; (2.584 \; \mathrm{au}, \; 1.1 \times 10^{-4} \; \mathrm{au}, \;  \sim0.9)$ as seen in the top panel of Fig.~\ref{fig.MariaAlpha} and is $\sim$5 standard deviations above the mean. The technique was repeated with the Maria family defined by \citet[][]{Milani2014} resulting in similar results as seen in Fig.~\ref{fig.MariaBorderAlphaMilani}.$\sim$1,800 Monte Carlo runs were completed with a mean of $\alpha$ is $\sim$0.87 $\pm$ 0.03 as seen in Fig.~\ref{fig.MariaMC}. The Maria family V-shape is better fit with $\alpha \; = \; 0.87$  than the V-shape with $\alpha \; = \; 1.0$ as seen in Fig.~\ref{fig.MariaTwoVs}.

 The family age of 1.16 $\pm$ 0.58 Gyrs is calculated using Eq.~\ref{eqn.familyagenoejectionpvgsolar}, with $C_{YE}$ = $6.5 \; \times 10^{-5}$ au calculated from Eq.~\ref{eqn.Ccombo} where $C\; = \; 1.1 \; \times 10^{-4}$ au in agreement with the results of \citep[][]{Aljbaae2017a} for the age of the Maria family. The value of $\mu_{\alpha} \; =$ 0.87 and $C_{EV} \; = \; 4.6 \; \times 10^{-5}$ au calculated using Eq.~\ref{eqn.VEVvsCalphaFinal} assuming $V_{EV}$ = 65 $\mps$ which is the escape speed of a 116 km diameter body with $\rho$ = 2.3 $\gpcmc$.
 
\subsubsection{Nemausa}
\label{s.Nemausa}

\begin{figure}
\centering
\hspace*{-0.7cm}
\ifincludeplots
\includegraphics[scale=0.455]{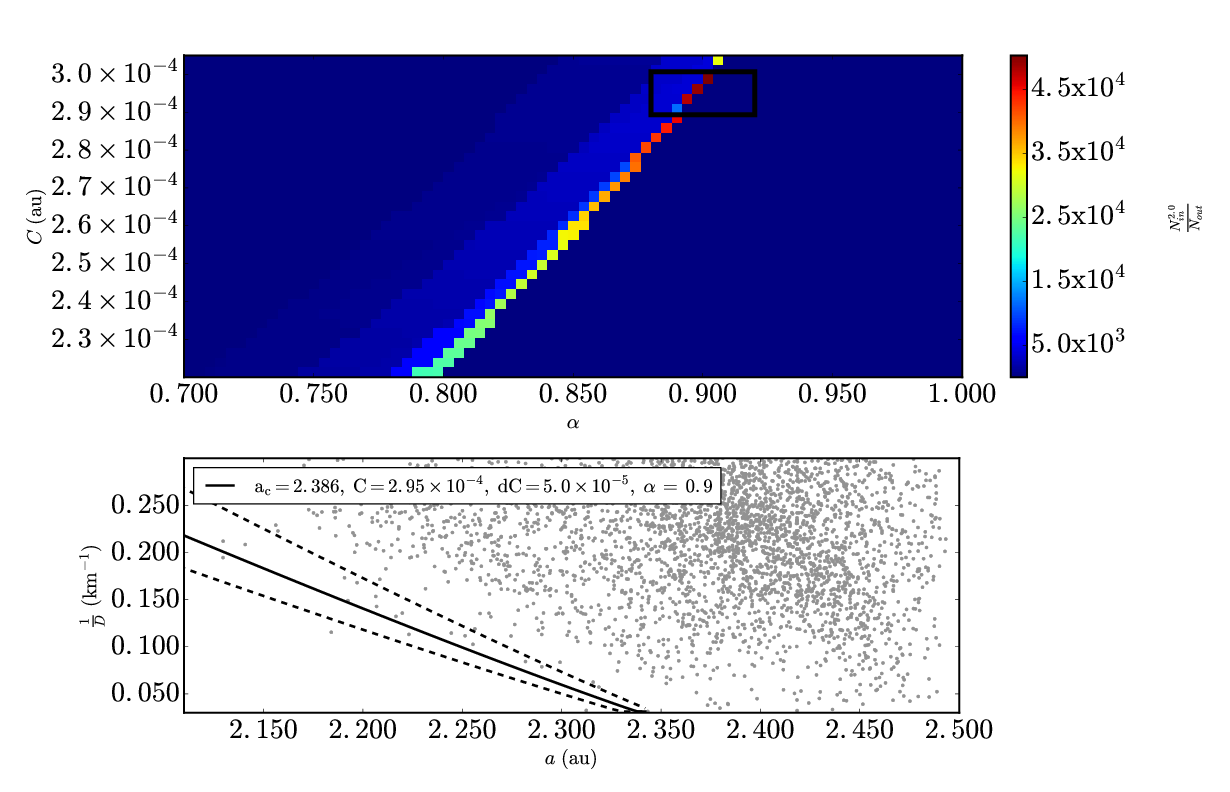}
\else
I am not enabling plots.
\fi
\caption{The same as Fig.~\ref{fig.synErig200Myrs} for Nemausa asteroid family data from \citet[][]{Nesvorny2015a}. (Top panel) $\Delta \alpha$ is equal to $4.0 \times 10^{-3}$ au and $\Delta C$, is equal to $2.5 \times 10^{-6}$ au. (Bottom Panel) $D_r(a,a_c,C\pm dC,\pv,\alpha)$ is plotted with $\pv = 0.05$, $a_c$ = 2.37 au and $dC \; = \; 5.0 \x 10^{-5}$ au.}
\label{fig.NemausaAlpha}
\end{figure}

\begin{figure}
\centering
\hspace*{-0.9cm}
\ifincludeplots
\includegraphics[scale=0.3225]{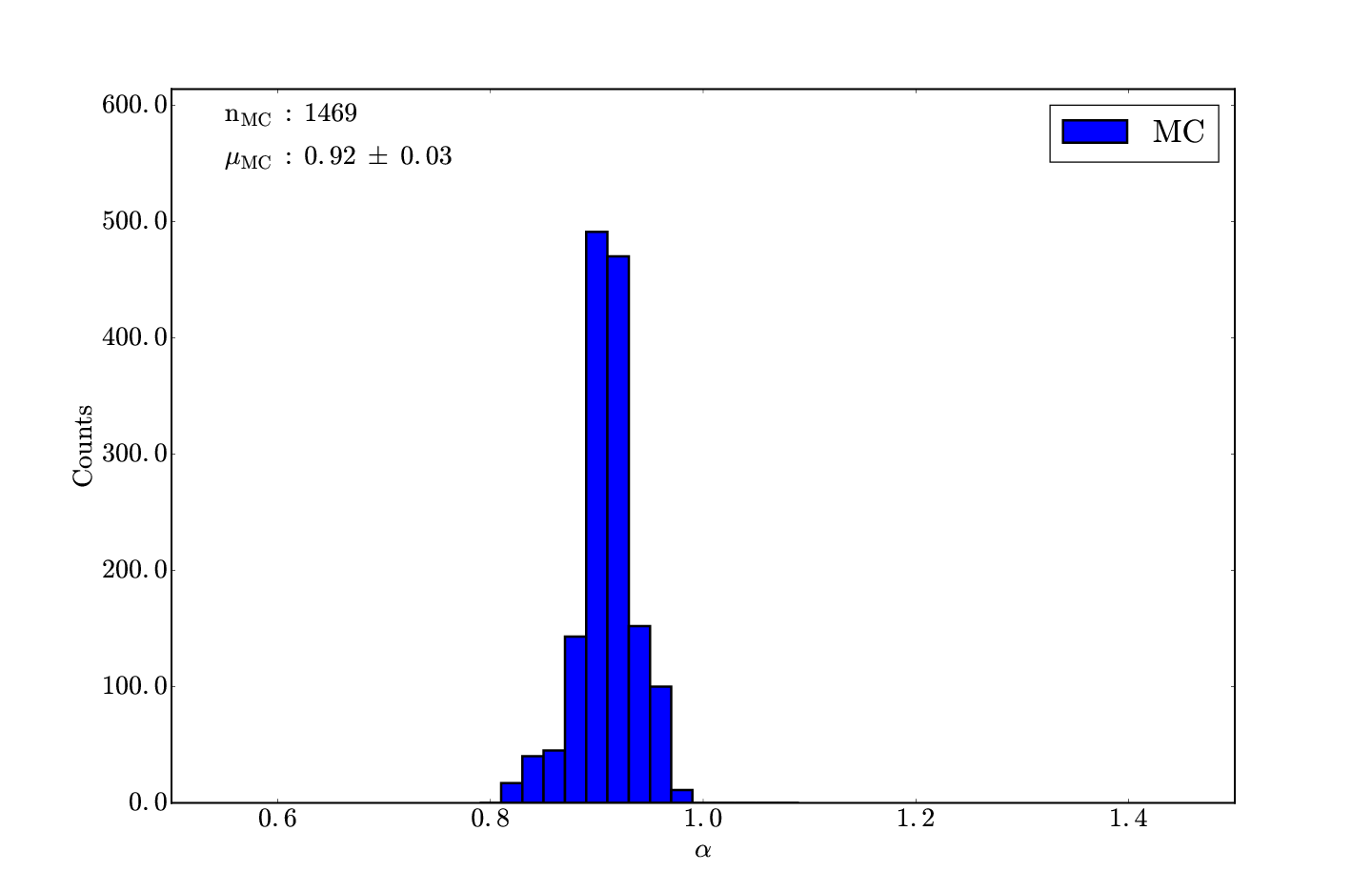}
\else
I am not enabling plots.
\fi
\caption{The same as Fig.~\ref{fig.ErigoneMC} with $\sim$1,500 trials repeating the V-shape technique for the Nemausa family. The mean of the distribution is centered at $\alpha$ = 0.92 $\pm$ 0.03 and the bin size in the histogram is 0.02.}
\label{fig.NemausaMC}
\end{figure}

The inner Main Belt contains a low albedo asteroid family with an age $\sim$4 Gyrs that borders the 3:1 MMR with Jupiter that we will call the Nemausa family \citep[][]{Delbo2017}. The V-shape identification technique was applied to 3,949 asteroids with 0.0 $<$ $e$ $<$ 0.35, 0.0$\tdeg$ $<$ $i$ $<$ 14.5$\tdeg$ and 0.0 $<$ $\pv$ $<$ 0.12. Only asteroids with known $D$ measurements from \citet[][]{Masiero2011} were used. Asteroid $\pv$ values were calculated with $H$ values from \citet[][]{Veres2015} and $D$ from \citet[][]{Masiero2011} according to Eq.~\ref{eq.DHtoPV}. The peak in $\frac{N_{in}^2}{N_{out}}$ at $(a_c, \; C, \; \alpha) \; = \; (2.386 \; \mathrm{au}, \; 2.95 \times 10^{-4} \; \mathrm{au}, \;  \sim0.9)$ as seen in the top panel of Fig.~\ref{fig.NemausaAlpha} and is $\sim$10 standard deviations above the mean. $\sim$1,500 Monte Carlo runs were completed with the mean value of $\alpha$ is $\sim$0.92 $\pm$ 0.03 as seen in Fig.~\ref{fig.NemausaMC}.

 The family age of 4.3 $\pm$ 2.1 Gyrs is calculated using Eq.~\ref{eqn.familyagenoejectionpvgsolar}, with $C_{YE}$ = $2.81 \; \times 10^{-4}$ au calculated from Eq.~\ref{eqn.Ccombo} where $C\; = \; 2.95 \; \times 10^{-4}$ au. The value of $\mu_{\alpha} \; =$ 0.92 and $C_{EV} \; = \; 1.40 \; \times 10^{-5}$ au calculated using Eq.~\ref{eqn.VEVvsCalphaFinal} assuming $V_{EV}$ = 60 $\mps$ which is the escape speed of a 140 km diameter body with $\rho$ = 1.4 $\gpcmc$.
 
\subsubsection{Nemesis}
\label{s.Nemesis}

\begin{figure}
\centering
\hspace*{-0.7cm}
\ifincludeplots
\includegraphics[scale=0.455]{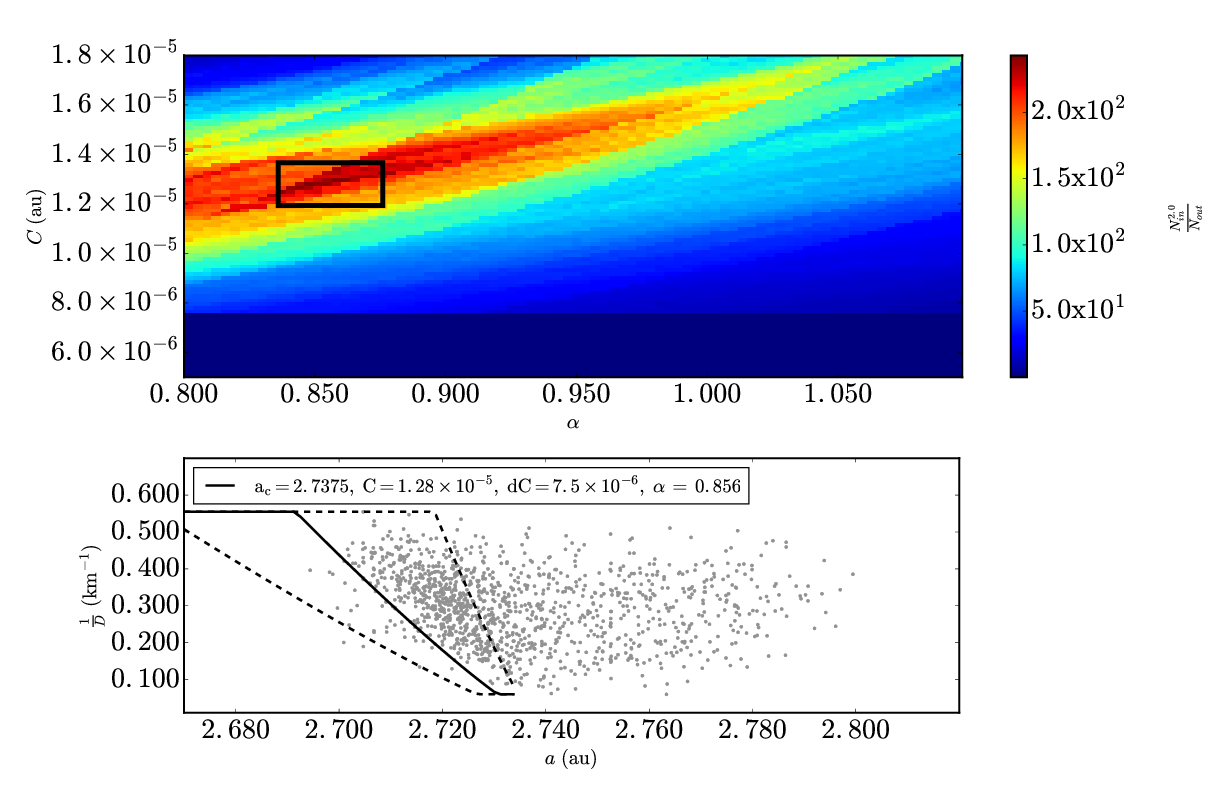}
\else
I am not enabling plots.
\fi
\caption{The same as Fig.~\ref{fig.synErig200Myrs} for Nemesis asteroid family data from \citet[][]{Nesvorny2015a}. (Top panel) $\Delta \alpha$ is equal to $3.5 \times 10^{-3}$ au and $\Delta C$, is equal to $1.5 \times 10^{-7}$ au. (Bottom Panel) $D_r(a,a_c,C\pm dC,\pv,\alpha)$ is plotted with $\pv = 0.05$, $a_c$ = 2.738 au and $dC \; = \; 7.5 \x 10^{-6}$ au.}
\label{fig.NemesisAlpha}
\end{figure}

\begin{figure}
\centering
\hspace*{-0.9cm}
\ifincludeplots
\includegraphics[scale=0.3225]{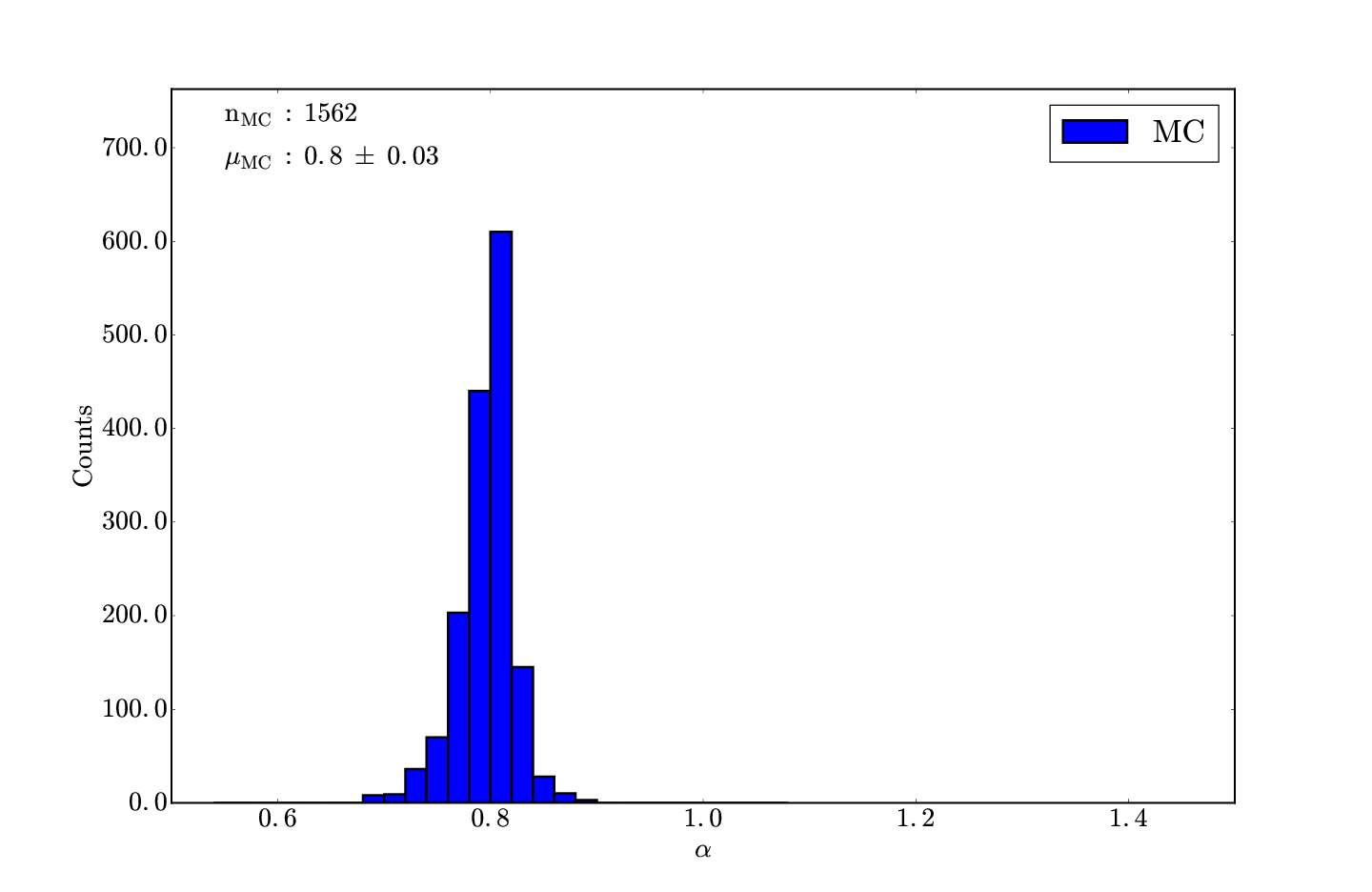}
\else
I am not enabling plots.
\fi
\caption{The same as Fig.~\ref{fig.ErigoneMC} with $\sim$1,600 trials repeating the V-shape technique for the Nemesis family. The mean of the distribution is centered at $\alpha$ = 0.80 $\pm$ 0.03 and the bin size in the histogram is 0.02.}
\label{fig.NemesisMC}
\end{figure}

The C-type Nemesis family is located in the central region of the MB \citep[][]{Bendjoya2002}. Family fragments in the outer V-shape half are depleted as a result of possible close encounters with Ceres \citep[][]{Spoto2015}. In addition, the nodal resonance between Nemsis family members and Ceres may play an important role in dynamically sculpting the asteroid family \citep[][]{Novakovic2015}.The V-shape identification technique was applied to 1,250 asteroids belonging to the Nemesis asteroid family as defined by \citet[][]{Nesvorny2015a}. The peak in $\frac{N_{in}^2}{N_{out}}$ at $(a_c, \; C, \; \alpha) \; = \; (2.738 \; \mathrm{au}, \; 1.28 \times 10^{-5} \; \mathrm{au}, \;  \sim0.856)$ as seen in the top panel of Fig.~\ref{fig.NemesisAlpha} and is $\sim$3 standard deviations above the mean. $\sim$1,600 Monte Carlo runs were completed with the mean value of $\alpha$ is $\sim$0.8 $\pm$ 0.03 as seen in Fig.~\ref{fig.NemesisMC}.

 The family age of 150 $\pm$ 80 Myrs is calculated using Eq.~\ref{eqn.familyagenoejectionpvg}, with $C_{YE}$ = $9.5 \; \times 10^{-6}$ au calculated from Eq.~\ref{eqn.Ccombo} where $C\; = \; 1.3 \; \times 10^{-5}$ au. The value of $\mu_{\alpha} \; =$ 0.8 and $C_{EV} \; = \; 3.3 \; \times 10^{-6}$ au calculated using Eq.~\ref{eqn.VEVvsCalphaFinal} assuming $V_{EV}$ = 12.8 $\mps$ which is the escape speed of a 29 km diameter body with $\rho$ = 1.4 $\gpcmc$. 
 
\subsubsection{New Polana}
\label{s.NewPolana}

\begin{figure}
\centering
\hspace*{-0.7cm}
\ifincludeplots
\includegraphics[scale=0.455]{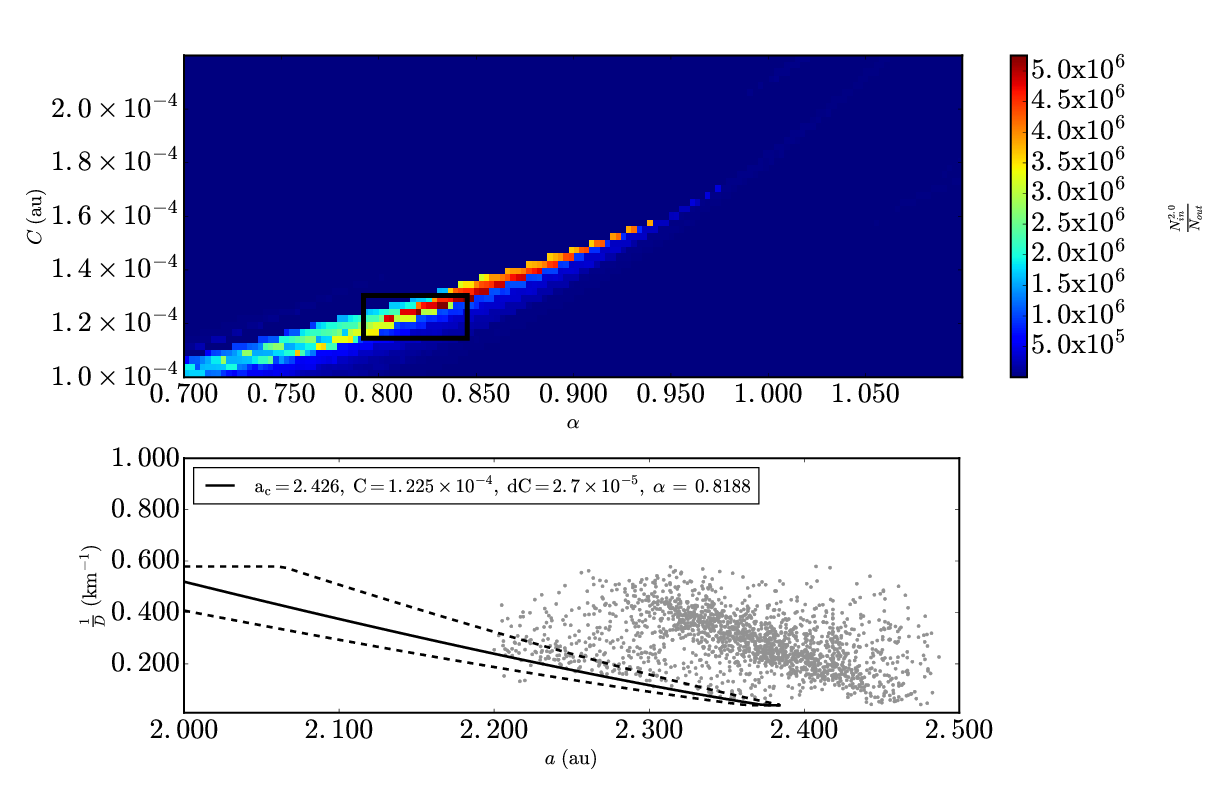}
\else
I am not enabling plots.
\fi
\caption{The same as Fig.~\ref{fig.synErig200Myrs} for New Polana asteroid family data from \citet[][]{Nesvorny2015a}. (Top panel) $\Delta \alpha$ is equal to $2.7 \times 10^{-3}$ au and $\Delta C$, is equal to $2.5 \times 10^{-6}$ au. (Bottom Panel) $D_r(a,a_c,C\pm dC,\pv,\alpha)$ is plotted with $\pv = 0.06$, $a_c$ = 2.426 au and $dC \; = \; 2.7 \x 10^{-5}$ au.}
\label{fig.NewPolanaAlpha}
\end{figure}

\begin{figure}
\centering
\hspace*{-0.9cm}
\ifincludeplots
\includegraphics[scale=0.3225]{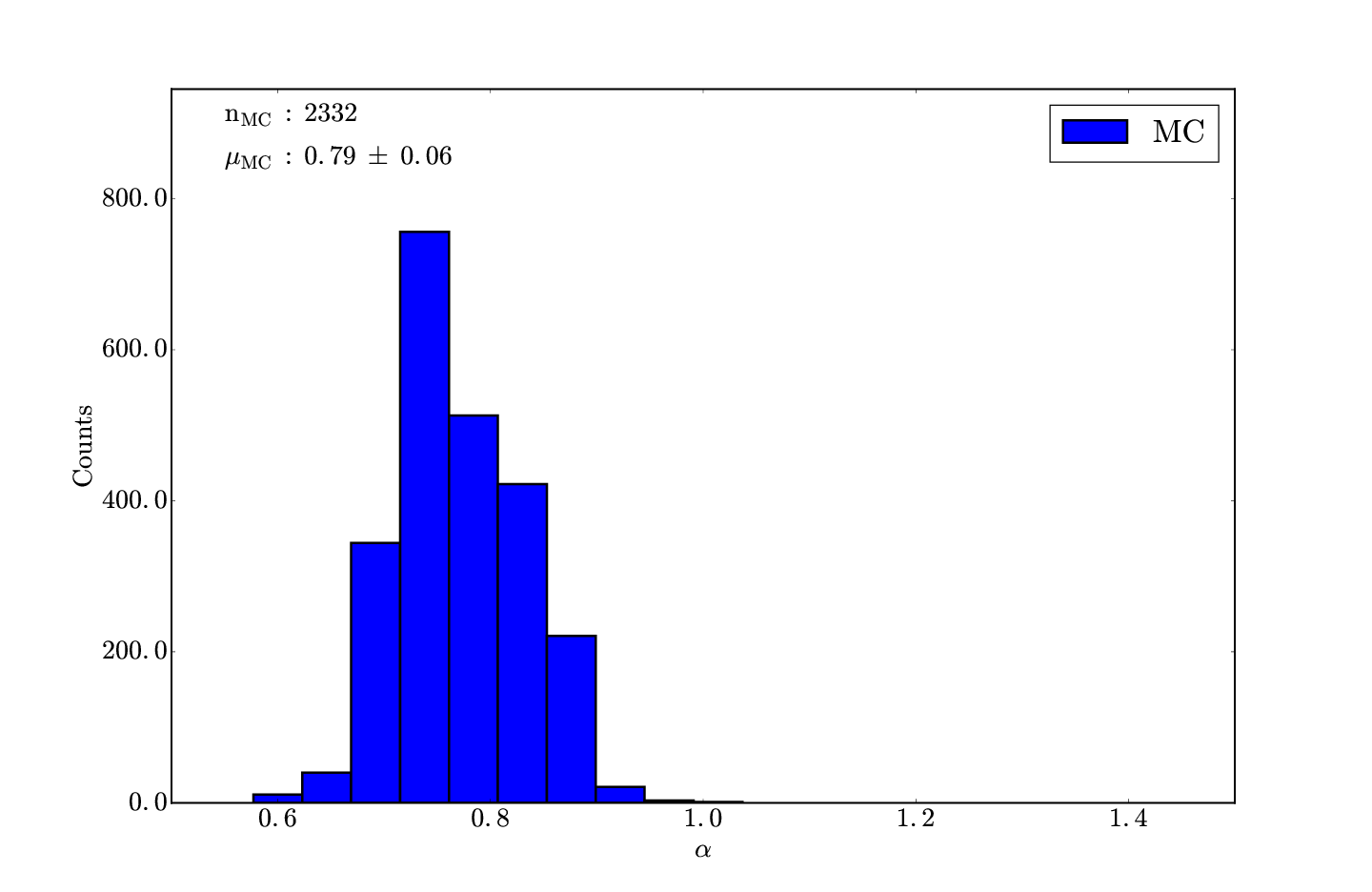}
\else
I am not enabling plots.
\fi
\caption{The same as Fig.~\ref{fig.ErigoneMC} with $\sim$2,300 trials repeating the V-shape technique for the New Polana family. The mean of the distribution is centered at $\alpha$ = 0.79 $\pm$ 0.06 and the bin size in the histogram is 0.05.}
\label{fig.NewPolanaMC}
\end{figure}

\begin{figure}
\centering
\hspace*{-1.1cm}
\ifincludeplots
\includegraphics[scale=0.55]{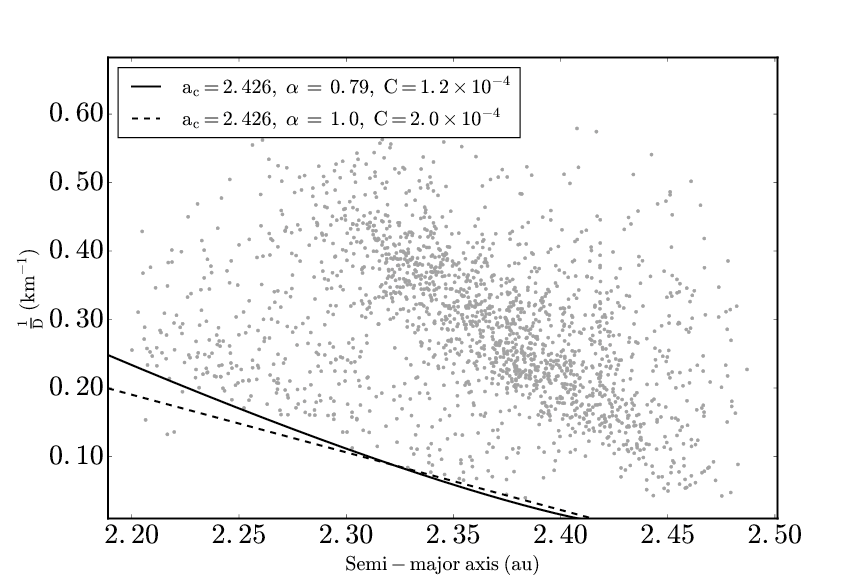}
\else
I am not enabling plots.
\fi
\caption{$a$ vs.$\frac{1}{D}$ plot for New Polana with V-shape borders that have $\alpha \; = \; 0.79$ and $\alpha \; = \; 1.0$}
\label{fig.NewPolanaTwoVs}
\end{figure}

The C-type New Polana family is located in the inner region of the MB and overlaps the 3:1 resonance with Jupiter \citep[][]{Walsh2013}. The V-shape identification technique was applied to 1,818 asteroids belonging to the Nysa-Polana asteroid family as defined by \citet[][]{Nesvorny2015a}. The peak in $\frac{N_{in}^2}{N_{out}}$ at $(a_c, \; C, \; \alpha) \; = \; (2.43 \; \mathrm{au}, \; 1.23 \times 10^{-4} \; \mathrm{au}, \;  \sim0.82)$ as seen in the top panel of Fig.~\ref{fig.NewPolanaAlpha} and is $\sim$9 standard deviations above the mean value. $\sim$2,300 Monte Carlo runs with a mean value of $\alpha$ is $\sim$0.79 $\pm$ 0.06 as seen in Fig.~\ref{fig.NewPolanaMC}. The New Polana family V-shape is better fit with $\alpha \; = \; 0.83$  than the V-shape with $\alpha \; = \; 1.0$ as seen in Fig.~\ref{fig.NewPolanaTwoVs}.

 The family age of 2.1 $\pm$ 1.0 Gyrs is calculated using Eq.~\ref{eqn.familyagenoejectionpvgsolar}, with $C_{YE}$ = $1.1 \; \times 10^{-4}$ au calculated from Eq.~\ref{eqn.Ccombo} where $C\; = \; 1.2 \; \times 10^{-5}$ au. The value of $\mu_{\alpha} \; =$ 0.79 and $C_{EV} \; = \; 1.3 \; \times 10^{-5}$ au calculated using Eq.~\ref{eqn.VEVvsCalphaFinal} assuming $V_{EV}$ = 57 $\mps$ which is the escape speed of a 130 km diameter body with $\rho$ = 1.4 $\gpcmc$. The calculation was repeating using the same parameters except with $\alpha$ = 1.0 and $C\; = \; 2.0 \; \times 10^{-4}$ obtaining a value of 2.6 $\pm$ 1.3 Gyrs.

\subsubsection{Rafita}
\label{s.Rafita}

\begin{figure}
\centering
\hspace*{-0.7cm}
\ifincludeplots
\includegraphics[scale=0.455]{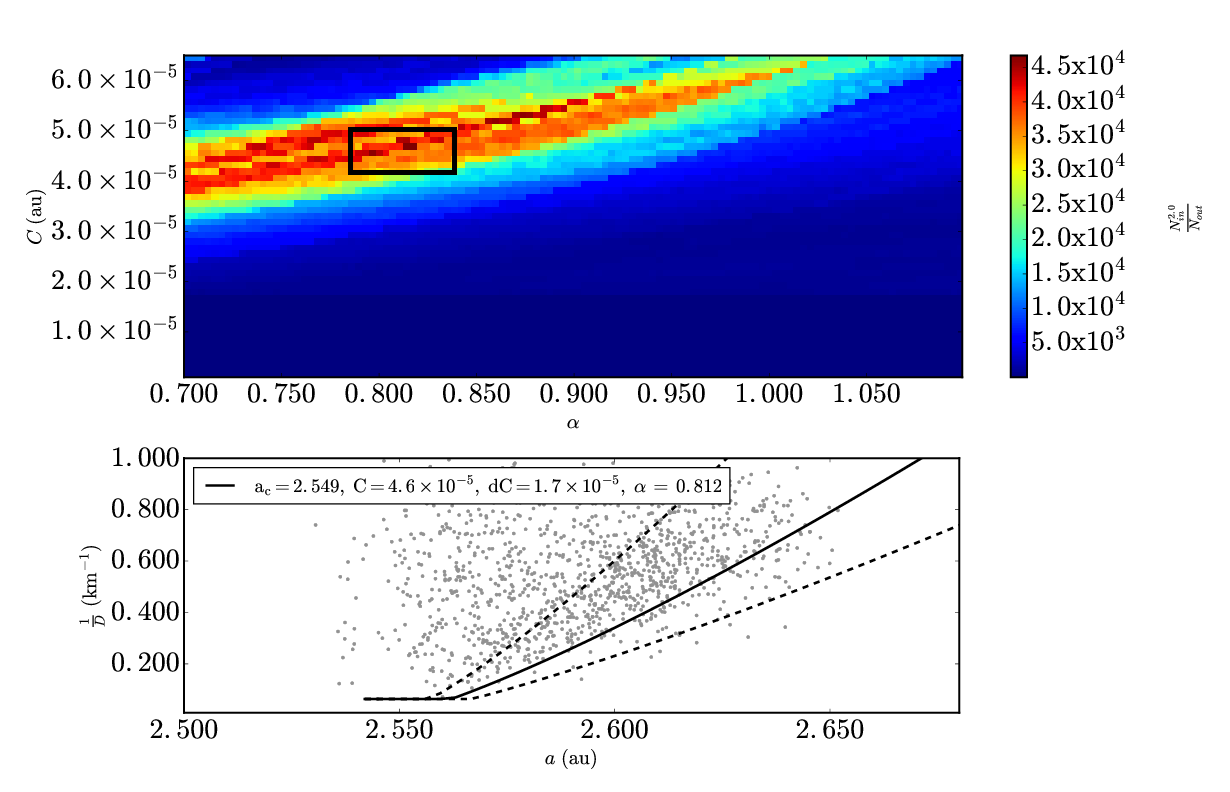}
\else
I am not enabling plots.
\fi
\caption{The same as Fig.~\ref{fig.synErig200Myrs} for Rafita asteroid family data from \citet[][]{Nesvorny2015a}. (Top panel) $\Delta \alpha$ is equal to $8.1 \times 10^{-3}$ au and $\Delta C$, is equal to $2.5 \times 10^{-6}$ au. (Bottom Panel) $D_r(a,a_c,C\pm dC,\pv,\alpha)$ is plotted with $\pv = 0.25$, $a_c$ = 2.549 au and $dC \; = \; 1.7 \times 10^{-5}$ au.}
\label{fig.RafitaAlpha}
\end{figure}

\begin{figure}
\centering
\hspace*{-0.9cm}
\ifincludeplots
\includegraphics[scale=0.3225]{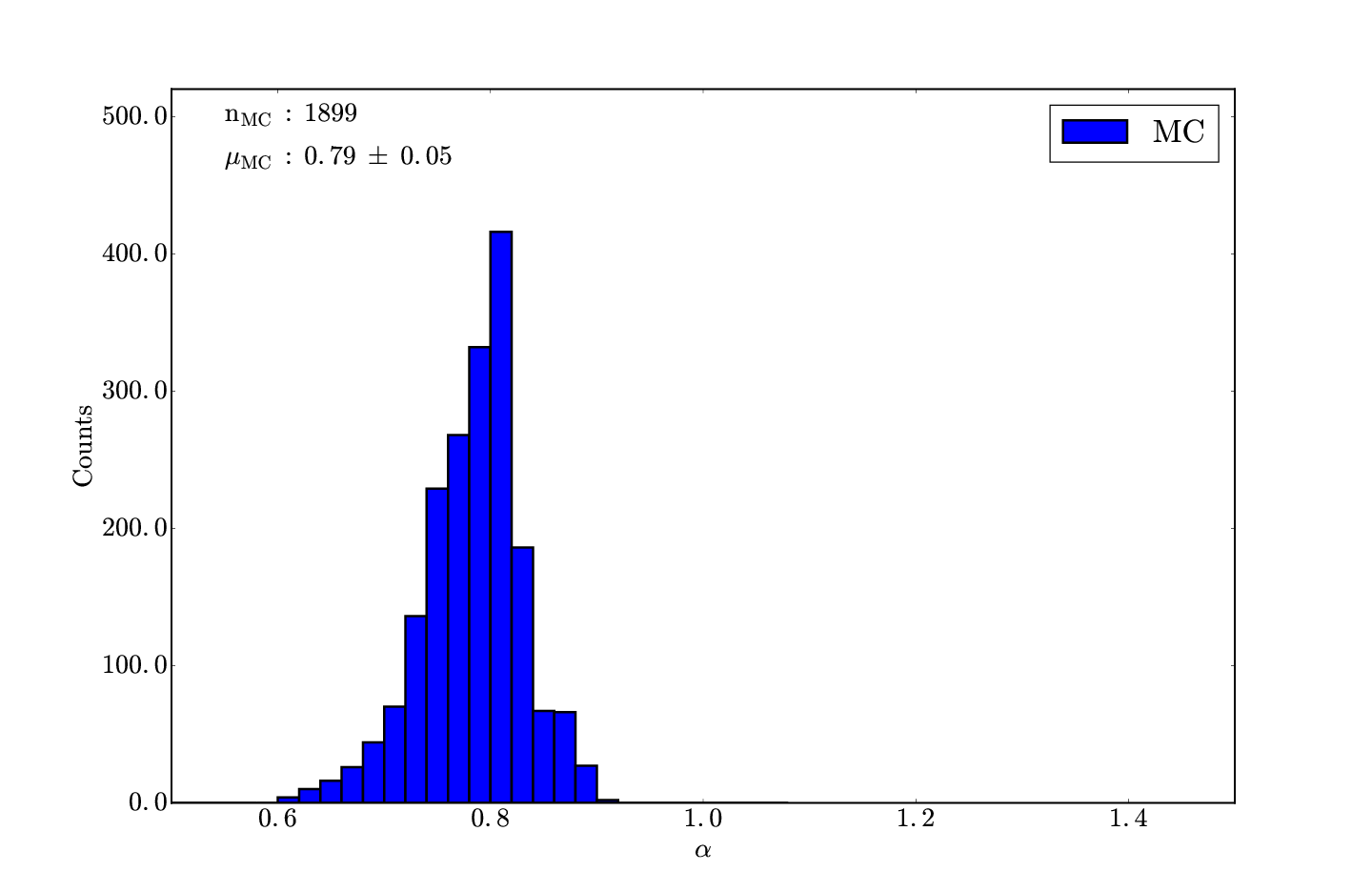}
\else
I am not enabling plots.
\fi
\caption{The same as Fig.~\ref{fig.ErigoneMC} with $\sim$1,900 trials repeating the V-shape technique for the Rafita family. The mean of the distribution is centered at $\alpha$ = 0.79 $\pm$ 0.05 and the bin size in the histogram is 0.02.}
\label{fig.RafitaMC}
\end{figure}

\begin{figure}
\centering
\hspace*{-1.1cm}
\ifincludeplots
\includegraphics[scale=0.55]{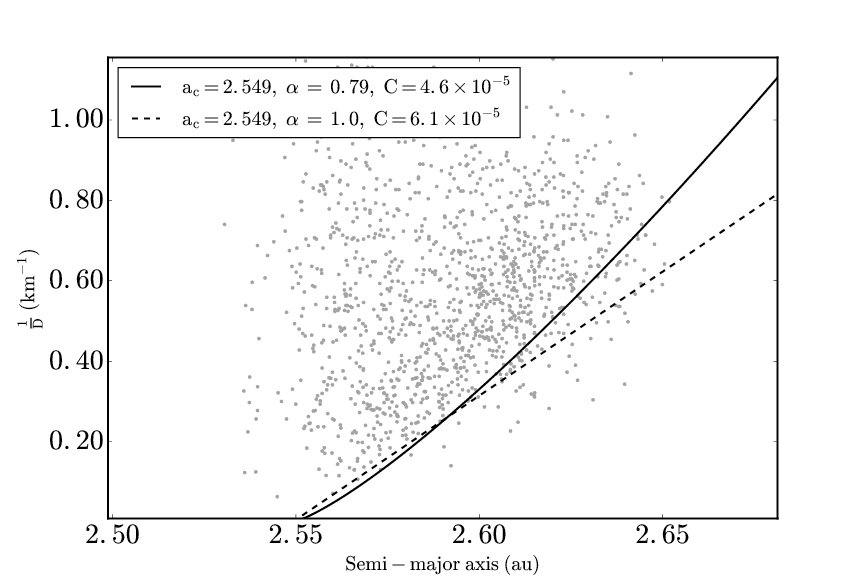}
\else
I am not enabling plots.
\fi
\caption{$a$ vs.$\frac{1}{D}$ plot for Rafita with V-shape borders that have $\alpha \; = \; 0.79$ and $\alpha \; = \; 1.0$}
\label{fig.RafitaTwoVs}
\end{figure}

The S-type Rafita family is located in the central region of the MB and borders the 3:1 MMR with Jupiter \citep[][]{Zappala1990}. The V-shape identification technique was applied to 1,251 asteroids belonging to the Rafita asteroid family as defined by \citet[][]{Nesvorny2015a}. The interval [$0.10,1.32$] for the Dirac delta function $\delta(D_{r,j}-D_r )$ is used and Eq.~\ref{eqn.apvDvsCfinal} is truncated to 0.06 km$^{-1}$ for $D_r$ $ <$ $0.06$ km$^{-1}$ and to 1.71 km$^{-1}$ for $D_r$ $>$ 1.71 km$^{-1}$. Asteroid $H$ values were converted to $D$ using Eq.~\ref{eq.HtoD} and $\pv$ = 0.25 typical for members of the Rafita family \citep[][]{Masiero2013,Spoto2015}. The peak in $\frac{N_{in}^2}{N_{out}}$ at $(a_c, \; C, \; \alpha) \; = \; (2.549 \; \mathrm{au}, \; 4.6 \times 10^{-5} \; \mathrm{au}, \;  \sim0.81)$ as seen in the top panel of Fig.~\ref{fig.RafitaAlpha} and is $\sim$3 standard deviations above the mean. $\sim$1,900 Monte Carlo runs were completed with a mean value of $\alpha$ is $\sim$0.79 $\pm$ 0.05 as seen in Fig.~\ref{fig.RafitaMC}. The Rafita family V-shape is better fit with $\alpha \; = \; 0.79$  than the V-shape with $\alpha \; = \; 1.0$ as seen in Fig.~\ref{fig.RafitaTwoVs}.

 The family age of 380 $\pm$ 190 Myrs is calculated using Eq.~\ref{eqn.familyagenoejectionpvg}, with $C_{YE}$ = $4.0 \; \times 10^{-5}$ au calculated from Eq.~\ref{eqn.Ccombo} where $C\; = \; 4.6 \; \times 10^{-5}$ au which overlaps with the 300-700 Myr estimate of \citep[][]{Aljbaae2017}. The value of $\mu_{\alpha} \; =$ 0.79 and $C_{EV} \; = \; 6.1 \; \times 10^{-6}$ au calculated using Eq.~\ref{eqn.VEVvsCalphaFinal} assuming $V_{EV}$ = 12 $\mps$ which is the escape speed of a 27 km diameter body with $\rho$ = 1.4 $\gpcmc$.
 
\subsubsection{Sulamitis}
\label{s.Sulamitis}

\begin{figure}
\centering
\hspace*{-0.7cm}
\ifincludeplots
\includegraphics[scale=0.455]{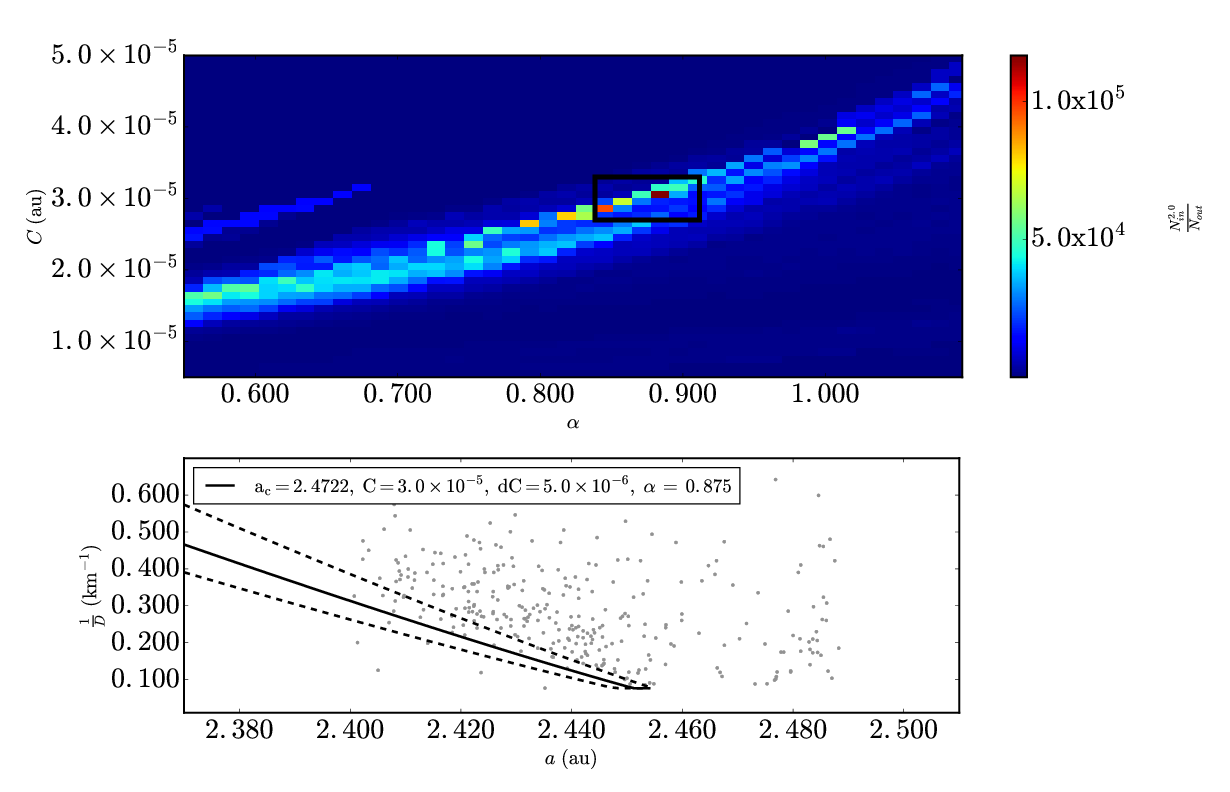}
\else
I am not enabling plots.
\fi
\caption{The same as Fig.~\ref{fig.synErig200Myrs} for Sulamitis asteroid family data from \citet[][]{Nesvorny2015a}. (Top panel) $\Delta \alpha$ is equal to $1.3 \times 10^{-2}$ au and $\Delta C$, is equal to $1.0 \times 10^{-6}$ au. (Bottom Panel) $D_r(a,a_c,C\pm dC,\pv,\alpha)$ is plotted with $\pv = 0.04$, $a_c$ = 2.472 au and $dC \; = \; 5.0 \x 10^{-6}$ au.}
\label{fig.SulamitisAlpha}
\end{figure}

\begin{figure}
\centering
\hspace*{-0.9cm}
\ifincludeplots
\includegraphics[scale=0.3225]{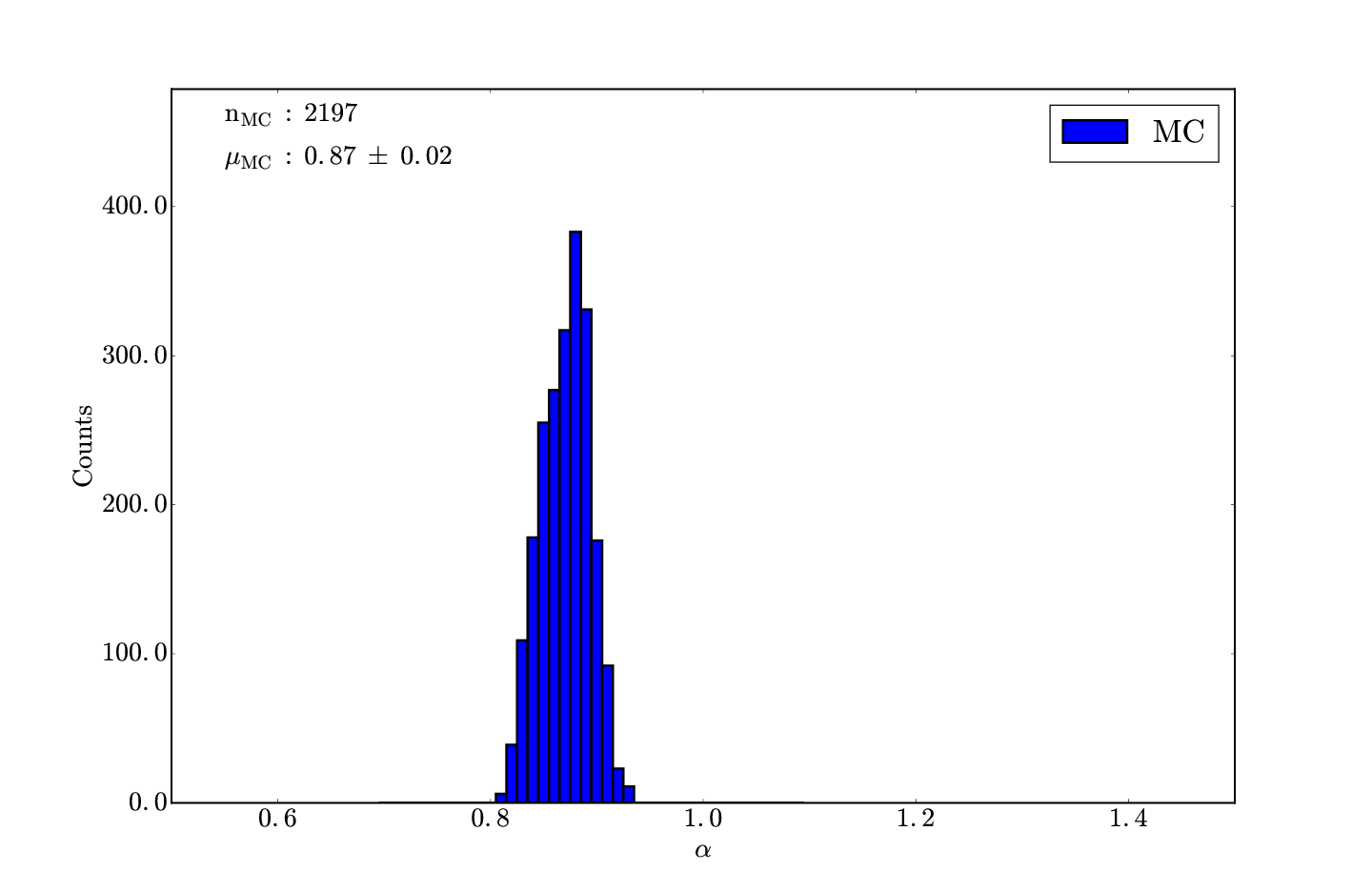}
\else
I am not enabling plots.
\fi
\caption{The same as Fig.~\ref{fig.ErigoneMC} with $\sim$2,200 trials repeating the V-shape technique for the Sulamitis family. The mean of the distribution is centered at $\alpha$ = 0.87 $\pm$ 0.02 and the bin size in the histogram is 0.01.}
\label{fig.SulamitisMC}
\end{figure}

The C-type Sulamitis family is located in the inner region of the MB and borders the 3:1 resonance with Jupiter \citep[][]{Zappala1995}. The V-shape identification technique was applied to 284 asteroids belonging to the Sulamitis asteroid family as defined by \citet[][]{Nesvorny2015a}. The peak in $\frac{N_{in}^2}{N_{out}}$ at $(a_c, \; C, \; \alpha) \; = \; (2.472 \; \mathrm{au}, \; 3.0 \times 10^{-5} \; \mathrm{au}, \;  \sim0.875)$ as seen in the top panel of Fig.~\ref{fig.SulamitisAlpha} and is $\sim$11 standard deviations above the mean value. $\sim$2,200 Monte Carlo runs with a mean value of $\alpha$ is $\sim$0.87 $\pm$ 0.02 as seen in Fig.~\ref{fig.SulamitisMC}. The family age of 470 $\pm$ 230 Myrs is calculated using Eq.~\ref{eqn.familyagenoejectionpvg}, with $C_{YE}$ = $2.4 \; \times 10^{-5}$ au calculated from Eq.~\ref{eqn.Ccombo} where $C\; = \; 3.0 \; \times 10^{-5}$ au.

\subsubsection{Ursula}
\label{s.Ursula}

\begin{figure}
\centering
\hspace*{-0.7cm}
\ifincludeplots
\includegraphics[scale=0.455]{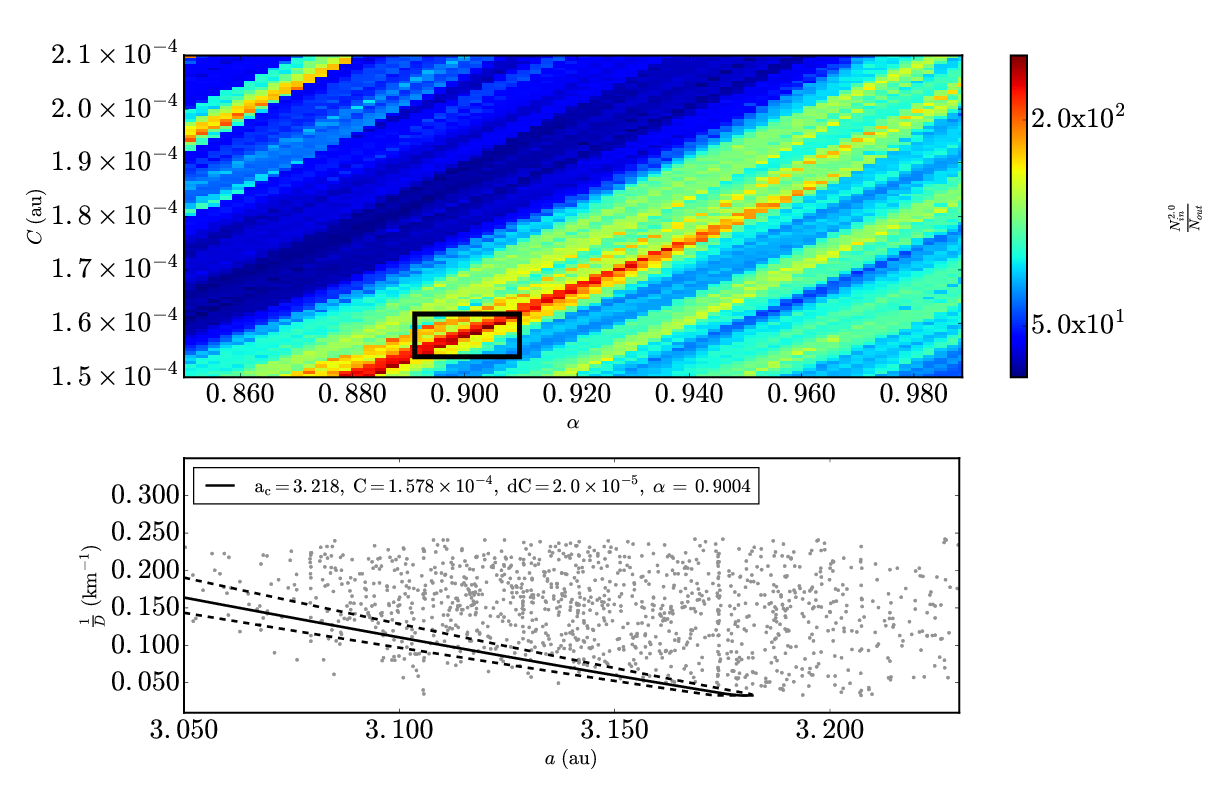}
\else
I am not enabling plots.
\fi
\caption{The same as Fig.~\ref{fig.synErig200Myrs} for Ursula asteroid family data from \citet[][]{Nesvorny2015a}. (Top panel) $\Delta \alpha$ is equal to $2.1 \times 10^{-3}$ au and $\Delta C$, is equal to $6.0 \times 10^{-7}$ au. (Bottom Panel) $D_r(a,a_c,C\pm dC,\pv,\alpha)$ is plotted with $\pv = 0.06$, $a_c$ = 3.218 au and $dC \; = \; 2.0 \x 10^{-5}$ au.}
\label{fig.UrsulaAlpha}
\end{figure}

\begin{figure}
\centering
\hspace*{-0.9cm}
\ifincludeplots
\includegraphics[scale=0.3225]{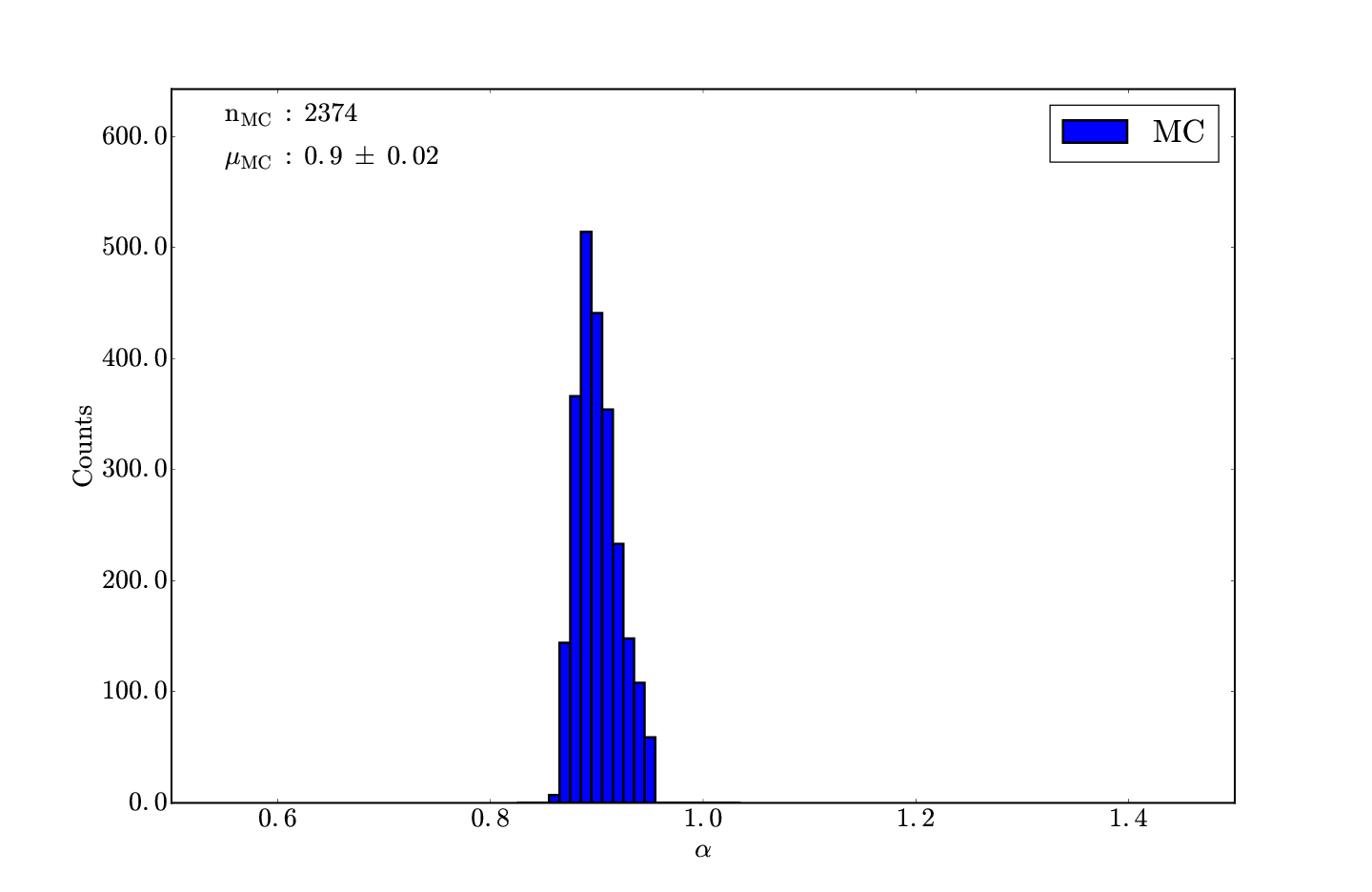}
\else
I am not enabling plots.
\fi
\caption{The same as Fig.~\ref{fig.ErigoneMC} with $\sim$2,300 trials repeating the V-shape technique for the Ursula family. The mean of the distribution is centered at $\alpha$ = 0.90 $\pm$ 0.02 and the bin size in the histogram is 0.01.}
\label{fig.UrsulaMC}
\end{figure}

The C-type Ursula family is located in the outer region of the MB and borders the 2:1 MMR with Jupiter at 3.2 au \citep[][]{Zappala1995}. The V-shape identification technique was applied to 1,209 asteroids belonging to the Ursula asteroid family as defined by \citet[][]{Nesvorny2015a}. The interval [$0.03,0.24$] for the Dirac delta function $\delta(D_{r,j}-D_r )$ is used and Eq.~\ref{eqn.apvDvsCfinal} is truncated to 0.03 km$^{-1}$ for $D_r$ $ <$ $0.03$ km$^{-1}$ and to 0.24 km$^{-1}$ for $D_r$ $>$ 0.24 km$^{-1}$. Asteroid $H$ values were converted to $D$ using Eq.~\ref{eq.HtoD} and $\pv$ = 0.06 typical for members of the Ursula family \citep[][]{Masiero2013,Spoto2015}. The peak in $\frac{N_{in}^2}{N_{out}}$ at $(a_c, \; C, \; \alpha) \; = \; (2.79 \; \mathrm{au}, \; 1.54 \times 10^{-5} \; \mathrm{au}, \;  \sim0.91)$ as seen in the top panel of Fig.~\ref{fig.UrsulaAlpha} and is $\sim$6 standard deviations above the mean.
$\sim$2,374 with a mean value of $\alpha$ is $\sim$0.90 $\pm$ 0.02 as seen in Fig.~\ref{fig.UrsulaMC}.  The family age of 2.3 $\pm$ 1.1 Gyrs is calculated using Eq.~\ref{eqn.familyagenoejectionpvgsolar}, with $C_{YE}$ = $1.2 \; \times 10^{-4}$ au calculated from Eq.~\ref{eqn.Ccombo} where $C\; = \; 1.6 \; \times 10^{-4}$ au and overlaps with the estimate from \citet[][]{Carruba2016c} of 1$\sim$4 Gyrs.

\begin{acknowledgements}
We would like to thank the reviewer of our manuscript, Valerio Carruba, for providing helpful comments and suggestions for improving the quality of the text. B.T. Bolin is supported by l'\`{E}cole Doctorale Sciences Fondatementales et Appliqu\'{e}es, ED.SFA (ED 364) at l'Universit\'{e} de Nice-Sophia Antipolis. K.J. Walsh was supported by the National Science Foundation, Grant 1518127. BTB would like to acknowledge J.W. Westover for thought-provoking discussions on the implementation of large-scale computing resources and algorithms that were used in the completion of this work.
\end{acknowledgements}


\bibliographystyle{aa}
\bibliography{references}

\end{document}